%% file: draft_arxiv.tex
\documentclass[aps,prl,twocolumn,superscriptaddress]{revtex4-2}
\input{preamble_arxiv}

\begin{document}

\title{Duality between dissipation-coherence trade-off and thermodynamic speed limit based on thermodynamic uncertainty relation for stochastic limit cycles}

\author{Ryuna Nagayama}
\email{ryuna.nagayama@ubi.s.u-tokyo.ac.jp}
\affiliation{Department of Physics, Graduate School of Science, The University of Tokyo, 7-3-1 Hongo, Bunkyo-ku, Tokyo 113-0033, Japan}

\author{Sosuke Ito}
\affiliation{Department of Physics, Graduate School of Science, The University of Tokyo, 7-3-1 Hongo, Bunkyo-ku, Tokyo 113-0033, Japan}
\affiliation{Universal Biology Institute, Graduate School of Science, The University of Tokyo, 7-3-1 Hongo, Bunkyo-ku, Tokyo 113-0033, Japan}
\date{\today}

\begin{abstract}
We derive two fundamental trade-offs for general stochastic limit cycles in the weak-noise limit. The first is the dissipation-coherence trade-off, which was \add{discovered and proved under additional assumptions} by Santolin and Falasco~[\href{https://doi.org/10.1103/vfdp-sdfm}{Phys. Rev. Lett. \textbf{135}, 057101 (2025)}]. This trade-off bounds the entropy production required for one oscillatory period using the number of oscillations that occur before steady-state correlations are disrupted. The second is the thermodynamic speed limit, which bounds the entropy production by the Euclidean length of the limit cycle. These trade-offs are obtained by substituting mutually dual observables, derived from the stability of the limit cycle, into the thermodynamic uncertainty relation. This fact allows us to regard the dissipation-coherence trade-off as the dual of the thermodynamic speed limit. We numerically demonstrate these trade-offs using the noisy R\"{o}ssler model. We also apply the trade-offs to stochastic chemical systems, where the diffusion coefficient matrix may contain zero eigenvalues. \add{Furthermore, we show that the dissipation-coherence trade-off is always achievable by appropriately modifying the diffusion coefficient matrix based on the phase reduction.}
\end{abstract}

\maketitle

\textit{Introduction}.---% 
The development of stochastic thermodynamics~\cite{schnakenberg1976network,sekimoto2010stochastic,seifert2012stochastic,Shiraishi2023,falasco2025macroscopic} allows us to clarify the relationship between a system's function and its associated energetic dissipation, measured with the entropy production (EP). The prominent examples are thermodynamic uncertainty relations (TURs)~\cite{barato2015thermodynamic,horowitz2020thermodynamic} and thermodynamic speed limits (TSLs)~\cite{sekimoto1997complementarity, aurell2012refined, shiraishi2018speed, chen2019stochastic, nakazato2021geometrical, ito2023geometric}. The TURs establish the trade-off relation between the EP and the precision of currents: higher precision requires a larger EP~\cite{Gingrich2016,horowitz2017proof,maes2017frenetic,proesmans2017discrete,PietzonkaFT2017,Dechantccurrent2018,DechantMulti2018,TimpanaroEFT2019,potts2019thermodynamic,Hasegawa2019,hasegawa2021thermodynamic,DechantImproving2021,shiraishi2021optimal,falasco2020unifying,DechantFRI2020,otsubo2020estimating,Liu2020,lee2021universal,shiraishi2021optimal,dechant2022geometric_E,yoshimura2021thermodynamic,yoshimura2023housekeeping,Hasegawa2023,kwon2024unified,aslyamov2025nonequilibrium,kwon2024fluctuation,van2025fundamental,kolchinsky2024generalized,nagayama2023geometric,yoshimura2024two,yoshimura2025force}. The TSLs provide another trade-off relation between the EP and the speed of the time evolution: faster transition requires a larger EP~\cite{dechant2019thermodynamic,falasco2020dissipation,van2021geometrical,lee2022speed,van2023topological,dechant2022minimum,van2023thermodynamic,miangolarra2023minimal,sabbagh2024wasserstein,nagayama2025infinite,dechant2022geometric_E,yoshimura2021thermodynamic,yoshimura2023housekeeping,kolchinsky2024generalized,nagayama2023geometric}. These two types of trade-off are closely related. We can derive TSLs and TURs from certain common inequalities~\cite{vo2020unified,van2022unified,Falasco2022,Hasegawa2023,kwon2024unified}. We can also obtain the TSL as a consequence of the short-time limit of the TUR (so called the short-time TUR~\cite{otsubo2020estimating})~\cite{nagayama2023geometric}.

Another function attracting attention in relation to thermodynamic costs is the maintenance of coherence, \add{i.e., the phase correlation,} in noisy oscillations~\cite{barkai2000circadian,gaspard2002correlation}. Noisy oscillations are ubiquitous in biological phenomena~\cite{ferrell2011modeling,goldbeter2008biological}. To maintain the biological functions, \add{the loss of phase correlation} must be avoided, which requires thermodynamic costs. These costs are actively studied~\cite{qian2000pumped,cao2015free,nguyen2018phase,oberreiter2019subharmonic,uhl2019affinity,del2020robust,del2020high,xu2025thermodynamic}. One prominent proposition is the dissipation-coherence trade-off, which states that the EP required for one oscillatory period has a lower bound proportional to the number of coherent oscillations. This trade-off was numerically conjectured in Markov jump processes~\cite{oberreiter2022universal} and has been used to estimate the dissipation in biological phenomena~\cite{chen2024energy}. Despite significant theoretical developments in this area~\cite{barato2017coherence,ohga2023thermodynamic,shiraishi2023entropy,kolchinsky2024thermodynamic,pietzonka2024thermodynamic,gao2024thermodynamic,zheng2024topological}, this dissipation-coherence trade-off remains unproven. The relationship between the dissipation-coherence trade-off and other trade-offs, such as TURs and TSLs, is also unclear.

\begin{figure}
    \centering
    \includegraphics[width=\linewidth]{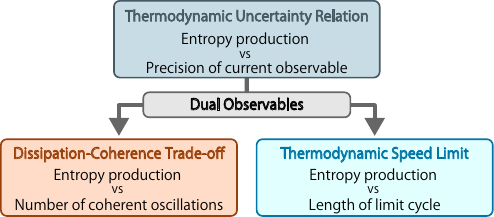}
    \caption{Duality of the dissipation-coherence trade-off and the TSL. We derive these trade-offs by substituting dual observables into the (short-time) TUR.}
    \label{fig:duality}
\end{figure}

Recently, Santolin and Falasco~\cite{santolin2025dissipation} discovered the same dissipation-coherence trade-off for the stochastic limit cycles in the weak-noise limit~\cite{kurrer1991effect,vance1996fluctuations,gaspard2002correlation,xiao2008entropy,cheng2021stochastic,remlein2022coherence} under one of the following conditions: (i) the diffusion coefficient matrix is proportional to the identity matrix, or (ii) the system is near the critical point of the Hopf bifurcation. However, an analytic proof for systems without these conditions is still lacking; only numerical validation was performed. The systems not covered by their proof are common in nature: The diffusion coefficient matrices that are not proportional to the identity matrix naturally arise in chemical reaction networks (CRNs) in the presence of noise~\cite{gillespie2000chemical,yoshimura2021thermodynamic} and the Brownian motion in anisotropic or inhomogeneous media~\cite{van1988diffusion,siggia2000diffusion,basser1994mr,hofling2013anomalous}. Oscillations far from the critical point of the bifurcation are also essential to achieve sufficient amplitude~\cite{Goldbeter1996,strogatz2018nonlinear,murray2003mathematical}. Therefore, it is important to determine whether the trade-off holds in the general case.

In this Letter, we derive the dissipation-coherence trade-off for the stochastic limit cycles in the weak-noise limit by substituting an observable into the short-time TUR~\cite{otsubo2020estimating}. \add{Our derivation is also applicable to systems without the conditions imposed in the proof by Santolin and Falasco~\cite{santolin2025dissipation}.} We also derive the TSL for the stochastic limit cycles in the weak-noise limit by substituting another observable, which is the dual of the one for the dissipation-coherence trade-off, into the TUR. This implies that, in the weak-noise limit, the dissipation-coherence trade-off is the dual of the TSL in terms of the observables used in the TUR as summarized in Fig.~\ref{fig:duality}. These results are based on the positive definiteness of the diffusion coefficient matrix. However, we extend the results to CRNs in the presence of noise that may not satisfy this condition. \add{In addition, we construct the diffusion coefficient matrix that saturates the dissipation-coherence trade-off based on the phase reduction~\cite{winfree1967biological,Kuramoto1984,monga2019phase, brown2004phase, Hoppensteadt1997, shirasaka2017phase}.}

%This proves the numerical conjecture by Santolin and Falasco~\cite{santolin2025dissipation} because our derivation is also applicable to systems without the conditions they imposed.

\textit{System and formulation}.---%
Here we describe our physical setup and define the quantities that appear in our result. In the following, a single dot and double dots over a variable imply the first and second time derivatives, respectively. We consider the time evolution of the state $\bm{x}_t\in\mathbb{R}^N$ according to the following overdamped Langevin equation:
\begin{align}
    \dot{\bm{x}}_t=\bm{F}(\bm{x}_t)+\sqrt{2\epsilon}\mathsf{G}(\bm{x}_t)\bullet\bm{\xi}_t.
    \label{Langevin eq}
\end{align}
Here $\bm{F}(\bm{x})$ denotes the value of the force field at position $\bm{x}$. We also let $\bm{\xi}_t$ be the $M$-dimensional white Gaussian noise, which satisfies $\langle(\bm{\xi}_t)_{\rho}\rangle=0$ and $\langle(\bm{\xi}_t)_{\rho}(\bm{\xi}_0)_{\rho'}\rangle=\delta_{\rho\rho'}\delta(t)$ for any $1\leq\rho,\rho'\leq M$. Here $\langle\cdots\rangle$ stands for the expected value. In general, $N$ and $M$ may differ. 
To relate the $M$-dimensional white Gaussian noise to the $N$-dimensional state, we use an $N\times M$ matrix $\mathsf{G}(\bm{x})$. The second term $\mathsf{G}(\bm{x}_t)\bullet\bm{\xi}_t$ implies that the matrix-vector multiplication is performed component-wise using the Ito interpretation~\footnote{Other interpretations may also apply if we consider some systems such as a Brownian particle in a medium with inhomogeneous mobility~\cite{van1988diffusion,Shiraishi2023}. However, this difference in interpretation does not affect our results in the weak-noise limit.}. We also use a \add{positive parameter $\epsilon>0$ to express the noise intensity.} 

%In the chemical Langevin equation~\cite{gillespie2000chemical}, for example, $N$ corresponds to the number of chemical species and $M$ corresponds to the number of reactions.

This stochastic system is also described by the time evolution of the probability distribution of the state being $\bm{x}=\left(x_1,x_2,\cdots,x_N\right)^{\top}\in\mathbb{R}^N$ at time $t$, $p_t(\bm{x})$. Here the transpose is denoted by $\top$. The time evolution of $p_t(\bm{x})$ is governed by the following Fokker--Planck equation~\cite{risken1989fokker,gardiner2009stochastic}:
\begin{align}
    \partial_t p_t(\bm{x})&=-\bm{\nabla}\cdot(p_t(\bm{x})\bm{\nu}_t(\bm{x})),\notag\\
    \bm{\nu}_t(\bm{x})&\coloneqq\bm{F}(\bm{x})-\epsilon\bm{\nabla}\cdot\mathsf{D}(\bm{x})-\epsilon\mathsf{D}(\bm{x})\bm{\nabla}\ln p_t(\bm{x}),
    \label{Fokker--Planck eq}
\end{align}
with a shorthand notation $[\bm{\nabla}\cdot\mathsf{D}(\bm{x})]_{\alpha}=\sum_{\beta}\partial_{x_{\beta}}[\mathsf{D}(\bm{x})]_{\beta\alpha}$. Here the symbol $\partial_{x_\beta}$ denotes the partial derivative with respect to $x_{\beta}$. We introduce the $N\times N$ matrix $\mathsf{D}(\bm{x})$ as $\mathsf{D}(\bm{x})\coloneqq\mathsf{G}(\bm{x})\mathsf{G}(\bm{x})^{\top}$, which is symmetric by definition. We also assume that $\mathsf{D}(\bm{x})$ is positive-definite. In Eq.~\eqref{Fokker--Planck eq}, $\epsilon\mathsf{D}(\bm{x})$ provides the diffusion coefficient matrix.

We assume that the deterministic time evolution induced by $\bm{F}(\bm{x})$, i.e., $\dot{\bm{x}}_t=\bm{F}(\bm{x}_t)$, has a globally stable limit cycle solution with period $\tau_{\mathrm{p}}$, denoted by $\{\bm{x}^{\mathrm{LC}}_t\}_{t\in[0,\infty)}$. Here $\dot{\bm{x}}^{\mathrm{LC}}_t=\bm{F}(\bm{x}^{\mathrm{LC}}_t)$ and $\bm{x}^{\mathrm{LC}}_t=\bm{x}^{\mathrm{LC}}_{t+\tau_{\mathrm{p}}}$ hold for all $t$. This assumption implies that the Langevin equation~\eqref{Langevin eq} and the Fokker--Planck equation~\eqref{Fokker--Planck eq} describe a stochastic limit cycle. We note that the stability of the limit cycle is essential for our results.

We characterize the limit cycle's stability~\cite{Anishchenko2014} by focusing on the deterministic time evolution. We consider the case where $\bm{x}_t$ is close enough to $\bm{x}^{\mathrm{LC}}_t$ and define the small deviation from the limit cycle as $\bm{z}_t\coloneqq\bm{x}_t-\bm{x}^{\mathrm{LC}}_t$. Neglecting the second- and higher-order terms in this small deviation, we obtain $\dot{\bm{z}}_t=\mathsf{K}_t\bm{z}_t$, where $[\mathsf{K}_t]_{\alpha\beta}\coloneqq\partial_{x_{\beta}}F_{\alpha}(\bm{x}_t^{\mathrm{LC}})$ is the Jacobian of $\bm{F}(\bm{x})$ on the limit cycle.
We introduce the fundamental matrix $\Phi_{t}$ as the solution of $\dot{\Phi}_{t}=\mathsf{K}_{t}\Phi_{t}$ under the initial condition that $\Phi_{0}$ is the identity matrix. By definition, this matrix is regular and solves the time evolution of $\bm{z}_t$ as $\bm{z}_t=\Phi_t\bm{z}_0$~\cite{Teschl2012}. In particular, the case of $t=\tau_{\mathrm{p}}$ shows how the deviation from the limit cycle behaves after one period as $\bm{z}_{\tau_{\mathrm{p}}}=\Phi_{\tau_{\mathrm{p}}}\bm{z}_0$. Here $\Phi_{\tau_{\mathrm{p}}}$ is called the monodromy matrix~\cite{Anishchenko2014, Teschl2012}. This matrix satisfies $\Phi_{t+\tau_{\mathrm{p}}}=\Phi_{t}\Phi_{\tau_{\mathrm{p}}}$ due to the periodicity of $\mathsf{K}_t$.

The stability of the limit cycle is characterized by the spectrum of $\Phi_{\tau_{\mathrm{p}}}$. Let $\{\lambda_i\}_{i=1}^N$ denote the eigenvalues of $\Phi_{\tau_{\mathrm{p}}}$.
Suppose for simplicity that $\Phi_{\tau_{\mathrm{p}}}$ is diagonalizable~\footnote{Otherwise, we obtain the same result using the Jordan normal form}.
We let $\bm{v}^{(i)}$ and $\bm{u}^{(i)}$ be the left and right eigenvectors of $\Phi_{\tau_{\mathrm{p}}}$ corresponding to $\lambda_i$, which are biorthogonalized so that $\bm{v}^{(i)\top}\bm{u}^{(j)}=\delta_{ij}$. We can find an $i$ such that $\lambda_i=1$ and $\bm{u}^{(i)}\propto\dot{\bm{x}}_0^{\mathrm{LC}}$. This fact is confirmed as follows: The time derivative of $\dot{\bm{x}}_t^{\mathrm{LC}}=\bm{F}(\bm{x}_t^{\mathrm{LC}})$ yields $\ddot{\bm{x}}_t^{\mathrm{LC}}=\mathsf{K}_t\dot{\bm{x}}_t^{\mathrm{LC}}$ due to the chain rule. The matrix $\Phi_{\tau_{\mathrm{p}}}$ solves this time evolution as $\Phi_{\tau_{\mathrm{p}}}\dot{\bm{x}}_0^{\mathrm{LC}}=\dot{\bm{x}}_{\tau_{\mathrm{p}}}^{\mathrm{LC}}$. The periodicity $\dot{\bm{x}}_{\tau_{\mathrm{p}}}^{\mathrm{LC}}=\dot{\bm{x}}_0^{\mathrm{LC}}$ concludes $\Phi_{\tau_{\mathrm{p}}}\dot{\bm{x}}_0^{\mathrm{LC}}=\dot{\bm{x}}_{0}^{\mathrm{LC}}$, which implies that $\dot{\bm{x}}_0^{\mathrm{LC}}$ is an eigenvector of $\Phi_{\tau_{\mathrm{p}}}$ corresponding to the eigenvalue $1$. 
In the following, we suppose $\lambda_1=1$ and $\bm{u}^{(1)}=\dot{\bm{x}}^{\mathrm{LC}}_0$ without loss of generality, which yield $\Phi_{\tau_{\mathrm{p}}}=\dot{\bm{x}}^{\mathrm{LC}}_0\bm{v}^{(1)\top}+\sum_{i=2}^{N}\lambda_{i}\bm{u}^{(i)}\bm{v}^{(i)\top}$. This spectral decomposition allows us to characterize the stability of the limit cycle by $|\lambda_i|<1$ for $i\geq2$~\cite{Anishchenko2014}; deviations perpendicular to the limit cycle decay after one period.

Below, we focus on the weak-noise limit $\epsilon\to0$. To quantify the EP required for an oscillatory period $\tau_{\mathrm{p}}$, we consider the time evolution of $p_t(\bm{x})$ as it concentrates on the limit cycle within the finite time interval $[0, \tau_{\mathrm{p}}]$. Due to the weak-noise limit, we can assume the large-deviation form of the probability distribution as $p_t(\bm{x})\asymp\mathrm{e}^{-I_t(\bm{x})/\epsilon}$ with the rate function $I_t(\bm{x})$~\cite{touchette2009large,Freidlin2012, van1992stochastic,falasco2025macroscopic}. The rate function is nonnegative and reaches its minimum value of $0$ at a single point. We also assume that the initial distribution $p_0(\bm{x})$ concentrates on a point belonging to the limit cycle, $\bm{x}^{\mathrm{LC}}_{0}$; $I_0(\bm{x})$ has the unique minimum $0$ at $\bm{x}^{\mathrm{LC}}_{0}$. Since the limit cycle is a solution of the deterministic time evolution, this assumption on $p_0(\bm{x})$ makes $p_t(\bm{x})$ concentrate on $\bm{x}^{\mathrm{LC}}_{t}$; $I_t(\bm{x})$ has the unique minimum $0$ at $\bm{x}^{\mathrm{LC}}_{t}$.  

Using the large-deviation form of $p_t(\bm{x})$, we can obtain the EP $\Sigma_{\tau_{\mathrm{p}}}$ required for one period of oscillation along the limit cycle solution in the weak-noise limit~\cite{santolin2025dissipation,falasco2025macroscopic} as
\begin{align}
    \Sigma_{\tau_{\mathrm{p}}}=\frac{1}{\epsilon}\int_{0}^{\tau_{\mathrm{p}}}dt\,\bm{F}(\bm{x}^{\mathrm{LC}}_t)^{\top}\mathsf{D}(\bm{x}^{\mathrm{LC}}_t)^{-1}\bm{F}(\bm{x}^{\mathrm{LC}}_t).
    \label{EP for Langevin}
\end{align}
We provide the derivation in End Matter. Here we set the Boltzmann constant to unity. Although $\Sigma_{\tau_{\mathrm{p}}}$ diverges as $\epsilon$ approaches $0$, $\lim_{\epsilon\to0}\epsilon\Sigma_{\tau_{\mathrm{p}}}$ is well defined. 
\add{The EP in Eq.~\eqref{EP for Langevin} was derived by considering the time-evolving distribution as done in Ref.~\cite{santolin2025dissipation}. However, this EP and the following results coincide with those of the steady state in the weak-noise limit. This is verified by considering the steady-state distribution, which has a crater-like profile along the limit cycle. See the Supplemental Materials (SM)~\cite{SuppMat} for details.}

%, where the distribution concentrates at the point $\bm{x}_t^{\mathrm{LC}}$ on the limit cycle at time $t$

We also introduce the correlation time to define a measure of coherence.
In the steady state, the correlation function of the observable $a(\bm{x})$ is defined as $\langle a(\bm{x}_t)a(\bm{x}_0)\rangle_{\mathrm{st}}$, where $\langle\cdots\rangle_{\mathrm{st}}$ denotes the expected value in the steady state. For sufficiently large $t$, the correlation function decays exponentially as $\langle a(\bm{x}_t)a(\bm{x}_0)\rangle_{\mathrm{st}}\sim\mathrm{e}^{-t/\tau_{\mathrm{c}}}$, where $\sim$ implies that we ignore the contribution of constant value remaining in the long-time limit and oscillatory behavior. The coefficient $\tau_{\mathrm{c}}$ is common to most observables, and we call it the correlation time. We explain this in more detail in SM~\cite{SuppMat}.

In the weak-noise limit, we can derive the explicit form of $\tau_{\mathrm{c}}$ by considering a linear Langevin equation describing a small deviation from the deterministic limit cycle~\cite{vance1996fluctuations,gaspard2002trace,gaspard2002correlation}. Below, we only present the explicit form, which is obtained by using the eigenvector of the monodromy matrix (see SM~\cite{SuppMat} for derivation). We define a vector-valued function $\bm{\zeta}_t$ as $\bm{\zeta}_t\coloneqq\Phi_t^{\top-1}\bm{v}^{(1)}$. Using this function, the correlation time is given by
\begin{align}
    \tau_{\mathrm{c}}=\frac{\tau_{\mathrm{p}}^3}{4\pi^2\epsilon}\left[\int_{0}^{\tau_{\mathrm{p}}}dt\,\bm{\zeta}_t^{\top}\mathsf{D}(\bm{x}^{\mathrm{LC}}_t)\bm{\zeta}_t\right]^{-1}.
    \label{correlation time}
\end{align}
We can regard $\bm{\zeta}_t$ as the dual of $\dot{\bm{x}}_t^{\mathrm{LC}}$ as shown below. Since $\Phi_t$ is regular, $\{\Phi_t\bm{u}^{(i)}\}_{i=1}^{N}$ is a basis of $\mathbb{C}^N$. The biorthogonality $(\Phi_t^{\top-1}\bm{v}^{(i)})^{\top}\Phi_t\bm{u}^{(j)}=\bm{v}^{(i)\top}\bm{u}^{(j)}=\delta_{ij}$ allows us to regard $\{\Phi_t^{\top-1}\bm{v}^{(i)}\}_{i=1}^{N}$ as the dual basis of $\{\Phi_t\bm{u}^{(i)}\}_{i=1}^{N}$. In this sense, $\bm{\zeta}_t=\Phi_t^{\top-1}\bm{v}^{(1)}$ is the dual of $\Phi_t\bm{u}^{(1)}=\Phi_t\dot{\bm{x}}^{\mathrm{LC}}_0=\dot{\bm{x}}^{\mathrm{LC}}_t$. In particular, the biorthogonality yields
\begin{align}
    \bm{\zeta}_t^{\top}\dot{\bm{x}}_t^{\mathrm{LC}}=1.
    \label{normalization}
\end{align}
We also note that $\bm{\zeta}_t$ is periodic as well as $\dot{\bm{x}}_t^{\mathrm{LC}}$. This fact is confirmed using $\Phi_{t+\tau_{\mathrm{p}}}=
\Phi_t\Phi_{\tau_{\mathrm{p}}}$ and $\Phi_{\tau_{\mathrm{p}}}^{\top}\bm{v}^{(1)}=\bm{v}^{(1)}$ as $\bm{\zeta}_{t+\tau_{\mathrm{p}}}=\Phi_{t+\tau_{\mathrm{p}}}^{\top-1}\bm{v}^{(1)}=\Phi_{t}^{\top-1}\Phi_{\tau_{\mathrm{p}}}^{\top-1}\bm{v}^{(1)}=\Phi_{t}^{\top-1}\bm{v}^{(1)}=\bm{\zeta}_{t}$.

The coherence of noisy oscillation is quantified by the number of oscillations that occur before the steady-state correlations are broken. Since we can regard $\tau_{\mathrm{c}}$ as the time it takes for the correlation to break down, this number is defined as $\mathcal{N}\coloneqq\tau_{\mathrm{c}}\times\tau_{\mathrm{p}}^{-1}$~\footnote{This definition is based on Ref.~\cite{oberreiter2022universal}. In Refs.~\cite{gaspard2002trace} and ~\cite{santolin2025dissipation}, the \textit{quality factor} $2\pi\tau_{\mathrm{c}}\times\tau_{\mathrm{p}}^{-1}$ is used instead of $\mathcal{N}$.}.
Combining this definition and Eq.~\eqref{correlation time}, we obtain the explicit form of $\mathcal{N}$ in the weak-noise limit as
\begin{align}
    \mathcal{N}=\frac{\tau_{\mathrm{p}}^2}{4\pi^2\epsilon}\left[\int_{0}^{\tau_{\mathrm{p}}}dt\,\bm{\zeta}_t^{\top}\mathsf{D}(\bm{x}^{\mathrm{LC}}_t)\bm{\zeta}_t\right]^{-1}.
    \label{Num of coherent weaknoise}
\end{align}
In the following, we call $\mathcal{N}$ the number of coherent oscillations.

\textit{Main results}.---%
We consider the short-time TUR in the weak-noise limit. Let us consider an $N$-dimensional vector field $\bm{\omega}(\bm{x})$. In the infinitesimal time interval $[t,t+\varDelta t]$, the current observable associated with $\bm{\omega}(\bm{x})$ is given by $J^{\bm{\omega}}_t\varDelta t=\bm{\omega}(\bm{x}_t)\circ d\bm{x}_t$. Here the increment of the state $\bm{x}_t$ is denoted by $d\bm{x}_t$, and the symbol $\circ$ implies that the vector-vector multiplication is performed component-wise using the Stratonovich interpretation. Letting $\mathbb{E}[J^{\bm{\omega}}_t]$ and $\mathrm{Var}[J^{\bm{\omega}}_t]$ denote the average and variance of the current, respectively, we obtain the short-time TUR~\cite{otsubo2020estimating} as
\begin{align}
    \dot{\Sigma}_t\geq\frac{2\mathbb{E}[J^{\bm{\omega}}_t]^2}{\varDelta t\mathrm{Var}[J^{\bm{\omega}}_t]},
    \label{short-time TUR}
\end{align}
where the EP rate is defined as $\dot{\Sigma}_t\coloneqq\epsilon^{-1}\int d\bm{x}\,p_t(\bm{x})\bm{\nu}_t(\bm{x})^{\top}\mathsf{D}(\bm{x})^{-1}\bm{\nu}_t(\bm{x})$. In the weak-noise limit, the short-time TUR yields the following bound on $\Sigma_{\tau_{\mathrm{p}}}$:
\begin{align}
    \Sigma_{\tau_{\mathrm{p}}}\geq\dfrac{\left[\displaystyle\int_{0}^{\tau_{\mathrm{p}}}dt\,\bm{\omega}(\bm{x}_t^{\mathrm{LC}})^{\top}\bm{F}(\bm{x}_t^{\mathrm{LC}})\right]^2}{\epsilon\displaystyle\int_{0}^{\tau_{\mathrm{p}}}dt\,\bm{\omega}(\bm{x}_t^{\mathrm{LC}})^{\top}\mathsf{D}(\bm{x}_t^{\mathrm{LC}})\bm{\omega}(\bm{x}_t^{\mathrm{LC}})}.
    \label{integrated TUR}
\end{align}
This inequality is derived by taking the time integration of the TUR and admitting the large deviation form $p_t(\bm{x})\asymp\mathrm{e}^{-I_t(\bm{x})/\epsilon}$ (see also End Matter for derivation). In the following, we derive the dissipation-coherence trade-off and the TSL by substituting different observables into the bound~\eqref{integrated TUR}.

First, we use the observable $\bm{\omega}_{\bm{\zeta}}(\bm{x})$ such that $\bm{\omega}_{\bm{\zeta}}(\bm{x}^{\mathrm{LC}}_t)=\bm{\zeta}_t$. We can find such $\bm{\omega}_{\bm{\zeta}}(\bm{x})$, since $\bm{\zeta}_t$ is periodic. In this case, the bound in Eq.~\eqref{integrated TUR} reduces to $\Sigma_{\tau_{\mathrm{p}}}\geq\frac{1}{\epsilon}\left[\int_{0}^{\tau_{\mathrm{p}}}dt\,\bm{\zeta}_t^{\top}\dot{\bm{x}}_t^{\mathrm{LC}}\right]^2\left[\int_{0}^{\tau_{\mathrm{p}}}dt\,\bm{\zeta}_t^{\top}\mathsf{D}(\bm{x}_t^{\mathrm{LC}})\bm{\zeta}_t\right]^{-1}$, where we used $\dot{\bm{x}}_t^{\mathrm{LC}}=\bm{F}(\bm{x}_t^{\mathrm{LC}})$.
Substituting Eqs.~\eqref{normalization} and ~\eqref{Num of coherent weaknoise} into this inequality, we obtain the trade-off relation between the EP and the number of coherent oscillations,
\begin{align}
    \frac{\Sigma_{\tau_{\mathrm{p}}}}{4\pi^2}\geq\mathcal{N}.
    \label{coherent bound}
\end{align}
This trade-off implies that more dissipation is required to remain coherent for a longer time. 

We emphasize that this derivation of the dissipation-coherence trade-off is more general than the one by Santolin and Falasco~\cite{santolin2025dissipation}, which is based on the projection of $\bm{x}_t$ onto the direction tangent to the limit cycle. \add{Although they numerically verify this trade-off for various systems, this projection restricts their derivation to systems that satisfy one of the following two conditions: (i) $\mathsf{D}(\bm{x})$ is always proportional to the identity matrix, or (ii) the system is in the vicinity of a Hopf bifurcation. The usage of the TUR enables us to prove the trade-off~\eqref{coherent bound} for general systems without these conditions.}

\begin{figure*}
    \centering
    \includegraphics[width=\linewidth]{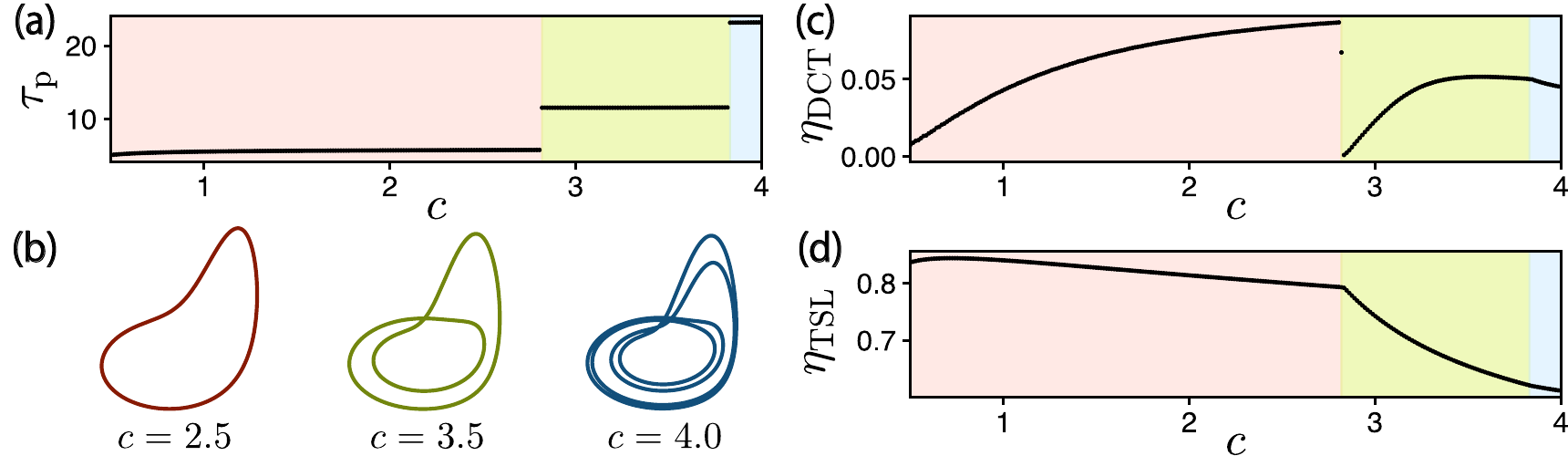}
    \caption{Numerical demonstration of the R\"{o}ssler model. \add{All quantities are computed using the deterministic solutions of the R\"{o}ssler model.} (a) $c$-dependence of the period $\tau_{\mathrm{p}}$. (b) Typical \add{deterministic} stable limit cycles in each phase of the period-doubling bifurcation and corresponding $c$. (c) $c$-dependence of the tightness of the dissipation-coherence trade-off. (d) $c$-dependence of the tightness of the TSL. In (a), (c), and (d), we color the range of $c$ corresponding to each phase of the period-doubling bifurcation in red, green, and blue.}
    \label{fig:Rossler}
\end{figure*}

Second, we use $\bm{F}(\bm{x})$ itself as the observable in the bound~\eqref{integrated TUR} to derive the TSL. To express the TSL, we define the Euclidean length of the limit cycle $l_{\mathrm{LC}}$ as $l_{\mathrm{LC}}\coloneqq\int_{0}^{\tau_{\mathrm{p}}}dt\,\|\dot{\bm{x}}^{\mathrm{LC}}_t\|$, where the Euclidean norm is denoted by $\|\cdots\|$. We also introduce the intensity of diffusion measured along the limit cycle as $D_{\mathrm{LC}}\coloneqq\epsilon\left[\int_{0}^{\tau_{\mathrm{p}}}dt\,\dot{\bm{x}}_t^{\mathrm{LC}\top}\mathsf{D}(\bm{x}_t^{\mathrm{LC}})\dot{\bm{x}}_t^{\mathrm{LC}}\right]/\left[\int_{0}^{\tau_{\mathrm{p}}}dt\,\|\dot{\bm{x}}_t^{\mathrm{LC}}\|^2\right]$. Substituting $\bm{\omega}(\bm{x})=\bm{F}(\bm{x})$ into Eq.~\eqref{integrated TUR} and using the Cauchy--Schwarz inequality $l_{\mathrm{LC}}^2\leq(\int_{0}^{\tau_{\mathrm{p}}}dt)(\int_{0}^{\tau_{\mathrm{p}}}dt\,\|\dot{\bm{x}}_t^{\mathrm{LC}}\|^2)$, we obtain the following TSL:
\begin{align}
    \Sigma_{\tau_{\mathrm{p}}}\geq\frac{l_{\mathrm{LC}}^2}{\tau_{\mathrm{p}}D_{\mathrm{LC}}}.
    \label{TSL}
\end{align}
The right-hand side of this inequality increases if we consider a large $l_{\mathrm{LC}}$ or a small $\tau_{\mathrm{p}}$. Thus, the TSL implies that more dissipation is required for large-amplitude or high-frequency oscillations.

This TSL differs from the existing ones for Langevin systems. The previous TSLs bound EP using the distance between the initial and final probability distributions~\cite{aurell2012refined} or the length of the path between them~\cite{nakazato2021geometrical}. In contrast, the TSL obtained here involves the length of the deterministic limit cycle. This TSL is rather similar to the one for deterministic CRNs, in which the EP is bounded by the taxicab length of the limit cycle~\cite{kolchinsky2024generalized}. These facts stem from the proximity to deterministic behavior via the weak-noise limit.
%これまでのところ、決定論的な長さと熱力学的な量が結びつくのは、決定論的なCRNにおけるTSLsに限られていた。我々のTSLは、弱ノイズ極限におけるEPを用いているため、CRNでない力学系に対しても適用できる。

Above, we derive the dissipation-coherence trade-off~\eqref{coherent bound} and the TSL~\eqref{TSL} by applying the consequence of the TUR~\eqref{integrated TUR} to $\bm{\omega}_{\bm{\zeta}}(\bm{x})$ and $\bm{F}(\bm{x})$, respectively. The duality between $\bm{\zeta}_t$ and $\dot{\bm{x}}^{\mathrm{LC}}_t$ allows us to regard the observable $\bm{\omega}_{\bm{\zeta}}(\bm{x})$ as the dual of the other observable $\bm{F}(\bm{x})$ on the limit cycle. Thus, from the viewpoint of the TUR, the dissipation-coherence trade-off is the dual bound of the TSL.

\textit{Example: Period-doubling bifurcation in the R\"{o}ssler model}.---%
We demonstrate our trade-offs using the noisy R\"{o}ssler model~\cite{rossler1976equation} in which each degree of freedom interacts with a heat bath at a distinct temperature. This system has three degrees of freedom ($N=3$) and is described by the following $\bm{F}(\bm{x})$ and $\mathsf{D}$:
\begin{align}
    \bm{F}(\bm{x})\coloneqq
\begin{pmatrix}
   -x_2-x_3 \\
   x_1 + ax_2 \\
   b + x_1x_3 - cx_3
\end{pmatrix},\;\;
\mathsf{D}\coloneqq
\begin{pmatrix}
   T_1 & 0 & 0\\
   0 & T_2 & 0 \\
   0 & 0 & T_3
\end{pmatrix}
.
\end{align}
Here $\epsilon T_{\alpha}$ is the temperature of the heat bath interacting with $x_{\alpha}$.

We consider the parameters where the system violates two conditions that were necessary for the previous proof of the dissipation-coherence trade-off in Ref.~\cite{santolin2025dissipation}. We set the temperatures divided by $\epsilon$ to $(T_1,T_2,T_3)=(1,2,3)$. This setting makes $\mathsf{D}$ not proportional to the identity matrix. We also set the parameters in $\bm{F}(\bm{x})$ to $a=b=0.2$ and $c\in[0.5,4.0]$. Under these parameters, the system is far from the Hopf bifurcation. We observe the period-doubling bifurcation as shown in Figs.~\ref{fig:Rossler}(a) and ~\ref{fig:Rossler}(b).

To observe the tightness of the dissipation-coherence trade-off~\eqref{coherent bound} and the TSL~\eqref{TSL}, we define the following efficiencies for the bounds:
\begin{align}
    \eta_{\mathrm{DCT}}\coloneqq\frac{4\pi^2\mathcal{N}}{\Sigma_{\tau_{\mathrm{p}}}},\;\eta_{\mathrm{TSL}}\coloneqq\frac{l_{\mathrm{LC}}^2}{\tau_{\mathrm{p}}D_{\mathrm{LC}}\Sigma_{\tau_{\mathrm{p}}}}.
\end{align}
These efficiencies range from $0$ to $1$. Values closer to $1$ imply that the corresponding trade-off is tighter. We show the $c$-dependence of these efficiencies in Figs.~\ref{fig:Rossler}(c) and ~\ref{fig:Rossler}(d). These figures show that both efficiencies are less than $1$, which confirms the trade-offs. \add{Although the dissipation-coherence trade-off becomes looser, the TSL provides a tight bound: $\eta_{\mathrm{TSL}}$ always remains above $0.5$.} We can also observe that the behavior of $\eta_{\mathrm{DCT}}$ and $\eta_{\mathrm{TSL}}$ differs before and after the first bifurcation. Before the first bifurcation, $\eta_{\mathrm{DCT}}$ increases with $c$, while $\eta_{\mathrm{TSL}}$ decreases with $c$. At the bifurcation point, $\eta_{\mathrm{DCT}}$ changes discontinuously, while $\eta_{\mathrm{TSL}}$ changes continuously with a kink. Following the bifurcation, $\eta_{\mathrm{DCT}}$ increases, while $\eta_{\mathrm{TSL}}$ decreases. These contrasting trends may be due to the fact that the dissipation-coherence trade-off and the TSL are dual trade-offs. We note that no such significant difference is observed near the second bifurcation.

\textit{Application to stochastic chemical systems}.---% 
We finally consider the chemical Langevin equation (CLE), which describes a stochastic CRN~\cite{gillespie2000chemical}. In general, the above results cannot be applied to the CLE because some eigenvalues of the diffusion coefficient matrix can be zero due to the conservation laws of the CRN~\cite{feinberg2019foundations,polettini2014irreversible,famili2003convex}. To apply the results, we derive a Langevin equation with a positive-definite diffusion coefficient matrix by reducing the degrees of freedom using the conservation laws (see End Matter for details). Then we can obtain the quantities in the trade-offs [Eqs.~\eqref{coherent bound} and ~\eqref{TSL}] by applying the definitions to this reduced Langevin equation: the EP $\Sigma^{\mathrm{RE}}_{\tau_{\mathrm{p}}}$, the number of coherent oscillations $\mathcal{N}^{\mathrm{RE}}$, the length of the limit cycle $l_{\mathrm{LC}}^{\mathrm{RE}}$, and the intensity of diffusion measured along the limit cycle $D^{\mathrm{RE}}_{\mathrm{LC}}$. For these \textit{reduced} quantities, the trade-offs hold.

\add{
Although the reduction obscures the physical meaning of $\Sigma^{\mathrm{RE}}_{\tau_{\mathrm{p}}}$, we can relate this EP to a physical quantity by considering the chemical master equation (CME)~\cite{mcquarrie1967stochastic,gillespie1992rigorous,schmiedl2007stochastic}. In the CME, we can introduce the parameter $\epsilon$ as the reciprocal of the system size. The CLE is obtained by truncating the Kramers--Moyal expansion of the CME at the second order in $\epsilon$~\cite{gillespie2000chemical}. In the weak-noise limit, we can also consider the steady state or the time evolution of probability distribution concentrating on $\bm{x}_t^{\mathrm{LC}}$ directly in the CME. In both cases, the EP for the CME scales as $\Sigma^{\mathrm{ME}}_{\tau_{\mathrm{p}}}=\epsilon^{-1}\int_{0}^{\tau_{\mathrm{p}}}dt\,\sigma_t$, where $\sigma_t$ is the EP rate at $\bm{x}_t^{\mathrm{LC}}$ associated with the deterministic rate equation~\cite{falasco2025macroscopic,rao2016nonequilibrium,ge2016mesoscopic,kondepudi2014modern}. This physical EP $\Sigma^{\mathrm{ME}}_{\tau_{\mathrm{p}}}$ provides an upper bound on $\Sigma^{\mathrm{RE}}_{\tau_{\mathrm{p}}}$ (see SM~\cite{SuppMat} for details).} This hierarchy leads to the following trade-offs for stochastic CRNs:
\begin{align}
    \frac{\Sigma^{\mathrm{ME}}_{\tau_{\mathrm{p}}}}{4\pi^2}\geq\frac{\Sigma^{\mathrm{RE}}_{\tau_{\mathrm{p}}}}{4\pi^2}\geq\mathcal{N}^{\mathrm{RE}},\;\;\Sigma^{\mathrm{ME}}_{\tau_{\mathrm{p}}}\geq\Sigma^{\mathrm{RE}}_{\tau_{\mathrm{p}}}\geq\frac{(l_{\mathrm{LC}}^{\mathrm{RE}}){}^2}{\tau_{\mathrm{p}}D^{\mathrm{RE}}_{\mathrm{LC}}}.
    \label{trade-offs stochastic CRNs}
\end{align}
We demonstrate these trade-offs using a CRN with a conservation law in End Matter. Note that the dissipation-coherence trade-off in Eq.~\eqref{trade-offs stochastic CRNs} was proposed by Santolin and Falasco~\cite{santolin2025dissipation}, but they did not consider CRNs with conservation laws explicitly. We also remark that the trade-offs for $\Sigma^{\mathrm{ME}}_{\tau_{\mathrm{p}}}$ in Eq.~\eqref{trade-offs stochastic CRNs} are also derived from the TUR for deterministic CRNs~\cite{yoshimura2021thermodynamic,yoshimura2023housekeeping}, \add{since $\lim_{\epsilon\to0}\epsilon\Sigma^{\mathrm{ME}}_{\tau_{\mathrm{p}}}$ converges to the EP for the deterministic rate equation (see SM~\cite{SuppMat} for details).}

\add{
\textit{Asymptotic phase and saturation of the bound}.---% 
Here we consider the relation between the phase reduction theory~\cite{winfree1967biological,Kuramoto1984,monga2019phase} and $\bm{\zeta}_t$ (see SM~\cite{SuppMat} for details). Since the deterministic time evolution $\dot{\bm{x}}_t=\bm{F}(\bm{x}_t)$ has a stable limit cycle solution, we can define the asymptotic phase $\Theta(\bm{x})\in[0,2\pi)$ up to modulo $2\pi$~\cite{monga2019phase}. This phase satisfies $[\bm{\nabla}\Theta(\bm{x})]^{\top}\bm{F}(\bm{x})=2\pi/\tau_{\mathrm{p}}$ for any $\bm{x}$. In the phase reduction, the gradient on the limit cycle $\bm{\nabla}\Theta(\bm{x}^{\mathrm{LC}}_t)$ characterizes the response of the phase to a weak perturbation on the deterministic system~\cite{winfree1967biological}, and is called \textit{infinitesimal phase response curve} (PRC)~\cite{monga2019phase,brown2004phase}. Considering the time derivative of $\bm{\nabla}\Theta(\bm{x}_t^{\mathrm{LC}})$~\cite{monga2019phase,brown2004phase,Hoppensteadt1997}, we can verify that the PRC corresponds to $\bm{\zeta}_t$ as $\bm{\nabla}\Theta(\bm{x}^{\mathrm{LC}}_t)=(2\pi/\tau_{\mathrm{p}})\bm{\zeta}_t$~\cite{shirasaka2017phase}. This relation implies that we can use $(2\pi)^{-1}\tau_{\mathrm{p}}\bm{\nabla}\Theta(\bm{x})$ as the observable $\bm{\omega}_{\bm{\zeta}}(\bm{x})$ in the derivation of the dissipation-coherence trade-off.
}

\add{
Based on the relation between the PRC and $\bm{\zeta}_t$, we can construct a diffusion coefficient matrix that saturates the dissipation-coherence trade-off (see SM~\cite{SuppMat} for details). Using the asymptotic phase, we introduce the following symmetric positive-definite matrix:
\begin{align}
    \mathsf{D}(\bm{x})=D\left[\mathsf{I}-\frac{\bm{\nabla}\Theta(\bm{x})[\bm{\nabla}\Theta(\bm{x})]^{\top}}{\|\bm{\nabla}\Theta(\bm{x})\|^2}+\bm{F}(\bm{x})\bm{F}(\bm{x})^{\top}\right],
    \label{optimal diffusion matrix main}
\end{align}
where $D$ is a positive constant. 
Due to $[\bm{\nabla}\Theta(\bm{x})]^{\top}\bm{F}(\bm{x})=2\pi/\tau_{\mathrm{p}}$, this matrix satisfies $\mathsf{D}(\bm{x})\bm{\nabla}\Theta(\bm{x})=(2\pi D/\tau_{\mathrm{p}})\bm{F}(\bm{x})$, which reduces to
$\mathsf{D}(\bm{x}_t^{\mathrm{LC}})\bm{\zeta}_t=D\bm{F}(\bm{x}_t^{\mathrm{LC}})$ on the limit cycle. Substituting this relation into Eqs.~\eqref{EP for Langevin} and ~\eqref{Num of coherent weaknoise}, we can verify that the dissipation-coherence trade-off is saturated as $\Sigma_{\tau_{\mathrm{p}}}=4\pi^2\mathcal{N}=\tau_{\mathrm{p}}/(\epsilon D)$. Because $D$ is arbitrary, we can achieve any desired $\mathcal{N}$ with minimum dissipation $4\pi^2\mathcal{N}$ only by modifying $\mathsf{D}(x)$ for a given force field. We also note that the TSL may not be achievable solely by tuning $\mathsf{D}(\bm{x})$. This is because the Cauchy--Schwarz inequality for $\bm{F}(\bm{x})$, $l_{\mathrm{LC}}^2\leq(\int_{0}^{\tau_{\mathrm{p}}}dt)(\int_{0}^{\tau_{\mathrm{p}}}dt\,\|\bm{F}(\bm{x}_t^{\mathrm{LC}})\|^2)$, is used in the derivation of the TSL (see SM~\cite{SuppMat} for details).
}

\textit{Discussion}.---% 
In this Letter, we derived the dissipation-coherence trade-off and the TSL for the general stochastic limit cycles in the weak-noise limit by applying the dual observables to the short-time TUR. These trade-offs were generalized to the CLE, which involves the diffusion coefficient matrix with zero eigenvalues. \add{We further revealed that the dissipation-coherence trade-off can always be saturated solely by adjusting the diffusion coefficient matrix. Our derivation of the dissipation-coherence trade-off extends its applicability beyond the specific cases proved by Santolin and Falasco~\cite{santolin2025dissipation}.} Unlike the dissipation-coherence trade-off, the TSL had not previously been considered in the weak-noise limit. This new TSL establishes a link between quantities derived from deterministic dynamical systems and thermodynamic quantities.

\add{
Verifying the validity of the dissipation-coherence trade-off for systems with finite $\epsilon$ is an important problem. As in the case of Markov jump processes~\cite{oberreiter2022universal}, $\mathcal{N}$ can be characterized by the eigenvalues of the Fokker--Planck operator. In SM~\cite{SuppMat}, we numerically obtain the steady-state distribution and the eigenvalues of the Fokker--Planck operator by discretizing its adjoint~\cite{houzelstein2025generalized} for several systems. Using the steady-state distribution and the eigenvalues, we compute the steady-state EP and $\mathcal{N}$ and verify the dissipation-coherence trade-off.
We confirm that the dissipation-coherence trade-off is satisfied for all considered systems. For certain systems, this trade-off is tighter in the finite-noise case than in the weak-noise limit.
}
%全ての場合でDCTは成立していたが、

We also note that developing thermodynamics in the weak-noise limit as a tool for investigating general dynamical systems appears promising. For example, we may classify some bifurcations based on the tightness of a thermodynamic trade-off. Indeed, the first and second bifurcations in Fig.~\ref{fig:Rossler}(a) may differ qualitatively from a thermodynamic perspective. This is because the tightness of the dissipation-coherence trade-off behaves discontinuously at the first bifurcation and continuously at the second, as shown in Fig.~\ref{fig:Rossler}(c).

\begin{acknowledgments}
The authors thank Kohei Yoshimura, Naruo Ohga, and Artemy Kolchinsky for their suggestive comments. R.N.\ thanks Davide Santolin, Kota Tokiwa, and Yoh Maekawa for discussion.
R.N.\ is supported by JSR Fellowship, the University of Tokyo, and JSPS KAKENHI Grants No.~25KJ0931.
S.I.\ is supported by JSPS KAKENHI Grants No.~22H01141, No.~23H00467, and No.~24H00834, 
JST ERATO Grant No.~JPMJER2302, 
and UTEC-UTokyo FSI Research Grant Program.
\end{acknowledgments}

\input{biblio.bbl}
\newpage

\onecolumngrid

\newpage

\section*{End Matter}

\twocolumngrid

\textit{Derivation of Eq.~\eqref{EP for Langevin}}.---%
We derive the EP required for one period of oscillation along the limit cycle solution in the weak-noise limit. In stochastic thermodynamics~\cite{sekimoto2010stochastic,seifert2012stochastic,Shiraishi2023}, the EP is given by $\Sigma_{\tau_{\mathrm{p}}}={\epsilon}^{-1}\int_{0}^{\tau_{\mathrm{p}}}dt\int d\bm{x}\,p_t(\bm{x})\bm{\nu}_t(\bm{x})^{\top}\mathsf{D}(\bm{x})^{-1}\bm{\nu}_t(\bm{x})$. Substituting the large-deviation form $p_t(\bm{x})\asymp\mathrm{e}^{-I_t(\bm{x})/\epsilon}$, we obtain $\Sigma_{\tau_{\mathrm{p}}}\asymp{\epsilon}^{-1}\int_{0}^{\tau_{\mathrm{p}}}dt\int d\bm{x}\,\mathrm{e}^{-I_t(\bm{x})/\epsilon}\bm{\nu}_t^{I}(\bm{x})^{\top}\mathsf{D}(\bm{x})^{-1}\bm{\nu}^{I}_t(\bm{x})$ with $\bm{\nu}^{I}_t(\bm{x})\coloneqq\bm{F}(\bm{x})-\epsilon\bm{\nabla}\cdot\mathsf{D}(\bm{x})+\mathsf{D}(\bm{x})\bm{\nabla}I_t(\bm{x})$. The weak-noise limit allows us to apply the Laplace approximation~\cite{touchette2009large} to this form, which yields
\begin{align}
    \Sigma_{\tau_{\mathrm{p}}}&=\frac{1}{\epsilon}\int_{0}^{\tau_{\mathrm{p}}}dt\,\bm{\nu}_t^{I}(\bm{x}^{\mathrm{LC}}_{t})^{\top}\mathsf{D}(\bm{x}^{\mathrm{LC}}_{t})^{-1}\bm{\nu}_t^{I}(\bm{x}^{\mathrm{LC}}_{t}).
    \label{Laplace approx}
\end{align}
Since $I_t(\bm{x})$ has the unique minimum at $\bm{x}^{\mathrm{LC}}_{t}$, we obtain $\bm{\nabla}I_t(\bm{x}^{\mathrm{LC}}_{t})=0$ and thus $\bm{\nu}^{I}_t(\bm{x}^{\mathrm{LC}}_{t})=\bm{F}(\bm{x}^{\mathrm{LC}}_{t})-\epsilon\bm{\nabla}\cdot\mathsf{D}(\bm{x}^{\mathrm{LC}}_{t})$. Substituting this equation into Eq.~\eqref{Laplace approx}, we obtain
\begin{align}
    \Sigma_{\tau_{\mathrm{p}}}&=\frac{1}{\epsilon}\int_{0}^{\tau_{\mathrm{p}}}dt\,\bm{F}(\bm{x}^{\mathrm{LC}}_{t})^{\top}\mathsf{D}(\bm{x}^{\mathrm{LC}}_{t})^{-1}\bm{F}(\bm{x}^{\mathrm{LC}}_{t})\notag\\
    &\quad-2\int_{0}^{\tau_{\mathrm{p}}}dt\,\bm{F}(\bm{x}^{\mathrm{LC}}_{t})^{\top}\mathsf{D}(\bm{x}^{\mathrm{LC}}_{t})^{-1}\bm{\nabla}\cdot\mathsf{D}(\bm{x}^{\mathrm{LC}}_{t})\notag\\
    &\quad+\epsilon\int_{0}^{\tau_{\mathrm{p}}}dt\,\{\bm{\nabla}\cdot\mathsf{D}(\bm{x}^{\mathrm{LC}}_{t})\}^{\top}\mathsf{D}(\bm{x}^{\mathrm{LC}}_{t})^{-1}\bm{\nabla}\cdot\mathsf{D}(\bm{x}^{\mathrm{LC}}_{t}).
\end{align}
In the weak-noise limit, the first term dominates, resulting in Eq.~\eqref{EP for Langevin}.

\textit{Derivation of Eq.~\eqref{integrated TUR}}.---%
The short-time TUR~\cite{otsubo2020estimating} in Eq.~\eqref{short-time TUR} leads to $\mathbb{E}[J^{\bm{\omega}}_t]\leq|\mathbb{E}[J^{\bm{\omega}}_t]|\leq\sqrt{\dot{\Sigma}_t\varDelta t\mathrm{Var}[J^{\bm{\omega}}_t]/2}.$ Integrating both sides and using the Cauchy--Schwarz inequality, we obtain
\begin{align}
    &\left(\int_{0}^{\tau_{\mathrm{p}}}dt\,\mathbb{E}[J^{\bm{\omega}}_t]\right)^2\leq\left(\int_{0}^{\tau_{\mathrm{p}}}dt\,\sqrt{\dot{\Sigma}_t\frac{\varDelta t\mathrm{Var}[J^{\bm{\omega}}_t]}{2}}\right)^2\notag\\
    &\leq\left(\int_{0}^{\tau_{\mathrm{p}}}dt\,\dot{\Sigma}_{t}\right)\left(\int_{0}^{\tau_{\mathrm{p}}}dt\,\frac{\varDelta t\mathrm{Var}[J^{\bm{\omega}}_t]}{2}\right).
    \label{CS integral TUR}
\end{align}
We can calculate the average and variance of $J^{\bm{\omega}}_t$ as $\mathbb{E}[J^{\bm{\omega}}_t]=\int d\bm{x}\,p_t(\bm{x})\bm{\omega}(\bm{x})^{\top}\bm{\nu}_t(\bm{x})$ and $\mathrm{Var}[J^{\bm{\omega}}_t]=2(\varDelta t)^{-1}\epsilon\int d\bm{x}\,p_t(\bm{x})\bm{\omega}(\bm{x})^{\top}\mathsf{D}(\bm{x})\bm{\omega}(\bm{x})$, respectively~\cite{otsubo2020estimating}. Using the large deviation form $p_t(\bm{x})\asymp\mathrm{e}^{-I_t(\bm{x})/\epsilon}$, we obtain $\dot{\Sigma}_t=\epsilon^{-1}\bm{F}(\bm{x}^{\mathrm{LC}}_t)^{\top}\mathsf{D}(\bm{x}^{\mathrm{LC}}_t)^{-1}\bm{F}(\bm{x}^{\mathrm{LC}}_t)$, $\mathbb{E}[J^{\bm{\omega}}_t]=\bm{\omega}(\bm{x}^{\mathrm{LC}}_t)^{\top}\bm{F}(\bm{x}^{\mathrm{LC}}_t)$, and $\mathrm{Var}[J^{\bm{\omega}}_t]=2(\varDelta t)^{-1}\epsilon\bm{\omega}(\bm{x}^{\mathrm{LC}}_t)^{\top}\mathsf{D}(\bm{x}^{\mathrm{LC}}_t)\bm{\omega}(\bm{x}^{\mathrm{LC}}_t)$. Here we used the fact that $I_t(\bm{x})$ has the unique minimum at $\bm{x}_{t}^{\mathrm{LC}}$ and collected the leading order as in the derivation of Eq.~\eqref{EP for Langevin}. We obtain Eq.~\eqref{integrated TUR} by substituting these expressions in the weak-noise limit into Eq.~\eqref{CS integral TUR}.

\textit{The CLE and its reduction}.---%
We consider a well-stirred mixture of chemically reacting
species, whose volume (system size) is given by $\epsilon^{-1}$. Here and below, we use the same symbols as in the general Langevin equation to represent quantities that play common roles in the CLE.
We assume that the reactions are described by a CRN consisting of $N$ internal species and $M$ reversible reactions~\cite{feinberg2019foundations}. We ignore external species that are exchanged with the outside of the system, since they do not affect our results. We index the internal species by $\alpha\in\{1,2,\cdots,N\}$ and the reversible reactions by $\rho\in\{1,2,\cdots,M\}$, respectively. Let $X_{\alpha}$ denote species $\alpha$ and $\ce{$\sum_{\alpha=1}^{N}n_{\alpha\rho}^+X_{\alpha}$ <=> $\sum_{\alpha=1}^{N}n_{\alpha\rho}^-X_{\alpha}$}$ denote reaction $\rho$. Here, we write the number of molecules of species $\alpha$ consumed (produced) through reaction $\rho$ as $n_{\alpha\rho}^+$ ($n_{\alpha\rho}^-$). 
This CRN is characterized by the $N\times M$ stoichiometric matrix $(\mathsf{S})_{\alpha\rho}\coloneqq n_{\alpha\rho}^--n_{\alpha\rho}^+$~\cite{feinberg2019foundations,famili2003convex}.

Here we introduce the CLE. Let $\bm{x}_t$ be the $N$-dimensional vector such that the $\alpha$-th element is the concentration of $X_{\alpha}$ at time $t$. We also let $j^+_{\rho}(\bm{x})$ and $j^-_{\rho}(\bm{x})$ denote the forward and reverse reaction rates, respectively. \add{If the system size is large, the following CLE~\cite{gillespie2000chemical} is often used to approximate the time evolution of $\bm{x}_t$ in the underlying CME:}
\begin{align}
    \dot{\bm{x}}_t=\mathsf{S}\bm{j}(\bm{x}_t)+\sqrt{2\epsilon}\mathsf{S}\mathsf{A}^{\frac{1}{2}}(\bm{x}_t)\bullet\bm{\xi}_t.
    \label{chemical Langevin eq.}
\end{align}
Here $\bm{j}(\bm{x})$ is the $M$-dimensional vector field such that the $\rho$-th element is the net current of reaction $\rho$, $j_{\rho}(\bm{x})\coloneqq j^+_{\rho}(\bm{x})-j^-_{\rho}(\bm{x})$. We also define a $M\times M$ positive-definite diagonal matrix $\mathsf{A}(\bm{x})$ as $[\mathsf{A}(\bm{x})]_{\rho\rho'}\coloneqq\delta_{\rho\rho'}(j^+_{\rho}(\bm{x})+j^-_{\rho}(\bm{x}))/2$, resulting in $[\mathsf{A}^{\frac{1}{2}}(\bm{x})]_{\rho\rho'}=\delta_{\rho\rho'}\sqrt{(j^+_{\rho}(\bm{x})+j^-_{\rho}(\bm{x}))/2}$. This CLE reduces to the rate equation $\dot{\bm{x}}_t=\mathsf{S}\bm{j}(\bm{x})$ with $\epsilon=0$. \add{Note that the validity of the CLE requires a time-scale separation~\cite{gillespie2000chemical} that is not solely determined by $\epsilon$. At least, however, $\epsilon$ must be chosen small enough to ensure that the probability of a concentration becoming negative is negligible.}

We can regard the CLE~\eqref{chemical Langevin eq.} as a special case of Eq.~\eqref{Langevin eq} by identifying $\mathsf{S}\bm{j}(\bm{x})$ and $\mathsf{S}\mathsf{A}^{\frac{1}{2}}(\bm{x})$ as $\bm{F}(\bm{x})$ and $\mathsf{G}(\bm{x})$, respectively. This identification also yields $\mathsf{D}(\bm{x})=\mathsf{S}\mathsf{A}(\bm{x})\mathsf{S}^{\top}$, which is called the scaled diffusion coefficient matrix~\cite{yoshimura2021thermodynamic}. Although this $\mathsf{D}(\bm{x})$ is always positive semi-definite, it may have $0$ as eigenvalues. This is because the kernel of $\mathsf{S}^{\top}$, denoted by $\ker\mathsf{S}^{\top}$, may include some nonzero vectors. Such vectors correspond to the conservation laws of the CRN~\cite{feinberg2019foundations,polettini2014irreversible,famili2003convex}. 

%Indeed, the positive definiteness of $\mathsf{D}(x)$ is equivalent to $\ker\mathsf{S}^{\top} = \{\bm{0}\}$, where $\bm{0}$ is the $N$-dimensional zero vector. This is verified as follows: If a nonzero vector $\bm{\psi}$ satisfies $\bm{\psi}^{\top}\mathsf{D}(\bm{x})\bm{\psi}\leq 0$, the positive definiteness of $\mathsf{A}(\bm{x})$ forces $\mathsf{S}^{\top}\bm{\psi}=\bm{0}$. This equality is achievable if and only if $\ker\mathsf{S}^{\top}$ contains a nonzero vector.

To obtain a Langevin equation with a positive definite diffusion coefficient matrix, we reduce the degrees of freedom in the CLE~\eqref{chemical Langevin eq.} using the basis of the conservation laws. We let $N_L(<N)$ denote the dimension of $\ker\mathsf{S}^{\top}$, and take a basis of $\ker\mathsf{S}^{\top}$ denoted by $\{\bm{l}^{(\mu)}\}_{\mu=1}^{N_L}$. Under the CLE~\eqref{chemical Langevin eq.}, the quantity $\bm{l}^{(\mu)\top}\bm{x}_t$ is conserved as $d_t(\bm{l}^{(\mu)\top}\bm{x}_t)=\bm{l}^{(\mu)\top}\dot{\bm{x}}_t=\bm{l}^{(\mu)\top}\mathsf{S}\{\bm{j}(\bm{x}_t)+\sqrt{2\epsilon}\mathsf{A}^{\frac{1}{2}}(\bm{x}_t)\bullet\bm{\xi}_t\}=0$, where $d_t(\cdots)$ stands for the time derivative. Letting $c_{\mu}$ denote $\bm{l}^{(\mu)\top}\bm{x}_0$, we obtain $\bm{l}^{(\mu)\top}\bm{x}_t=c_{\mu}$. We can rewrite this relation as the following linear equation:
\begin{align}
    \mathsf{L}\bm{x}_t=\bm{c},
    \label{conservation linear eq.}
\end{align}
where $\mathsf{L}$ denotes an $N_L\times N$ matrix whose $\mu$-th row is $\bm{l}^{(\mu)\top}$, and $\bm{c}$ denotes a $N_L$-dimensional vector whose $\mu$-th element is $c_{\mu}$. This linear equation constrains the time evolution of $\bm{x}_t$.

We split the set of species to solve Eq.~\eqref{conservation linear eq.}. Since $\{\bm{l}^{(\mu)}\}_{\mu=1}^{N_L}$ is a basis of $\ker\mathsf{S}^{\top}$, the rank of $\mathsf{L}$ is $N_L$. This fact implies that there exists a choice of $N_L$ columns of $\mathsf{L}$ such that the resulting $N_L \times N_L$ submatrix $\mathsf{L}_{\mathrm{sq}}$ is invertible. Without loss of generality, we relabel the species so that the $N_L$ columns forming the invertible $\mathsf{L}_{\mathrm{sq}}$ correspond to the last $N_L$ columns of $\mathsf{L}$. Following this relabeling, we can split $\mathsf{L}$ as $\mathsf{L}=(\mathsf{L}_{\mathrm{rem}},\mathsf{L}_{\mathrm{sq}})$, where $\mathsf{L}_{\mathrm{rem}}$ denotes the remaining $N_L\times (N-N_L)$ submatrix. We define $\bm{x}^{\mathrm{ind}}_t$ as the $N-N_L$-dimensional vector consisting of
the first $N-N_L$ elements of $\bm{x}_t$. We also define $\bm{x}^{\mathrm{dep}}_t$ as the $N_L$-dimensional vector consisting of the remaining $N_L$ elements.
These vectors split $\bm{x}_t$ as $\bm{x}_t=(\bm{x}^{\mathrm{ind}\top}_t,\bm{x}^{\mathrm{dep}\top}_t)^{\top}$. The partitions of $\mathsf{L}$ and $\bm{x}_t$ rewrite Eq.~\eqref{conservation linear eq.} as $\mathsf{L}_{\mathrm{rem}}\bm{x}^{\mathrm{ind}}_t+\mathsf{L}_{\mathrm{sq}}\bm{x}^{\mathrm{dep}}_t=\bm{c}$. The invertibility of $\mathsf{L}_{\mathrm{sq}}$ enables us to determine $\bm{x}^{\mathrm{dep}}_t$ by $\bm{x}^{\mathrm{ind}}_t$ as $\bm{x}^{\mathrm{dep}}_t=\mathsf{L}_{\mathrm{sq}}^{-1}(\bm{c}-\mathsf{L}_{\mathrm{rem}}\bm{x}^{\mathrm{ind}}_t)$. This equation implies that we only need to consider the time evolution of $\bm{x}^{\mathrm{ind}}_t$ if we fix the initial state $\bm{x}_0$.

\begin{figure}[!t]
    \centering
    \includegraphics[width=\linewidth]{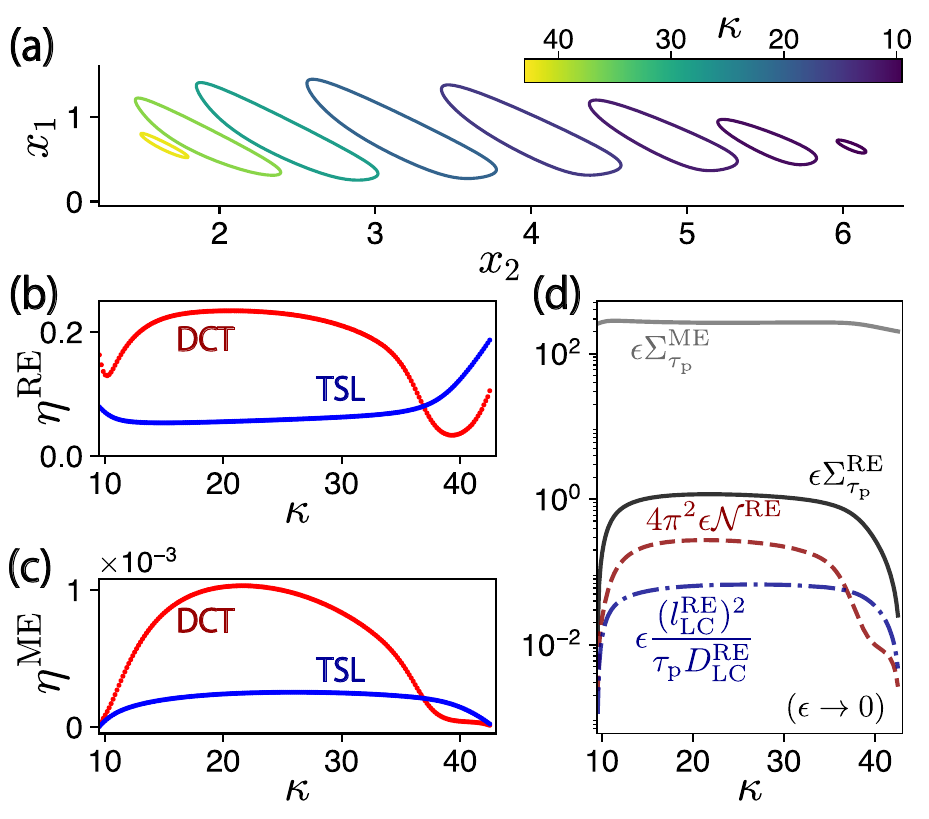}
    \caption{Numerical demonstration of the chemical oscillator with a conservation law. \add{All quantities are computed using the deterministic solutions of the chemical oscillator. See also SM~\cite{SuppMat} for the finite-noise demonstrations with $\kappa=10,20,30,$ and $40$.} (a) The stable limit cycles on $x_2$-$x_1$ plane for several $\kappa$. (b) $\kappa$-dependence of $\eta^{\mathrm{RE}}_{\mathrm{DCT}}$ (red line) and $\eta^{\mathrm{RE}}_{\mathrm{TSL}}$ (blue line). (c) $\kappa$-dependence of $\eta^{\mathrm{ME}}_{\mathrm{DCT}}$ (red line) and $\eta^{\mathrm{ME}}_{\mathrm{TSL}}$ (blue line). (d) The EPs multiplied by $\epsilon$ and their lower bounds. We plot the values in $\epsilon\to0$.}
    \label{fig:chemicaloscillator}
\end{figure}

To consider the time evolution of $\bm{x}^{\mathrm{ind}}_t$, we define $\mathsf{S}'$ as the $(N - N_L)\times M$ submatrix consisting of the first $N-N_L$ rows of $\mathsf{S}$. We also define $\bm{X}[\bm{y}]$ for $N-N_L$-dimensional vector $\bm{y}$ as
\begin{align}
    \bm{X}[\bm{y}]\coloneqq
    \begin{pmatrix}
        \bm{y}\\
        \mathsf{L}_{\mathrm{sq}}^{-1}(\bm{c}-\mathsf{L}_{\mathrm{rem}}\bm{y})
    \end{pmatrix},
\end{align}
which recovers the original $\bm{x}_t$ with $\bm{y}=\bm{x}^{\mathrm{ind}}_t$. Using these quantities, we obtain the Langevin equation describing the time evolution of $\bm{x}^{\mathrm{ind}}_t$ as
\begin{align}
    \dot{\bm{x}}^{\mathrm{ind}}_t=\mathsf{S}'\bm{j}(\bm{X}[\bm{x}^{\mathrm{ind}}_t])+\sqrt{2\epsilon}\mathsf{S}'\mathsf{A}^{\frac{1}{2}}(\bm{X}[\bm{x}^{\mathrm{ind}}_t])\bullet\bm{\xi}_t.
    \label{reduced Langevin eq.}
\end{align}
The corresponding scaled diffusion coefficient matrix is given by $\mathsf{D}'(\bm{y})\coloneqq\mathsf{S}'\mathsf{A}(\bm{X}[\bm{y}])\mathsf{S}'^{\top}$. This matrix is positive-definite because $\ker\mathsf{S}'^{\top}$ contains only the zero vector. Otherwise, we could reduce the degrees of freedom further from $\bm{x}^{\mathrm{ind}}_t$. However, this contradicts the fact that degrees of freedom can be reduced by no more than the number of the independent conservation laws.

\textit{Example: Chemical oscillator with a conservation law}.---%
We consider the CRN, which consists of three internal species $\{X_{\alpha}\}_{\alpha=1}^{3}$, an external species $A$, and the following three reactions: {$\ce{$A$ <=> $X_1 + X_3$}$, $\ce{$X_1 + X_3$ <=> $X_2$}$, and $\ce{$2X_1 + X_2$ <=> $3X_1 + X_3$}$.} 
We label the reactions as $\rho=1$, $2$, and $3$ from left to right. We assume the law of mass action and let $\kappa^{\pm}_{\rho}$ denote the reaction rate constant for forward ($+$) and reverse ($-$) reaction $\rho$. We fix the concentration of $A$ to $1$. We set the reaction rate constants as $(\kappa^+_1,\kappa^+_2,\kappa^+_3,\kappa^-_1,\kappa^-_2,\kappa^-_3)=(1, 30, \kappa, 1, 1, 1)$ with $\kappa\in[9.55, 42.5]$, which provides a stable limit cycle as shown in Fig.~\ref{fig:chemicaloscillator}(a). Under these parameters, the quantities in the CLE, $\mathsf{S}$, $\bm{j}(\bm{x})$, and $\mathsf{A}(\bm{x})$, are given by 
\begin{align}
&\mathsf{S}=
\begin{pmatrix}
1 & -1 & 1 \\
0 &  1 & -1\\
1 & -1 & 1 \\
\end{pmatrix}
,\;\;\bm{j}(\bm{x})=
\begin{pmatrix}
1-x_1x_3\\
30x_1x_3-x_2\\
\kappa x_1^2x_2-x_1^3x_3\\
\end{pmatrix}
,
\end{align}
and $\mathsf{A}(\bm{x})=(1/2)\mathrm{diag}(1+x_1x_3,30x_1x_3+x_2, \kappa x_1^2x_2+x_1^3x_3)$. Here $\ker\mathsf{S}^{\top}$ includes the nonzero vector $(1,0,-1)^{\top}$, which corresponds to the conserved quantity $x_1-x_3$. We introduce $\bm{x}^{\mathrm{ind}}=(x_1,x_2)^{\top}$. Letting $c_1$ denote the value of $x_1-x_3$ at the initial time, we obtain $\bm{X}[\bm{x}^{\mathrm{ind}}]=(x_1,x_2,x_1-c_1)^{\top}$. We also define $\mathsf{S}'$ as the submatrix consisting of the first two rows of $\mathsf{S}$. Then, we obtain the reduced Langevin equation~\eqref{reduced Langevin eq.}. In the numerical demonstration, we set $c_1$ as $-0.9$.

To observe the tightness of the trade-offs in Eq.~\eqref{trade-offs stochastic CRNs}, we introduce the efficiencies for the bounds on $\Sigma^{\mathrm{RE}}_{\tau_{\mathrm{p}}}$ and $\Sigma^{\mathrm{ME}}_{\tau_{\mathrm{p}}}$ as $\eta^{\mathrm{RE}}_{\mathrm{DCT}}\coloneqq4\pi^2\mathcal{N}^{\mathrm{RE}}/{\Sigma^{\mathrm{RE}}_{\tau_{\mathrm{p}}}}$, $\eta^{\mathrm{RE}}_{\mathrm{TSL}}\coloneqq{l_{\mathrm{LC}}^{\mathrm{RE}2}}/(\tau_{\mathrm{p}}D^{\mathrm{RE}}_{\mathrm{LC}}\Sigma^{\mathrm{RE}}_{\tau_{\mathrm{p}}})$, $\eta^{\mathrm{ME}}_{\mathrm{DCT}}\coloneqq4\pi^2\mathcal{N}^{\mathrm{RE}}/{\Sigma^{\mathrm{ME}}_{\tau_{\mathrm{p}}}}$, and $\eta^{\mathrm{ME}}_{\mathrm{TSL}}\coloneqq{l_{\mathrm{LC}}^{\mathrm{RE}2}}/(\tau_{\mathrm{p}}D^{\mathrm{RE}}_{\mathrm{LC}}\Sigma^{\mathrm{ME}}_{\tau_{\mathrm{p}}})$.
We show the $\kappa$-dependence of these efficiencies in Figs.~\ref{fig:chemicaloscillator}(b) and (c). These figures confirm the trade-offs for both EPs. 
In both cases of the reduced Langevin equation and the CME, the behaviors of the dissipation-coherence trade-off and the TSL are different; which trade-off becomes tighter depends on $\kappa$. We also compare the trade-offs for $\Sigma^{\mathrm{RE}}_{\tau_{\mathrm{p}}}$ and for $\Sigma^{\mathrm{ME}}_{\tau_{\mathrm{p}}}$; the difference is most pronounced at the edges of the parameter region of $\kappa$. The limit cycle becomes small and approaches a fixed point in this region. 
Figure~\ref{fig:chemicaloscillator}(d) shows that, in this parameter region, $\epsilon\Sigma^{\mathrm{RE}}_{\tau_{\mathrm{p}}}$ and its lower bounds in the trade-offs approach zero, while $\epsilon\Sigma^{\mathrm{ME}}_{\tau_{\mathrm{p}}}$ remains finite. Thus, the difference in the efficiencies is due to this qualitative difference in the EPs.

%%%%%%%%%% Merge with supplemental materials %%%%%%%%%%
\pagebreak
\setcounter{secnumdepth}{2}
\onecolumngrid
\begin{center}
\textbf{\large Supplemental Material for\\``Duality between dissipation-coherence trade-off and thermodynamic speed limit based on thermodynamic uncertainty relation for stochastic limit cycles''}
\end{center}
%%%%%%%%%% Merge with supplemental materials %%%%%%%%%%
%%%%%%%%%% Prefix a "S" to all equations, figures, tables and reset the counter %%%%%%%%%%
\setcounter{section}{0}
\setcounter{equation}{0}
\setcounter{figure}{0}
\setcounter{table}{0}
\setcounter{page}{1}
\makeatletter
\renewcommand{\theequation}{S\arabic{equation}}
\renewcommand{\thefigure}{S\arabic{figure}}
\renewcommand{\thesection}{\Roman{section}}
%\renewcommand{\bibnumfmt}[1]{[S#1]}
%\renewcommand{\citenumfont}[1]{S#1}
%%%%%%%%%% Prefix a "S" to all equations, figures, tables and reset the counter %%%%%%%%%%

\add{
\section{Equivalence of the steady-state entropy production and $\Sigma_{\tau_{\mathrm{p}}}$ in the weak-noise limit}
\label{secSM:steady-state EP}
}

\add{
\subsection{Steady-state entropy production rate for general $\epsilon$}
\label{subsecSM:steady-state EPR}
We consider the steady-state EP rate (EPR) $\dot{\Sigma}^{\mathrm{st}}$ for general cases. Let $p^{\mathrm{st}}(\bm{x})$ denote the steady-state distribution of the Fokker--Planck equation in Eq.~(2). We also introduce
the mean local velocity in the steady-state as 
\begin{align}
    \bm{\nu}^{\mathrm{st}}(\bm{x})\coloneqq\bm{F}(\bm{x})-\epsilon\bm{\nabla}\cdot\mathsf{D}(\bm{x})-\epsilon\mathsf{D}(\bm{x})\bm{\nabla}\ln p^{\mathrm{st}}(\bm{x}).
    \label{st mean local velocity}
\end{align}
Using these quantities, the steady-state EPR is given by
\begin{align}
    \dot{\Sigma}^{\mathrm{st}}\coloneqq\frac{1}{\epsilon}\int d\bm{x}\,p^{\mathrm{st}}(\bm{x})\bm{\nu}^{\mathrm{st}}(\bm{x})^{\top}\mathsf{D}(\bm{x})^{-1}\bm{\nu}^{\mathrm{st}}(\bm{x}).
\end{align}
Since the steady-state distribution satisfies $-\bm{\nabla}\cdot\{p^{\mathrm{st}}(\bm{x})\bm{\nu}^{\mathrm{st}}(\bm{x})\}=0$, we can rewrite $\dot{\Sigma}^{\mathrm{st}}$ as
\begin{align}
    \dot{\Sigma}^{\mathrm{st}}&=\frac{1}{\epsilon}\int d\bm{x}\,p^{\mathrm{st}}(\bm{x})\{\bm{F}(\bm{x})-\epsilon\bm{\nabla}\cdot\mathsf{D}(\bm{x})\}^{\top}\mathsf{D}(\bm{x})^{-1}\bm{\nu}^{\mathrm{st}}(\bm{x})-\int d\bm{x}\,p^{\mathrm{st}}(\bm{x})\{\bm{\nabla}\ln p^{\mathrm{st}}(\bm{x})\}^{\top}\bm{\nu}^{\mathrm{st}}(\bm{x})\notag\\
    &=\frac{1}{\epsilon}\int d\bm{x}\,p^{\mathrm{st}}(\bm{x})\{\bm{F}(\bm{x})-\epsilon\bm{\nabla}\cdot\mathsf{D}(\bm{x})\}^{\top}\mathsf{D}(\bm{x})^{-1}\bm{\nu}^{\mathrm{st}}(\bm{x})+\int d\bm{x}\,\ln p^{\mathrm{st}}(\bm{x})\bm{\nabla}\cdot\{p^{\mathrm{st}}(\bm{x})\bm{\nu}^{\mathrm{st}}(\bm{x})\}\notag\\
    &=\frac{1}{\epsilon}\int d\bm{x}\,p^{\mathrm{st}}(\bm{x})\{\bm{F}(\bm{x})-\epsilon\bm{\nabla}\cdot\mathsf{D}(\bm{x})\}^{\top}\mathsf{D}(\bm{x})^{-1}\bm{\nu}^{\mathrm{st}}(\bm{x}),
    \label{steady-state EPR 1}
\end{align}
where we performed the integration by parts in the first transform. 
}

\subsection{Steady-state distribution in the weak-noise limit}
\label{subsecSM:steady state weak-noise}
In the weak-noise limit, the steady-state distribution $p^{\mathrm{st}}(\bm{x})$ concentrates on the limit cycle because
we consider the case where the limit cycle is stable. The distribution $p^{\mathrm{st}}(\bm{x})$ is given by the following form:
\begin{align}
    p^{\mathrm{st}}(\bm{x})=\frac{1}{\tau_{\mathrm{p}}}\int_{0}^{\tau_{\mathrm{p}}}dt\,\delta(\bm{x}-\bm{x}_t^{\mathrm{LC}}),
    \label{steady state distribution}
\end{align}
This fact is confirmed as follows: Substituting $\epsilon=0$, the Fokker-Planck equation [Eq.~(2)] reduces to the Liouville equation,
\begin{align}
    \partial_tp_t(\bm{x})=-\bm{\nabla}\cdot(p_t(\bm{x})\bm{F}(\bm{x})).
    \label{Liouville eq}
\end{align}
For any observable $a(\bm{x})$, the distribution in Eq.~\eqref{steady state distribution} satisfies
\begin{align}
    \int d\bm{x}\,a(\bm{x})[-\bm{\nabla}\cdot(p^{\mathrm{st}}(\bm{x})\bm{F}(\bm{x}))]&=\int d\bm{x}\,(\bm{\nabla}a(\bm{x}))\cdot p^{\mathrm{st}}(\bm{x})\bm{F}(\bm{x})\notag\\
    &=\frac{1}{\tau_{\mathrm{p}}}\int_0^{\tau_{\mathrm{p}}} dt\,(\bm{\nabla}a(\bm{x}_t^{\mathrm{LC}}))\cdot\dot{\bm{x}}_t^{\mathrm{LC}}\notag\\
    &=\frac{1}{\tau_{\mathrm{p}}}\{a(\bm{x}_{\tau_{\mathrm{p}}}^{\mathrm{LC}})-a(\bm{x}_{0}^{\mathrm{LC}})\}\notag\\
    &=0.
    \label{confirm Liouville}
\end{align}
In the first transform, we ignore the surface term, since $p^{\mathrm{st}}(\bm{x})$ concentrates on the limit cycle. This equation implies that $p^{\mathrm{st}}(\bm{x})$ in Eq.~\eqref{steady state distribution} is the steady-state distribution of the Liouville equation~\eqref{Liouville eq}.

\add{
\subsection{Steady-state entropy production rate in the weak-noise limit}
\label{subsecSM:steady-state EPR weak-noise}
To obtain the steady-state EPR in the weak-noise limit, we need to substitute Eq.~\eqref{steady state distribution} into Eq.~\eqref{steady-state EPR 1}. However, calculating $\bm{\nabla}\ln p^{\mathrm{st}}(\bm{x})$ in $\bm{\nu}^{\mathrm{st}}(\bm{x})$ is difficult due to the delta function in Eq.~\eqref{steady state distribution}. To avoid this difficulty, we further transform Eq.~\eqref{steady-state EPR 1}. In the following, we may omit arguments for simplicity. Substituting Eq.~\eqref{st mean local velocity} into Eq.~\eqref{steady-state EPR 1}, we obtain
\begin{align}
    \dot{\Sigma}^{\mathrm{st}}&=\frac{1}{\epsilon}\int d\bm{x}\,p^{\mathrm{st}}(\bm{x})\{\bm{F}(\bm{x})-\epsilon\bm{\nabla}\cdot\mathsf{D}(\bm{x})\}^{\top}\mathsf{D}(\bm{x})^{-1}\left\{\bm{F}(\bm{x})-\epsilon\bm{\nabla}\cdot\mathsf{D}(\bm{x})-\epsilon\mathsf{D}(\bm{x})\bm{\nabla}\ln p^{\mathrm{st}}(\bm{x})\right\}\notag\\
    &=\frac{1}{\epsilon}\int d\bm{x}\,p^{\mathrm{st}}(\bm{F}-\epsilon\bm{\nabla}\cdot\mathsf{D})^{\top}\mathsf{D}^{-1}(\bm{F}-\epsilon\bm{\nabla}\cdot\mathsf{D})-\int d\bm{x}\,p^{\mathrm{st}}(\bm{F}-\epsilon\bm{\nabla}\cdot\mathsf{D})^{\top}\bm{\nabla}\ln p^{\mathrm{st}}\notag\\
    &=\frac{1}{\epsilon}\int d\bm{x}\,p^{\mathrm{st}}(\bm{F}-\epsilon\bm{\nabla}\cdot\mathsf{D})^{\top}\mathsf{D}^{-1}(\bm{F}-\epsilon\bm{\nabla}\cdot\mathsf{D})-\int d\bm{x}\,(\bm{F}-\epsilon\bm{\nabla}\cdot\mathsf{D})^{\top}\bm{\nabla} p^{\mathrm{st}}\notag\\
    &=\frac{1}{\epsilon}\int d\bm{x}\,p^{\mathrm{st}}(\bm{F}-\epsilon\bm{\nabla}\cdot\mathsf{D})^{\top}\mathsf{D}^{-1}(\bm{F}-\epsilon\bm{\nabla}\cdot\mathsf{D})+\int d\bm{x}\,p^{\mathrm{st}}\bm{\nabla}\cdot(\bm{F}-\epsilon\bm{\nabla}\cdot\mathsf{D}),
    \label{steady state EPR 2}
\end{align}
where we performed the integration by parts in the last transform. This expression does not include the gradient of the steady-state distribution.
}

\add{
Following the spirit of the Laplace approximation, we substitute Eq.~\eqref{steady state distribution} into Eq.~\eqref{steady state EPR 2}. This yields
\begin{align}
    \dot{\Sigma}^{\mathrm{st}}&=\frac{1}{\epsilon\tau_{\mathrm{p}}}\int_{0}^{\tau_{\mathrm{p}}} dt\,\{\bm{F}(\bm{x}_t^{\mathrm{LC}})-\epsilon\bm{\nabla}\cdot\mathsf{D}(\bm{x}_t^{\mathrm{LC}})\}^{\top}\mathsf{D}(\bm{x}_t^{\mathrm{LC}})^{-1}\{\bm{F}(\bm{x}_t^{\mathrm{LC}})-\epsilon\bm{\nabla}\cdot\mathsf{D}(\bm{x}_t^{\mathrm{LC}})\}\notag\\
    &\qquad\qquad+\frac{1}{\tau_{\mathrm{p}}}\int_{0}^{\tau_{\mathrm{p}}} dt\,[\bm{\nabla}\cdot\bm{F}(\bm{x}_t^{\mathrm{LC}})-\epsilon\bm{\nabla}\cdot\{\bm{\nabla}\cdot\mathsf{D}(\bm{x}_t^{\mathrm{LC}})\}].
\end{align}
Here, the argument $\bm{x}_t^{\mathrm{LC}}$ is substituted after applying all differential operators. Collecting the dominant term at the weak-noise limit, namely the term proportional to $\epsilon^{-1}$, we obtain
\begin{align}
    \dot{\Sigma}^{\mathrm{st}}=\frac{1}{\epsilon\tau_{\mathrm{p}}}\int_{0}^{\tau_{\mathrm{p}}} dt\,\bm{F}(\bm{x}_t^{\mathrm{LC}})^{\top}\mathsf{D}(\bm{x}_t^{\mathrm{LC}})^{-1}\bm{F}(\bm{x}_t^{\mathrm{LC}}).
    \label{steady state EPR weak noise}
\end{align}
As in the case of $\Sigma_{\tau_{\mathrm{p}}}$, this equation strictly means $\lim_{\epsilon\to0}\epsilon\dot{\Sigma}^{\mathrm{st}}=\int_{0}^{\tau_{\mathrm{p}}} dt\,\bm{F}(\bm{x}_t^{\mathrm{LC}})^{\top}\mathsf{D}(\bm{x}_t^{\mathrm{LC}})^{-1}\bm{F}(\bm{x}_t^{\mathrm{LC}})$.
Using Eq.~\eqref{steady state EPR weak noise}, the EP required for one period in the steady state is given by
\begin{align}
    \tau_{\mathrm{p}}\dot{\Sigma}^{\mathrm{st}}=\frac{1}{\epsilon}\int_{0}^{\tau_{\mathrm{p}}} dt\,\bm{F}(\bm{x}_t^{\mathrm{LC}})^{\top}\mathsf{D}(\bm{x}_t^{\mathrm{LC}})^{-1}\bm{F}(\bm{x}_t^{\mathrm{LC}}).
    \label{steady state EP}
\end{align}
This is equivalent to $\Sigma_{\tau_{\mathrm{p}}}$ in Eq.~(3), which was derived by considering the non-stationary time evolution of $p_t(\bm{x})$, where the distribution concentrates at the point $\bm{x}_t^{\mathrm{LC}}$ on the limit cycle at time $t$.
}

\add{
\subsection{Short-time TUR for steady state in the weak-noise limit}
\label{subsecSM:TUR weak-noise}
In the main text, the key inequality [Eq.~(8)] is derived by integrating the short-time TUR~\cite{otsubo2020estimating} for the non-stationary $p_t(\bm{x})$. Here, we show that the short-time TUR for the steady-state distribution [Eq.~\eqref{steady state distribution}] provides the same result as shown below. 
In the steady state, the average of $J^{\bm{\omega}}_t$ is calculated as
\begin{align}
    \mathbb{E}[J^{\bm{\omega}}_t]&=\int d\bm{x}\,p^{\mathrm{st}}(\bm{x})\bm{\omega}(\bm{x})^{\top}\bm{\nu}^{\mathrm{st}}(\bm{x})\notag\\
    &=\int d\bm{x}\,p^{\mathrm{st}}(\bm{x})\bm{\omega}(\bm{x})^{\top}\bm{F}(\bm{x}) - \epsilon\int d\bm{x}\,p^{\mathrm{st}}(\bm{x})\bm{\omega}(\bm{x})^{\top}\bm{\nabla}\cdot\mathsf{D}(\bm{x}) -\epsilon\int d\bm{x}\,\bm{\omega}(\bm{x})^{\top}\mathsf{D}(\bm{x})\bm{\nabla}p^{\mathrm{st}}(\bm{x})\notag\\
    &=\int d\bm{x}\,p^{\mathrm{st}}(\bm{x})\bm{\omega}(\bm{x})^{\top}\bm{F}(\bm{x}) - \epsilon\int d\bm{x}\,p^{\mathrm{st}}(\bm{x})\bm{\omega}(\bm{x})^{\top}\bm{\nabla}\cdot\mathsf{D}(\bm{x}) +\epsilon\int d\bm{x}\,p^{\mathrm{st}}(\bm{x})\bm{\nabla}\cdot(\mathsf{D}(\bm{x})\bm{\omega}(\bm{x})).
    \label{average of current}
\end{align}
where we used Eq.~\eqref{st mean local velocity}. In the weak-noise limit, we can reduce this to
\begin{align}
    \mathbb{E}[J^{\bm{\omega}}_t]=\frac{1}{\tau_{\mathrm{p}}}\int_{0}^{\tau_{\mathrm{p}}} d\bm{t}\,\bm{\omega}(\bm{x}^{\mathrm{LC}}_t)^{\top}\bm{F}(\bm{x}^{\mathrm{LC}}_t),
    \label{average of current weak noise}
\end{align}
where we substituted Eq.~\eqref{steady state distribution} into Eq.~\eqref{average of current} and collected the dominant term. The steady-state variance in the weak-noise limit is also calculated as
\begin{align}
    \mathrm{Var}[J^{\bm{\omega}}_t]&=2(\varDelta t)^{-1}\epsilon\int d\bm{x}\,p^{\mathrm{st}}(\bm{x})\bm{\omega}(\bm{x})^{\top}\mathsf{D}(\bm{x})\bm{\omega}(\bm{x})\notag\\
    &=\frac{2(\varDelta t)^{-1}\epsilon}{\tau_{\mathrm{p}}}\int_{0}^{\tau_{\mathrm{p}}} d\bm{t}\,\bm{\omega}(\bm{x}^{\mathrm{LC}}_t)^{\top}\mathsf{D}(\bm{x}^{\mathrm{LC}}_t)\bm{\omega}(\bm{x}^{\mathrm{LC}}_t).
    \label{variance of current}
\end{align}
Using Eqs.~\eqref{average of current weak noise} and ~\eqref{variance of current}, we obtain the short-time TUR as
\begin{align}
    \dot{\Sigma}^{\mathrm{st}}\geq\frac{2\mathbb{E}[J^{\bm{\omega}}_t]^2}{\varDelta t\mathrm{Var}[J^{\bm{\omega}}_t]}=\dfrac{\left[\displaystyle\int_{0}^{\tau_{\mathrm{p}}} d\bm{t}\,\bm{\omega}(\bm{x}^{\mathrm{LC}}_t)^{\top}\bm{F}(\bm{x}^{\mathrm{LC}}_t)\right]^2}{\epsilon\tau_{\mathrm{p}}\displaystyle\int_{0}^{\tau_{\mathrm{p}}} d\bm{t}\,\bm{\omega}(\bm{x}^{\mathrm{LC}}_t)^{\top}\mathsf{D}(\bm{x}^{\mathrm{LC}}_t)\bm{\omega}(\bm{x}^{\mathrm{LC}}_t)}.
\end{align}
Multiplying both sides of this equation by $\tau_{\mathrm{p}}$, we obtain the counterpart of Eq.~(8),
\begin{align}
    \tau_{\mathrm{p}}\dot{\Sigma}^{\mathrm{st}}\geq\dfrac{\left[\displaystyle\int_{0}^{\tau_{\mathrm{p}}} d\bm{t}\,\bm{\omega}(\bm{x}^{\mathrm{LC}}_t)^{\top}\bm{F}(\bm{x}^{\mathrm{LC}}_t)\right]^2}{\epsilon\displaystyle\int_{0}^{\tau_{\mathrm{p}}} d\bm{t}\,\bm{\omega}(\bm{x}^{\mathrm{LC}}_t)^{\top}\mathsf{D}(\bm{x}^{\mathrm{LC}}_t)\bm{\omega}(\bm{x}^{\mathrm{LC}}_t)}.
\end{align}
This implies that the trade-offs are derived as consequences of the short-time TUR, even when considering the steady state rather than the time-evolving probability distribution used in the main text.
}

\section{Observable-independence of correlation time}
\label{secSM:correlation time}

Here we show that the correlation time $\tau_{\mathrm{c}}$ is common to most observables except in some special cases. We introduce the conditional probability distribution of the state being $\bm{x}$ at time $t$, given that the state was $\bm{x}'$ at time $0$, denoted by $p_t(\bm{x}|\bm{x}')$.
Using this quantity and the steady-state distribution $p^{\mathrm{st}}(\bm{x})$, we can obtain the correlation function as
\begin{align}
    \langle a(\bm{x}_t)a(\bm{x}_0)\rangle_{\mathrm{st}}=\int d\bm{x}\int d\bm{x}'\,a(\bm{x})a(\bm{x}')p_t(\bm{x}|\bm{x}')p^{\mathrm{st}}(\bm{x}').
    \label{correlation function}
\end{align}

To transform Eq.~\eqref{correlation function}, we introduce the spectral decomposition of $p_t(\bm{x}|\bm{x}')$~\cite{gardiner2009stochastic}. We can rewrite the Fokker--Planck equation in Eq.~(2) as 
\begin{align}
    \partial_t p_t(\bm{x})&=-\sum_{\alpha}\partial_{x_{\alpha}}(F_{\alpha}(\bm{x})p_t(\bm{x}))+\epsilon\sum_{\alpha,\beta}\partial_{x_{\alpha}}\partial_{x_{\beta}}\{[\mathsf{D}(\bm{x})]_{\alpha\beta}p_t(\bm{x})\}\notag\\
    &=\mathcal{L}[p_t(\bm{x})],
\end{align}
where we introduced the Fokker--Planck operator $\mathcal{L}$ based on the linearity of the Fokker--Planck equation. We also define the formal adjoint of the Fokker--Planck operator (Kolmogorov’s backward operator) $\mathcal{L}^{\dag}$ as
\begin{align}
    \mathcal{L}^{\dag}[q(\bm{x})]\coloneqq\sum_{\alpha}F_{\alpha}(\bm{x})\partial_{x_{\alpha}}q(\bm{x})+\epsilon\sum_{\alpha,\beta}[\mathsf{D}(\bm{x})]_{\alpha\beta}\partial_{x_{\alpha}}\partial_{x_{\beta}}q(\bm{x}).
\end{align}
For simplicity, suppose that the operators $\mathcal{L}$ and $\mathcal{L}^{\dag}$ possess a discrete set of eigenvalues $\{\Lambda_{n}\}_{n=0}^{\infty}$. We arrange the eigenvalues so that $\mathrm{Re}(\Lambda_0)\geq\mathrm{Re}(\Lambda_1)\geq\mathrm{Re}(\Lambda_2)\geq\cdots$, where $\mathrm{Re}(\Lambda_n)$ denotes the real part of $\Lambda_n$. Let $P_{n}(\bm{x})$ and $Q_{n}(\bm{x})$ denote the eigenfunctions of $\mathcal{L}$ and $\mathcal{L}^{\dag}$ corresponding to the eigenvalue $\Lambda_n$, i.e., $\mathcal{L}P_{n}(\bm{x})=\Lambda_nP_{n}(\bm{x})$ and $\mathcal{L}^{\dag}Q_n(\bm{x})=\Lambda_nQ_n(\bm{x})$. These eigenfunctions satisfy the following biorthogonality:
\begin{align}
    \int d\bm{x}\,Q_n(\bm{x})P_{m}(\bm{x})=\delta_{nm}.
\end{align}
To ensure the unique existence of the steady state, we impose $\mathrm{Re}(\Lambda_n)<0$ for all $n\geq1$. Then, we can obtain $\Lambda_0=0$ and set $P_0(\bm{x})=p^{\mathrm{st}}(\bm{x})$ and $Q_0(\bm{x})=1$.
The eigenvalues and eigenfunctions lead to the spectral decomposition of $p_t(\bm{x}|\bm{x}')$ as
\begin{align}
    p_t(\bm{x}|\bm{x}')=p^{\mathrm{st}}(\bm{x})+\sum_{n\geq1}\mathrm{e}^{\Lambda_n t}Q_n(\bm{x}')P_{n}(\bm{x}),
    \label{decomp of propagator}
\end{align}
with the completeness of the eigenfunctions~\cite{gardiner2009stochastic}.

We calculate the correlation time using the eigenvalues of $\mathcal{L}$.
We introduce the set $\mathbb{N}_1\coloneqq\{n\mid\mathrm{Re}(\Lambda_n)=\mathrm{Re}(\Lambda_1)\}$.
Substituting Eq.~\eqref{decomp of propagator} into Eq.~\eqref{correlation function}, we obtain
\begin{align}
    \langle a(\bm{x}_t)a(\bm{x}_0)\rangle_{\mathrm{st}}&=\int d\bm{x}\int d\bm{x}'\,a(\bm{x})a(\bm{x}')\left[p^{\mathrm{st}}(\bm{x})+\sum_{n\geq1}\mathrm{e}^{\Lambda_n t}Q_n(\bm{x}')P_{n}(\bm{x})\right]p^{\mathrm{st}}(\bm{x}')\notag\\
    &=\left\{\int d\bm{x}\,a(\bm{x})p^{\mathrm{st}}(\bm{x})\right\}^2+\sum_{n\geq1}\mathrm{e}^{\Lambda_n t}\int d\bm{x}\,a(\bm{x})P_{n}(\bm{x})\int d\bm{x}'\,a(\bm{x}')Q_{n}(\bm{x}')p^{\mathrm{st}}(\bm{x}')\notag\\
    &=\left\{\int d\bm{x}\,a(\bm{x})p^{\mathrm{st}}(\bm{x})\right\}^2+\mathrm{e}^{\mathrm{Re}(\Lambda_1) t}\sum_{n\in\mathbb{N}_1}\mathrm{e}^{\mathrm{i}\mathrm{Im}(\Lambda_n) t}\int d\bm{x}\,a(\bm{x})P_{n}(\bm{x})\int d\bm{x}'\,a(\bm{x}')Q_{n}(\bm{x}')p^{\mathrm{st}}(\bm{x}')\notag\\
    &\phantom{=\left\{\int d\bm{x}\,a(\bm{x})p^{\mathrm{st}}(\bm{x})\right\}^2}+\sum_{n\notin\mathbb{N}_1\cup\{0\}}\mathrm{e}^{\mathrm{Re}(\Lambda_n) t}\mathrm{e}^{\mathrm{i}\mathrm{Im}(\Lambda_n) t}\int d\bm{x}\,a(\bm{x})P_{n}(\bm{x})\int d\bm{x}'\,a(\bm{x}')Q_{n}(\bm{x}')p^{\mathrm{st}}(\bm{x}'),
\end{align}
where $\mathrm{Im}(\Lambda_n)$ denotes the imaginary part of $\Lambda_n$.
Considering the long-time limit, $\mathrm{Re}(\Lambda_1)\geq\mathrm{Re}(\Lambda_n)$ for $n\geq2$ yields
\begin{align}
    \langle a(\bm{x}_t)a(\bm{x}_0)\rangle_{\mathrm{st}}\simeq\left\{\int d\bm{x}\,a(\bm{x})p^{\mathrm{st}}(\bm{x})\right\}^2+\mathrm{e}^{\mathrm{Re}(\Lambda_1) t}\sum_{n\in\mathbb{N}_1}\mathrm{e}^{\mathrm{i}\mathrm{Im}(\Lambda_n) t}\int d\bm{x}\,a(\bm{x})P_{n}(\bm{x})\int d\bm{x}'\,a(\bm{x}')Q_{n}(\bm{x}')p^{\mathrm{st}}(\bm{x}'),
    \label{general correlation function in long time}
\end{align}
except when $\int d\bm{x}\,a(\bm{x})P_{n}(\bm{x})=0$ or $\int d\bm{x}'\,a(\bm{x}')Q_{n}(\bm{x}')p^{\mathrm{st}}(\bm{x}')=0$ holds for each $n\in\mathbb{N}_1$.
Thus, for observables that do not satisfy this special condition, the correlation time is common and given by $\tau_{\mathrm{c}}=-1/\mathrm{Re}(\Lambda_1)$.

\section{Correlation time in the weak-noise limit}
\label{secSM:correlation time in the weak-noise regeme}
We derive the explicit form of the correlation time in the weak-noise limit [Eq.~(4)]. This quantity was obtained using the large deviation and the WKB approximation in Refs.~\cite{vance1996fluctuations,gaspard2002trace,gaspard2002correlation}. Here we consider the linear Langevin equation that governs small deviations from the limit cycle~\cite{cheng2021stochastic} instead of large deviation.

\subsection{Gaussian approximation of the conditional probability distribution}

Using the steady-state distribution in the weak-noise limit in Eq.~\eqref{steady state distribution}, the correlation function~\eqref{correlation function} reduces to
\begin{align}
    \langle a(\bm{x}_t)a(\bm{x}_0)\rangle_{\mathrm{st}}=\frac{1}{\tau_{\mathrm{p}}}\int d\bm{x}\int_{0}^{\tau_{\mathrm{p}}} dt'\,a(\bm{x})a(\bm{x}^{\mathrm{LC}}_{t'})p_t(\bm{x}|\bm{x}^{\mathrm{LC}}_{t'}).
    \label{correlation function with ssd}
\end{align}
In the following, we approximate the conditional probability distribution $p_t(\bm{x}|\bm{x}^{\mathrm{LC}}_{t'})$ as the probability distribution for the linear Langevin equation governing the small deviations from the limit cycle.

Let us consider the original Langevin equation in Eq.~(1). If the initial state is given as $\bm{x}_0=\bm{x}_{t'}^{\mathrm{LC}}$, the stability of the limit cycle and $\epsilon\ll1$ make the state at time $t$ in the vicinity of $\bm{x}_{t'+t}^{\mathrm{LC}}$. We can introduce a small deviation from the limit cycle $\bm{z}_{t|t'}\coloneqq\bm{x}_t-\bm{x}_{t'+t}^{\mathrm{LC}}$. This deviation satisfies $\|\bm{z}_{t|t'}\|=O(\sqrt{\epsilon})$ because the noise term is $O(\sqrt{\epsilon})$ in the original Langevin equation. Substituting $\bm{x}_t=\bm{x}_{t'+t}^{\mathrm{LC}}+\bm{z}_{t|t'}$ into the Langevin equation and collecting terms of $O(\sqrt{\epsilon})$, we obtain the following linear Langevin equation:
\begin{align}
    \dot{\bm{z}}_{t|t'}=\mathsf{K}_{t'+t}\bm{z}_{t|t'}+\sqrt{2\epsilon}\mathsf{G}(\bm{x}_{t'+t}^{\mathrm{LC}})\cdot\bm{\xi}_t.
    \label{linear Langevin eq.}
\end{align}
Here, the matrix $\mathsf{K}_{t'+t}$ stands for the Jacobian of $\bm{F}(\bm{x})$ at $\bm{x}^{\mathrm{LC}}_{t'+t}$.
Let $g_{t|t'}(\bm{z}|\bm{z}')$ denote the conditional probability distribution of the state being $\bm{z}$ at time $t$, given that the state was $\bm{z}'$ at time $0$ for the linear Langevin equation~\eqref{linear Langevin eq.}. Using this distribution, we can approximate $p_t(\bm{x}|\bm{x}^{\mathrm{LC}}_{t'})$ as
\begin{align}
    p_t(\bm{x}|\bm{x}^{\mathrm{LC}}_{t'})\simeq g_{t|t'}(\bm{x}-\bm{x}_{t'+t}^{\mathrm{LC}}|\bm{0}).
    \label{Gaussian approximation}
\end{align}
Note that the distribution $g_{t|t'}(\bm{z}|\bm{0})$ is a Gaussian distribution, since the underlying Langevin equation~\eqref{linear Langevin eq.} is linear and $g_{0|t'}(\bm{z}|\bm{0})=\delta(\bm{z})$.

To obtain the Gaussian distribution $g_{t|t'}(\bm{z}|\bm{0})$, we introduce the mean and covariance matrix of $\bm{z}_{t|t'}$ as $\bm{\mu}_{t|t'}\coloneqq\langle\bm{z}_{t|t'}\rangle$ and $\mathsf{V}_{t|t'}\coloneqq\langle(\bm{z}_{t|t'}-\bm{\mu}_{t|t'})(\bm{z}_{t|t'}-\bm{\mu}_{t|t'})^{\top}\rangle$, respectively. Combining these definitions and the linear Langevin equation~\eqref{linear Langevin eq.}, the mean and covariance matrix evolve according to the following equations~\cite{risken1989fokker}:
\begin{align}
    \dot{\bm{\mu}}_{t|t'}&=\mathsf{K}_{t'+t}\bm{\mu}_{t|t'},\notag\\
    \dot{\mathsf{V}}_{t|t'}&=\mathsf{K}_{t'+t}\mathsf{V}_{t|t'}+\mathsf{V}_{t|t'}\mathsf{K}_{t'+t}^{\top}+2\epsilon\mathsf{D}(\bm{x}^{\mathrm{LC}}_{t'+t}).
    \label{mean covariance time evolution}
\end{align}
We can obtain the mean and covariance matrix of $g_{t|t'}(\bm{z}|\bm{0})$ by solving Eq.~\eqref{mean covariance time evolution} with the initial conditions $\bm{\mu}_{0|t'}=\bm{0}$ and $\mathsf{V}_{0|t'}=\mathsf{O}$ ($\mathsf{O}$ is the zero matrix). 
This problem is solved by the fundamental matrix $\Phi_t$ as
\begin{align}
    \bm{\mu}_{t|t'}&=\bm{0},\notag\\
    \mathsf{V}_{t|t'}&=2\epsilon\Phi_{t'+t}\left\{\int_{0}^{t}ds\,\Phi_{t'+s}^{-1}\mathsf{D}(\bm{x}^{\mathrm{LC}}_{t'+s})\Phi_{t'+s}^{-1\top}\right\}\Phi_{t'+t}^{\top}.
    \label{mean covariance}
\end{align}
We can easily verify that these solutions satisfy Eq.~\eqref{mean covariance time evolution} and the initial conditions.

\subsection{Long time regime}
In the following, we focus on a sufficiently large $t$ to obtain the correlation time. Due to Eqs.~\eqref{correlation function with ssd}, ~\eqref{Gaussian approximation}, and ~\eqref{mean covariance}, the behavior of the correlation function at large $t$ is determined by that of $\Phi_{t'+t}$.

Since we consider a large $t$, we can take a large integer $N_{\mathrm{p}}\gg1$ and $t^{\star}\in[0,\tau_{\mathrm{p}})$ such that $t=N_{\mathrm{p}}\tau_{\mathrm{p}}+t^{\star}$. Using these quantities, we can express $\Phi_{t'+t}$ in $\mathsf{V}_{t|t'}$ as
\begin{align}
    \Phi_{t'+t}&=\Phi_{t'+N_{\mathrm{p}}\tau_{\mathrm{p}}+t^{\star}}=\Phi_{t'+t^{\star}}(\Phi_{\tau_{\mathrm{p}}})^{N_{\mathrm{p}}}=\Phi_{t'+t^{\star}}\sum_{i=1}^{N}\lambda_i^{N_{\mathrm{p}}}\bm{u}^{(i)}\bm{v}^{(i)\top}.
\end{align}
Here, we also used $\Phi_{t+\tau_{\mathrm{p}}}=\Phi_{t}\Phi_{\tau_{\mathrm{p}}}$ and the spectral decomposition of $\Phi_{\tau_{\mathrm{p}}}$. Since $\lambda_1=1$ and $|\lambda_i|<1$ for $i\geq2$ hold, the large $N_{\mathrm{p}}$ leads to
\begin{align}
    \Phi_{t'+t}\simeq\Phi_{t'+t^{\star}}\bm{u}^{(1)}\bm{v}^{(1)\top}=\dot{\bm{x}}_{t'+t^{\star}}^{\mathrm{LC}}\bm{v}^{(1)\top}=\dot{\bm{x}}_{t'+t}^{\mathrm{LC}}\bm{v}^{(1)\top},
\end{align}
where we used $\bm{u}^{(1)}=\dot{\bm{x}}_0^{\mathrm{LC}}$ and the periodicity $\dot{\bm{x}}_{t'+t^{\star}}^{\mathrm{LC}}=\dot{\bm{x}}_{t'+N_{\mathrm{p}}\tau_{\mathrm{p}}+t^{\star}}^{\mathrm{LC}}=\dot{\bm{x}}_{t'+t}^{\mathrm{LC}}$. Substituting this approximation into the covariance matrix in Eq.~\eqref{mean covariance}, we obtain the form of $\mathsf{V}_{t|t'}$ at large $t$ as
\begin{align}
    \mathsf{V}_{t|t'}&\simeq2\epsilon\dot{\bm{x}}_{t'+t}^{\mathrm{LC}}\bm{v}^{(1)\top}\left\{\int_{0}^{t}ds\,\Phi_{t'+s}^{-1}\mathsf{D}(\bm{x}^{\mathrm{LC}}_{t'+s})\Phi_{t'+s}^{-1\top}\right\}\bm{v}^{(1)}\dot{\bm{x}}_{t'+t}^{\mathrm{LC}\top}\notag\\
    &=2\epsilon\dot{\bm{x}}_{t'+t}^{\mathrm{LC}}\left\{\int_{0}^{t}ds\,\bm{\zeta}_{t'+s}^{\top}\mathsf{D}(\bm{x}^{\mathrm{LC}}_{t'+s})\bm{\zeta}_{t'+s}\right\}\dot{\bm{x}}_{t'+t}^{\mathrm{LC}\top}\notag\\
    &=2\epsilon\dot{\bm{x}}_{t'+t}^{\mathrm{LC}}\left\{N_{\mathrm{p}}\int_{0}^{\tau_{\mathrm{p}}}ds\,\bm{\zeta}_{t'+s}^{\top}\mathsf{D}(\bm{x}^{\mathrm{LC}}_{t'+s})\bm{\zeta}_{t'+s}+\int_{0}^{t^{\star}}ds\,\bm{\zeta}_{t'+s}^{\top}\mathsf{D}(\bm{x}^{\mathrm{LC}}_{t'+s})\bm{\zeta}_{t'+s}\right\}\dot{\bm{x}}_{t'+t}^{\mathrm{LC}\top}\notag\\
    &\simeq2\epsilon N_{\mathrm{p}}\dot{\bm{x}}_{t'+t}^{\mathrm{LC}}\left\{\int_{0}^{\tau_{\mathrm{p}}}ds\,\bm{\zeta}_{t'+s}^{\top}\mathsf{D}(\bm{x}^{\mathrm{LC}}_{t'+s})\bm{\zeta}_{t'+s}\right\}\dot{\bm{x}}_{t'+t}^{\mathrm{LC}\top}\notag\\
    &\simeq\frac{2\epsilon t}{\tau_{\mathrm{p}}}\dot{\bm{x}}_{t'+t}^{\mathrm{LC}}\left\{\int_{0}^{\tau_{\mathrm{p}}}ds\,\bm{\zeta}_{s}^{\top}\mathsf{D}(\bm{x}^{\mathrm{LC}}_{s})\bm{\zeta}_{s}\right\}\dot{\bm{x}}_{t'+t}^{\mathrm{LC}\top}.
    \label{V large t}
\end{align}
In the second line, we used $\bm{\zeta}_t=\Phi_t^{-1\top}\bm{v}^{(1)}$. In the fourth line, we used $N_{\mathrm{p}}\gg1$, which also yields the approximation $N_{\mathrm{p}}\simeq t/\tau_{\mathrm{p}}$ used in the last line. We also used the periodicity of $\bm{\zeta}_t$ and $\bm{x}^{\mathrm{LC}}_{t}$ in the last line.

We obtain the form of $g_{t|t'}(\bm{z}|\bm{0})$ at large $t$ using the form of $\mathsf{V}_{t|t'}$ in Eq.~\eqref{V large t}. We define the projection matrix onto the direction parallel to the limit cycle as $\Pi^{\mathrm{LC}}_{t}\coloneqq\dot{\bm{x}}_{t}^{\mathrm{LC}}\dot{\bm{x}}_{t}^{\mathrm{LC}\top}/\|\dot{\bm{x}}_{t}^{\mathrm{LC}}\|^2$. This projection matrix rewrites $\mathsf{V}_{t|t'}$ as
\begin{align}
    \mathsf{V}_{t|t'}\simeq\frac{2\epsilon t}{\tau_{\mathrm{p}}}\|\dot{\bm{x}}_{t'+t}^{\mathrm{LC}}\|^2\mathcal{I}_{\bm{\zeta}}\Pi^{\mathrm{LC}}_{t'+t},
\end{align}
where we introduced the abbreviation
\begin{align}
    \mathcal{I}_{\bm{\zeta}}\coloneqq\int_{0}^{\tau_{\mathrm{p}}}dt\,\bm{\zeta}_{t}^{\top}\mathsf{D}(\bm{x}^{\mathrm{LC}}_{t})\bm{\zeta}_{t}.
\end{align}
This expression of $\mathsf{V}_{t|t'}$ implies that the fluctuations perpendicular to the limit cycle disappear, leaving only those parallel to it. Thus, we obtain
\begin{align}
    g_{t|t'}(\bm{z}|\bm{0})\simeq\delta((\mathsf{I}-\Pi_{t'+t}^{\mathrm{LC}})\bm{z})\sqrt{\frac{\tau_{\mathrm{p}}}{4\pi\epsilon t\|\dot{\bm{x}}_{t'+t}^{\mathrm{LC}}\|^2\mathcal{I}_{\bm{\zeta}}}}\exp\left[{-\frac{\tau_{\mathrm{p}}\bm{z}^{\top}\Pi^{\mathrm{LC}}_{t'+t}\bm{z}}{4\epsilon t\|\dot{\bm{x}}_{t'+t}^{\mathrm{LC}}\|^2\mathcal{I}_{\bm{\zeta}}}}\right],
\end{align}
where $\mathsf{I}$ denotes the identity matrix. For simplicity, we introduce
\begin{align}
    z_{\parallel}\coloneqq\frac{\bm{z}^{\top}\dot{\bm{x}}^{\mathrm{LC}}_{t'+t}}{\|\dot{\bm{x}}^{\mathrm{LC}}_{t'+t}\|},\;\;\bm{z}_{\perp}\coloneqq(\mathsf{I}-\Pi_{t'+t}^{\mathrm{LC}})\bm{z}.
\end{align}
These quantities yield
\begin{align}
    \bm{z}=z_{\parallel}\frac{\dot{\bm{x}}^{\mathrm{LC}}_{t'+t}}{\|\dot{\bm{x}}^{\mathrm{LC}}_{t'+t}\|}+\bm{z}_{\perp},
\end{align}
and
\begin{align}
    g_{t|t'}(\bm{z}|\bm{0})\simeq\delta(\bm{z}_{\perp})\sqrt{\frac{\tau_{\mathrm{p}}}{4\pi\epsilon t\|\dot{\bm{x}}_{t'+t}^{\mathrm{LC}}\|^2\mathcal{I}_{\bm{\zeta}}}}\exp\left[{-\frac{\tau_{\mathrm{p}} z_{\parallel}^2}{4\epsilon t\|\dot{\bm{x}}_{t'+t}^{\mathrm{LC}}\|^2\mathcal{I}_{\bm{\zeta}}}}\right].
    \label{detail Gaussian}
\end{align}

\subsection{Calculation of correlation function}
Here we perform the integration in Eq.~\eqref{correlation function with ssd} using the above approximations. Substituting Eq.~\eqref{Gaussian approximation} into Eq.~\eqref{correlation function with ssd}, we obtain
\begin{align}
    \langle a(\bm{x}_t)a(\bm{x}_0)\rangle_{\mathrm{st}}&\simeq\frac{1}{\tau_{\mathrm{p}}}\int d\bm{x}\int_{0}^{\tau_{\mathrm{p}}} dt'\,a(\bm{x})a(\bm{x}^{\mathrm{LC}}_{t'})g_{t|t'}(\bm{x}-\bm{x}^{\mathrm{LC}}_{t'+t}|\bm{0})\notag\\
    &=\frac{1}{\tau_{\mathrm{p}}}\int d\bm{z}\int_{0}^{\tau_{\mathrm{p}}} dt'\,a(\bm{x}^{\mathrm{LC}}_{t'+t}+\bm{z})a(\bm{x}^{\mathrm{LC}}_{t'})g_{t|t'}(\bm{z}|\bm{0}).
\end{align}
We substitute Eq.~\eqref{detail Gaussian} into this equation and perform the integration on $\bm{z}_{\perp}$. Then, we obtain
\begin{align}
    \langle a(\bm{x}_t)a(\bm{x}_0)\rangle_{\mathrm{st}}&\simeq\frac{1}{\tau_{\mathrm{p}}}\int dz_{\parallel}\int_{0}^{\tau_{\mathrm{p}}} dt'\,a(\bm{x}^{\mathrm{LC}}_{t'+t}+z_{\parallel}\|\dot{\bm{x}}_{t'+t}^{\mathrm{LC}}\|^{-1}\dot{\bm{x}}_{t'+t}^{\mathrm{LC}})a(\bm{x}^{\mathrm{LC}}_{t'})\sqrt{\frac{\tau_{\mathrm{p}}}{4\pi\epsilon t\|\dot{\bm{x}}_{t'+t}^{\mathrm{LC}}\|^2\mathcal{I}_{\bm{\zeta}}}}\exp\left[-\frac{\tau_{\mathrm{p}} z_{\parallel}^2}{4\epsilon t\|\dot{\bm{x}}_{t'+t}^{\mathrm{LC}}\|^2\mathcal{I}_{\bm{\zeta}}}\right]\notag\\
    &=\frac{1}{\tau_{\mathrm{p}}}\int dz\int_{0}^{\tau_{\mathrm{p}}} dt'\,a(\bm{x}^{\mathrm{LC}}_{t'+t}+z\dot{\bm{x}}_{t'+t}^{\mathrm{LC}})a(\bm{x}^{\mathrm{LC}}_{t'})\sqrt{\frac{\tau_{\mathrm{p}}}{4\pi\epsilon t\mathcal{I}_{\bm{\zeta}}}}\exp\left[-\frac{\tau_{\mathrm{p}} z^2}{4\epsilon t\mathcal{I}_{\bm{\zeta}}}\right],
\end{align}
where we introduced a new variable $z=z_{\parallel}\|\dot{\bm{x}}_{t'+t}^{\mathrm{LC}}\|^{-1}$. Due to the weak-noise limit, only small $z$ contributes to the integral. This fact allows us to approximate $\bm{x}^{\mathrm{LC}}_{t'+t}+z\dot{\bm{x}}_{t'+t}^{\mathrm{LC}}$ as $\bm{x}^{\mathrm{LC}}_{t'+t+z}$, which yields
\begin{align}
    \langle a(\bm{x}_t)a(\bm{x}_0)\rangle_{\mathrm{st}}\simeq\frac{1}{\tau_{\mathrm{p}}}\int dz\int_{0}^{\tau_{\mathrm{p}}} dt'\,a(\bm{x}^{\mathrm{LC}}_{t'+t+z})a(\bm{x}^{\mathrm{LC}}_{t'})\sqrt{\frac{\tau_{\mathrm{p}}}{4\pi\epsilon t\mathcal{I}_{\bm{\zeta}}}}\exp\left[-\frac{\tau_{\mathrm{p}} z^2}{4\epsilon t\mathcal{I}_{\bm{\zeta}}}\right],
\end{align}
Since the function $a(\bm{x}^{\mathrm{LC}}_{t})$ is periodic, we can expand it into the Fourier series as
\begin{align}
    a(\bm{x}^{\mathrm{LC}}_{t})=\sum_{n\in\mathbb{Z}}\hat{a}_n\exp\left[\mathrm{i}\frac{2\pi t}{\tau_{\mathrm{p}}}n\right],
\end{align}
where $\hat{a}_n$ is the corresponding Fourier coefficient. Using this Fourier series, we obtain
\begin{align}
    \langle a(\bm{x}_t)a(\bm{x}_0)\rangle_{\mathrm{st}}
    &\simeq\sum_{n\in\mathbb{Z}}\sum_{m\in\mathbb{Z}}\hat{a}_n\hat{a}_m\exp\left[\mathrm{i}\frac{2n\pi t}{\tau_{\mathrm{p}}}\right]\left[\frac{1}{\tau_{\mathrm{p}}}\int_{0}^{\tau_{\mathrm{p}}} dt'\,\exp\left\{\mathrm{i}\frac{2\pi}{\tau_{\mathrm{p}}}(n+m)t'\right\}\right]\notag\\
    &\phantom{\sum_{n\in\mathbb{Z}}\sum_{m\in\mathbb{Z}}\hat{a}_n\hat{a}_m\exp\left[\mathrm{i}\frac{2n\pi t}{\tau_{\mathrm{p}}}\right]}\qquad\times\left[\sqrt{\frac{\tau_{\mathrm{p}}}{4\pi\epsilon t\mathcal{I}_{\bm{\zeta}}}}\int dz\,\exp\left\{-\frac{\tau_{\mathrm{p}} z^2}{4\epsilon t\mathcal{I}_{\bm{\zeta}}}+\mathrm{i}\frac{2n\pi z}{\tau_{\mathrm{p}}}\right\}\right]\notag\\
    &=\sum_{n\in\mathbb{Z}}\sum_{m\in\mathbb{Z}}\hat{a}_n\hat{a}_m\delta_{-nm}\exp\left[-\frac{4n^2\pi^2\epsilon\mathcal{I}_{\bm{\zeta}}}{\tau_{\mathrm{p}}^3}t+\mathrm{i}\frac{2n\pi}{\tau_{\mathrm{p}}}t\right]\notag\\
    &=\sum_{n\in\mathbb{Z}}\hat{a}_n\hat{a}_{-n}\exp\left[-\frac{4n^2\pi^2\epsilon\mathcal{I}_{\bm{\zeta}}}{\tau_{\mathrm{p}}^3}t+\mathrm{i}\frac{2n\pi}{\tau_{\mathrm{p}}}t\right]\notag\\
    &=\hat{a}_0\hat{a}_{0}+\sum_{n\geq1}2\mathrm{Re}\left(\hat{a}_n\hat{a}_{-n}\exp\left[\mathrm{i}\frac{2n\pi}{\tau_{\mathrm{p}}}t\right]\right)\exp\left[-\frac{4n^2\pi^2\epsilon\mathcal{I}_{\bm{\zeta}}}{\tau_{\mathrm{p}}^3}t\right],
\end{align}
where we also used the fact that $a(\bm{x})$ is the real-valued function. In the long-time regime, we obtain
\begin{align}
    \langle a(\bm{x}_t)a(\bm{x}_0)\rangle_{\mathrm{st}}\simeq|\hat{a}_0|^2+2|\hat{a}_1|^2\exp\left[-\frac{4\pi^2\epsilon\mathcal{I}_{\bm{\zeta}}}{\tau_{\mathrm{p}}^3}t\right]\cos\left(\frac{2\pi}{\tau_{\mathrm{p}}}t\right),
    \label{asymp correlation function}
\end{align}
except when $\hat{a}_1=0$ holds. This exception may correspond to the one introduced in Sec.~\ref{secSM:correlation time}. Thus, the correlation time is obtained as
\begin{align}
    \tau_{\mathrm{c}}=\left[\frac{4\pi^2\epsilon\mathcal{I}_{\bm{\zeta}}}{\tau_{\mathrm{p}}^3}\right]^{-1}=\frac{\tau_{\mathrm{p}}^3}{4\pi^2\epsilon}\left[\int_{0}^{\tau_{\mathrm{p}}}dt\,\bm{\zeta}_{t}^{\top}\mathsf{D}(\bm{x}^{\mathrm{LC}}_{t})\bm{\zeta}_{t}\right]^{-1}.
\end{align}

\section{Trade-offs for stochastic chemical reaction networks}

This section provides additional details of the trade-offs for stochastic CRNs discussed in the main text and End Matter: In Sec.~\ref{sec:trade-offs for reduced LE}, we define the quantities in the trade-offs for the reduced Langevin equation [Eq.~(21)]. In Sec.~\ref{sec:hierarchy of EPs}, we prove the hierarchy between EPs $\Sigma_{\tau_{\mathrm{p}}}^{\mathrm{ME}}\geq\Sigma_{\tau_{\mathrm{p}}}^{\mathrm{RE}}$.
\add{
Section~\ref{sec:hierarchy of EPs} also shows the equivalence between the steady-state EP associated with the CME and $\Sigma_{\tau_{\mathrm{p}}}^{\mathrm{ME}}$.
}
In Sec.~\ref{sec:TUR for CRN}, we show the derivation of the trade-offs for $\Sigma_{\tau_{\mathrm{p}}}^{\mathrm{ME}}$ based on the TUR for deterministic CRNs~\cite{yoshimura2021thermodynamic,yoshimura2023housekeeping}.

\subsection{Quantities in the trade-offs for the reduced Langevin equation}
\label{sec:trade-offs for reduced LE}

Here, we introduce the quantities in the trade-offs by applying the definitions to the reduced Langevin equation in Eq.~(21). In the following, we split $\bm{x}^{\mathrm{LC}}_t$ as $\bm{x}^{\mathrm{LC}}_t=(\bm{x}^{\mathrm{LC},\mathrm{ind}\top}_t,\bm{x}^{\mathrm{LC},\mathrm{dep}\top}_t)^{\top}$ as in the case of general $\bm{x}_t$. We note that the time derivative $\dot{\bm{x}}^{\mathrm{LC},\mathrm{ind}}_t$ implies $\mathsf{S}'\bm{j}(\bm{X}[\bm{x}^{\mathrm{LC},\mathrm{ind}}_t])$.

We first introduce the EP associated with the reduced Langevin equation. The positive definiteness of $\mathsf{D}'(\bm{y})$ allows us to apply Eq.~(3) to Eq.~(21) as
\begin{align}
    \Sigma_{\tau_{\mathrm{p}}}^{\mathrm{RE}}&\coloneqq\frac{1}{\epsilon}\int_{0}^{\tau_{\mathrm{p}}}dt\,\bm{j}(\bm{X}[\bm{x}^{\mathrm{LC},\mathrm{ind}}_t])^{\top}\mathsf{S}'^{\top}\mathsf{D}'(\bm{x}^{\mathrm{LC},\mathrm{ind}}_t)^{-1}\mathsf{S}'\bm{j}(\bm{X}[\bm{x}^{\mathrm{LC},\mathrm{ind}}_t])\notag\\
    &=\frac{1}{\epsilon}\int_{0}^{\tau_{\mathrm{p}}}dt\,\bm{j}(\bm{x}^{\mathrm{LC}}_t)^{\top}\mathsf{S}'^{\top}[\mathsf{S}'\mathsf{A}(\bm{x}^{\mathrm{LC}}_t)\mathsf{S}'^{\top}]^{-1}\mathsf{S}'\bm{j}(\bm{x}^{\mathrm{LC}}_t)
    \label{EP for reduced Langevin eq.}
\end{align}
in the weak-noise (macroscopic) limit. Here we also used $\bm{X}[\bm{x}^{\mathrm{LC},\mathrm{ind}}_t]=\bm{x}^{\mathrm{LC}}_t$.

We also introduce the number of coherent oscillations $\mathcal{N}^{\mathrm{RE}}$. 
To define this quantity, we define the $(N-N_L)\times(N-N_L)$ Jacobian $\mathsf{K}'_{t}$ as $[\mathsf{K}'_{t}]_{\alpha\beta}\coloneqq\left.\partial_{y_{\beta}}[\mathsf{S}'\bm{j}(\bm{X}[\bm{y}])]_{\alpha}\right|_{\bm{y}=\bm{x}^{\mathrm{LC},\mathrm{ind}}_t}$ for $1\leq\alpha,\beta\leq N-N_L$. 
Note that we need to consider the $y$-derivative of $\bm{X}[\bm{y}]$ due to the chain rule. 
We also introduce the fundamental matrix corresponding to this Jacobian $\Phi'_{t}$ as the solution of $\dot{\Phi}_t'=\mathsf{K}'_{t}\Phi'_t$ with the initial condition that $\Phi'_{0}$ is the identity matrix. 
As in the generic case, the corresponding monodromy matrix $\Phi'_{\tau_{\mathrm{p}}}$ is supposed to be diagonalizable. Let $\{\lambda_i'\}_{i=1}^{N-N_L}$ denote the set of eigenvalues. 
The stability of the limit cycle allows us to assume $\lambda'_1=1$ and $|\lambda'_i|<1$ for $i\geq2$. 
We also let $\bm{u}'^{(1)}$ and $\bm{v}'^{(1)}$ denote the right and left eigenvectors of $\Phi_{\tau_{\mathrm{p}}}'$ corresponding to the eigenvalue $\lambda'_1=1$, which are normalized by $\bm{v}'^{(1)\top}\bm{u}'^{(1)}=1$. We set $\bm{u}'^{(1)}$ to $\dot{\bm{x}}^{\mathrm{LC},\mathrm{ind}}_0$. 
Under this setting, we introduce a vector-valued function $\bm{\zeta}_t'\coloneqq\Phi_{t}'^{-1\top}\bm{v}'^{(1)}$. 
This function allows us to represent the correlation time of the reduced Langevin equation in Eq.~(21) as
\begin{align}
    \mathcal{N}^{\mathrm{RE}}\coloneqq\frac{\tau_{\mathrm{p}}^2}{4\pi^2\epsilon}\left[\int_{0}^{\tau_{\mathrm{p}}}dt\,{\bm{\zeta}'_t}^{\top}\mathsf{D}'(\bm{x}^{\mathrm{LC},\mathrm{ind}}_t)\bm{\zeta}_t'\right]^{-1}.
    \label{Num of coherent reduced LE}
\end{align}

In addition, we introduce the quantities in the TSL, i.e., 
the length of the limit cycle and the intensity of diffusion measured along the limit cycle.
These quantities are defined with $\dot{\bm{x}}_t^{\mathrm{LC,ind}}$ as
\begin{align}
    l_{\mathrm{LC}}^{\mathrm{RE}}\coloneqq\int_{0}^{\tau_{\mathrm{p}}}dt\,\|\dot{\bm{x}}_t^{\mathrm{LC},\mathrm{ind}}\|,\;\;D_{\mathrm{LC}}^{\mathrm{RE}}\coloneqq\frac{\epsilon\displaystyle\int_{0}^{\tau_{\mathrm{p}}}dt\,\dot{\bm{x}}_t^{\mathrm{LC},\mathrm{ind}\top}\mathsf{D}'(\bm{x}_t^{\mathrm{LC},\mathrm{ind}})\dot{\bm{x}}_t^{\mathrm{LC},\mathrm{ind}}}{\displaystyle\int_{0}^{\tau_{\mathrm{p}}}dt\,\|\dot{\bm{x}}_t^{\mathrm{LC},\mathrm{ind}}\|^2}.
\end{align}

Applying the derivation for the trade-offs in the main text to the quantities introduced here, we obtain a part of Eq.~(13):
\begin{align}
    \frac{\Sigma_{\tau_{\mathrm{p}}}^{\mathrm{RE}}}{4\pi^2}\geq\mathcal{N}^{\mathrm{RE}},\;\;\Sigma^{\mathrm{RE}}_{\tau_{\mathrm{p}}}\geq\frac{{(l^{\mathrm{RE}}_{\mathrm{LC}}})^2}{\tau_{\mathrm{p}} D^{\mathrm{RE}}_{\mathrm{LC}}}.
    \label{trade-off for reduced LE}
\end{align}

\subsection{Hierarchy of entropy productions}
\label{sec:hierarchy of EPs}

Here we prove that the EP associated with the reduced Langevin equation provides a lower bound of the one associated with the corresponding CME.

\add{
\subsubsection{Chemical master equation and entropy production rate}
}

\add{
We introduce the CME. We express the microstate of this stochastic CRN by a vector $\bm{m}=(m_1,m_2,\cdots,m_N)^{\top}\in\mathbb{Z}_{\geq0}^{N}$, whose $\alpha$-th element is the number of molecules of $X_{\alpha}$. Here, $\mathbb{Z}_{\geq0}^{N}$ denotes the set of $N$-dimensional vectors of nonnegative integers. We also define $\bm{S}_{\rho}$ as a vector whose $\alpha$-th element is $(\mathsf{S})_{\alpha\rho}$. This vector corresponds to the change in the number of molecules when reaction $\rho$ occurs once in the forward direction.
Let $P_{t}(\bm{m})$ be the probability that the particle number distribution is given by $\bm{m}$ at time $t$. The time evolution of this probability distribution is described by the following CME:
\begin{align}
    \dot{P}_t(\bm{m})=&\sum_{\rho=1}^M\left[R_{\rho}^{+}(\bm{m}-\bm{S}_{\rho})P_t(\bm{m}-\bm{S}_{\rho})-R_{\rho}^{-}(\bm{m})P_t(\bm{m})\right]\notag\\
    &+\sum_{\rho=1}^M\left[R_{\rho}^{-}(\bm{m}+\bm{S}_{\rho})P_t(\bm{m}+\bm{S}_{\rho})-R_{\rho}^{+}(\bm{m})P_t(\bm{m})\right].
    \label{CME}
\end{align}
Here, $R_{\rho}^{+}(\bm{m})$ ($R_{\rho}^{-}(\bm{m})$) is the rate at which reaction $\rho$ occurs in the forward (reverse) direction when the microstate is $\bm{m}$. To ensure the nonnegativity of the number of molecules, we impose the following two assumptions: (i) $R_{\rho}^{\pm}(\bm{m})$ is zero if $\bm{m}\pm\bm{S}_{\rho}$ has at least one negative component and (ii)  $P_0(\bm{m})$ is zero if $\bm{m}$ has at least one negative component. These conditions lead to $P_t(\bm{m})=0$ for $\bm{m}\notin\mathbb{Z}_{\geq0}^N$. 
}

\add{
In stochastic thermodynamics, the EPR associated with this CME is defined by
\begin{align}
    \dot{\Sigma}_t^{\mathrm{ME}}=\sum_{\bm{m}\in\mathbb{Z}_{\geq0}^N}\sum_{\rho=1}^M[R_{\rho}^{+}(\bm{m})P_t(\bm{m})-R_{\rho}^{-}(\bm{m}+\bm{S}_{\rho})P_t(\bm{m}+\bm{S}_{\rho})]\ln\frac{R_{\rho}^{+}(\bm{m})P_t(\bm{m})}{R_{\rho}^{-}(\bm{m}+\bm{S}_{\rho})P_t(\bm{m}+\bm{S}_{\rho})}.
    \label{EPR CME general}
\end{align}
Here, if $\bm{m}+\bm{S}_{\rho}$ has negative elements, we can regard the summand as $0$. This is because the assumptions lead to $R^{+}_{\rho}(\bm{m})=0$ and $P_t(\bm{m}+\bm{S}_{\rho})=0$, which conclude $R_{\rho}^{+}(\bm{m})P_t(\bm{m})=R_{\rho}^{-}(\bm{m}+\bm{S}_{\rho})P_t(\bm{m}+\bm{S}_{\rho})=0$. 
We can rewrite this EPR as
\begin{align}
    \dot{\Sigma}_t^{\mathrm{ME}}=\frac{d}{dt}S[P_t] + \sum_{\bm{m}\in\mathbb{Z}_{\geq0}^N}\sum_{\rho=1}^M[R_{\rho}^{+}(\bm{m})P_t(\bm{m})-R_{\rho}^{-}(\bm{m}+\bm{S}_{\rho})P_t(\bm{m}+\bm{S}_{\rho})]\ln\frac{R_{\rho}^{+}(\bm{m})}{R_{\rho}^{-}(\bm{m}+\bm{S}_{\rho})},
    \label{EPR CME general with Shannon}
\end{align}
where we define the Shannon entropy as $S[P_t]\coloneqq-\sum_{\bm{m}\in\mathbb{Z}_{\geq0}^N}P_t(\bm{m})\ln P_t(\bm{m})$. This expression is obtained by the following calculation:
\begin{align}
    \frac{d}{dt}S[P_t]&=-\sum_{\bm{m}\in\mathbb{Z}_{\geq0}^N}\dot{P}_t(\bm{m})\ln P_t(\bm{m})-\sum_{\bm{m}\in\mathbb{Z}_{\geq0}^N}\dot{P}_t(\bm{m})\notag\\
    &=\sum_{\bm{m}\in\mathbb{Z}_{\geq0}^N}\sum_{\rho=1}^M[R_{\rho}^{+}(\bm{m})P_t(\bm{m})-R_{\rho}^{-}(\bm{m}+\bm{S}_{\rho})P_t(\bm{m}+\bm{S}_{\rho})]\ln P_t(\bm{m})\notag\\
    &\qquad-\sum_{\bm{m}\in\mathbb{Z}_{\geq0}^N}\sum_{\rho=1}^M[R_{\rho}^{+}(\bm{m}-\bm{S}_{\rho})P_t(\bm{m}-\bm{S}_{\rho})-R_{\rho}^{-}(\bm{m})P_t(\bm{m})]\ln P_t(\bm{m})\notag\\
    &=\sum_{\bm{m}\in\mathbb{Z}_{\geq0}^N}\sum_{\rho=1}^M[R_{\rho}^{+}(\bm{m})P_t(\bm{m})-R_{\rho}^{-}(\bm{m}+\bm{S}_{\rho})P_t(\bm{m}+\bm{S}_{\rho})]\ln\frac{P_t(\bm{m})}{P_t(\bm{m}+\bm{S}_{\rho})},
\end{align}
where we used the conservation of probability $\sum_{\bm{m}\in\mathbb{Z}_{\geq0}^N}\dot{P}_t(\bm{m})=0$ and the CME in Eq.~\eqref{CME}. 
}

\add{
Since $\epsilon^{-1}$ is the volume of the system, we can regard $\bm{x}=\epsilon\bm{m}\in\mathbb{R}_{\geq0}^N$ as the concentration distribution. In the limit of weak-noise (large system-size), $\epsilon\to0$, we can regard $\bm{x}$ as a continuous variable. We can also define the probability distribution that the concentration distribution becomes $\bm{x}$ at time $t$ as $p^{\mathrm{ME}}_{t}(\bm{x})\coloneqq\epsilon^{-N}P_t(\epsilon^{-1}\bm{x})$.
To ensure thermodynamic consistency, we also impose the following scaling on $R^{\pm}_{\rho}(\bm{m})$~\cite{falasco2025macroscopic}: 
\begin{align}
    \lim_{\epsilon\to 0}\epsilon R^{\pm}_{\rho}(\epsilon^{-1}\bm{x})=j^{\pm}_{\rho}(\bm{x}).
    \label{scaling of jump rate}
\end{align}
}

\add{
\subsubsection{Entropy production for the time evolution along the limit cycle}
}

\add{
We consider the weak-noise limit and the time evolution of $p^{\mathrm{ME}}_t(\bm{x})$ as it concentrates on the limit cycle within the finite time interval $[0, \tau_{\mathrm{p}}]$. We assume that the initial distribution $p_0(\bm{x})$ is concentrated on $\bm{x}^{\mathrm{LC}}_{0}$. Due to the weak-noise limit, for $t\in[0, \tau_{\mathrm{p}}]$, we can assume the large-deviation form of the probability distribution as $p^{\mathrm{ME}}_t(\bm{x})\asymp\mathrm{e}^{-I^{\mathrm{ME}}_t(\bm{x})/\epsilon}$. Here, the rate function $I^{\mathrm{ME}}_t(\bm{x})$ is nonnegative and has the unique minimum $0$ at $\bm{x}^{\mathrm{LC}}_{t}$. 
}

\add{
We derive the EP associated with the above time evolution of $p^{\mathrm{ME}}_t(\bm{x})$~\cite{falasco2025macroscopic,santolin2025dissipation}. In general, we can rewrite Eq.~\eqref{EPR CME general} as
\begin{align}
    \dot{\Sigma}_t^{\mathrm{ME}}=\sum_{\bm{m}\in\mathbb{Z}_{\geq0}^N}\sum_{\rho=1}^M P_t(\bm{m})\left[R_{\rho}^{+}(\bm{m})\ln\frac{R_{\rho}^{+}(\bm{m})P_t(\bm{m})}{R_{\rho}^{-}(\bm{m}+\bm{S}_{\rho})P_t(\bm{m}+\bm{S}_{\rho})}-R_{\rho}^{-}(\bm{m})\ln\frac{R_{\rho}^{+}(\bm{m}-\bm{S}_{\rho})P_t(\bm{m}-\bm{S}_{\rho})}{R_{\rho}^{-}(\bm{m})P_t(\bm{m})}\right].
    \label{rearrange EPR CME}
\end{align}
We consider the asymptotic behaviors of quantities in Eq.~\eqref{rearrange EPR CME} in $\epsilon\to0$. Equation~\eqref{scaling of jump rate} leads to 
\begin{align}
    R_{\rho}^{\pm}(\bm{m})=\frac{1}{\epsilon}j^{\pm}_{\rho}(\bm{x})+O(1)
    \label{asymptotic jump rate}
\end{align}
and
\begin{align}
    \ln\frac{R^{\pm}(\bm{m})}{R^{\mp}(\bm{m}\pm\bm{S}_{\rho})}=\ln\frac{\epsilon^{-1}j^{\pm}_{\rho}(\bm{x})+O(1)}{\epsilon^{-1}j^{\mp}_{\rho}(\bm{x}\pm\epsilon\bm{S}_{\rho})+O(1)}=\ln\frac{j^{\pm}_{\rho}(\bm{x})}{j^{\mp}_{\rho}(\bm{x})}+O(\epsilon).
    \label{asymptotic ratio jumps}
\end{align}
Because the large-deviation form yields $\bm{\nabla}\ln p_t^{\mathrm{ME}}(\bm{x})=-\epsilon^{-1}\bm{\nabla}I_t^{\mathrm{ME}}(\bm{x})$, we also obtain
\begin{align}
    \ln\frac{P_t(\bm{m}\pm\bm{S}_{\rho})}{P_t(\bm{m})}=\ln p_t^{\mathrm{ME}}(\bm{x}\pm\epsilon\bm{S}_{\rho})-\ln p_t^{\mathrm{ME}}(\bm{x})=\mp\bm{S}_{\rho}^{\top}\bm{\nabla}I_t^{\mathrm{ME}}(\bm{x})+O(\epsilon).
    \label{asymptotic ratio probs}
\end{align}
Substituting Eqs.~\eqref{asymptotic jump rate}, ~\eqref{asymptotic ratio jumps}, and ~\eqref{asymptotic ratio probs} into Eq.~\eqref{rearrange EPR CME}, we obtain
\begin{align}
    \dot{\Sigma}_t^{\mathrm{ME}}=\frac{1}{\epsilon}\int d\bm{x}\, p_t^{\mathrm{ME}}(\bm{x})\sum_{\rho=1}^{M}\left[\{j^{+}_{\rho}(\bm{x})-j^{-}_{\rho}(\bm{x})\}\left\{\ln\frac{j^{+}_{\rho}(\bm{x})}{j^{-}_{\rho}(\bm{x})}+\bm{S}_{\rho}^{\top}\bm{\nabla}I_t^{\mathrm{ME}}(\bm{x})\right\}\right],
    \label{asymptotic EPR CME}
\end{align}
where we replaced the sum $\sum_{\bm{m}\in\mathbb{Z}_{\geq0}^N}P_t(\bm{m})\cdots=\sum_{\bm{m}\in\mathbb{Z}_{\geq0}^N}\epsilon^{N}p_t^{\mathrm{ME}}(\epsilon\bm{m})\cdots$ with an integral $\int d\bm{x}\,p^{\mathrm{ME}}_t(\bm{x})\cdots$. We also ignored $O(1)$ terms. The large deviation form $p^{\mathrm{ME}}_t(\bm{x})\asymp\mathrm{e}^{-I^{\mathrm{ME}}_t(\bm{x})/\epsilon}$ allows us to apply the Laplace approximation to Eq.~\eqref{asymptotic EPR CME}, which leads to
\begin{align}
    \dot{\Sigma}_t^{\mathrm{ME}}&=\frac{1}{\epsilon}\sum_{\rho=1}^{M}\left[\{j^{+}_{\rho}(\bm{x}^{\mathrm{LC}}_t)-j^{-}_{\rho}(\bm{x}^{\mathrm{LC}}_t)\}\left\{\ln\frac{j^{+}_{\rho}(\bm{x}^{\mathrm{LC}}_t)}{j^{-}_{\rho}(\bm{x}^{\mathrm{LC}}_t)}+\bm{S}_{\rho}^{\top}\bm{\nabla}I_t^{\mathrm{ME}}(\bm{x}^{\mathrm{LC}}_t)\right\}\right]\notag\\
    &=\frac{1}{\epsilon}\sum_{\rho=1}^{M}\{j^{+}_{\rho}(\bm{x}^{\mathrm{LC}}_t)-j^{-}_{\rho}(\bm{x}^{\mathrm{LC}}_t)\}\ln\frac{j^{+}_{\rho}(\bm{x}^{\mathrm{LC}}_t)}{j^{-}_{\rho}(\bm{x}^{\mathrm{LC}}_t)}.
\end{align}
Here we also used $\bm{\nabla}I_t^{\mathrm{ME}}(\bm{x}^{\mathrm{LC}}_t)=\bm{0}$, which follows from the fact that $I_t^{\mathrm{ME}}(\bm{x}^{\mathrm{LC}}_t)$ has the unique minimum $0$ at $\bm{x}^{\mathrm{LC}}_{t}$. Integrating this EPR over $t\in[0,\tau_{\mathrm{p}}]$, we obtain the asymptotic behaviors of the EP for the time evolution along the limit cycle as
\begin{align}
    \Sigma_{\tau_{\mathrm{p}}}^{\mathrm{ME}}=\frac{1}{\epsilon}\int_{0}^{\tau_{\mathrm{p}}}dt\,\sum_{\rho=1}^{M}\{j^{+}_{\rho}(\bm{x}^{\mathrm{LC}}_t)-j^{-}_{\rho}(\bm{x}^{\mathrm{LC}}_t)\}\ln\frac{j^{+}_{\rho}(\bm{x}^{\mathrm{LC}}_t)}{j^{-}_{\rho}(\bm{x}^{\mathrm{LC}}_t)}.
    \label{EP for CME}
\end{align}
Due to the weak-noise limit, this form equals the EP for the deterministic time evolution~\cite{rao2016nonequilibrium,ge2016mesoscopic,kondepudi2014modern} multiplied by the system size $1/\epsilon$.
}

\add{
\subsubsection{Steady-state entropy production}
}

\add{
We consider the steady-state EPR. Let $P^{\mathrm{st}}(\bm{m})$ denote the steady-state distribution of the CME in Eq.~\eqref{CME}. In the steady state, the change in the Shannon entropy in Eq.~\eqref{EPR CME general with Shannon} vanishes. Thus, we obtain the steady-state EPR $\dot{\Sigma}^{\mathrm{ME,st}}$ as
\begin{align}
    \dot{\Sigma}^{\mathrm{ME,st}}&=\sum_{\bm{m}\in\mathbb{Z}_{\geq0}^N}\sum_{\rho=1}^M[R_{\rho}^{+}(\bm{m})P^{\mathrm{st}}(\bm{m})-R_{\rho}^{-}(\bm{m}+\bm{S}_{\rho})P^{\mathrm{st}}(\bm{m}+\bm{S}_{\rho})]\ln\frac{R_{\rho}^{+}(\bm{m})}{R_{\rho}^{-}(\bm{m}+\bm{S}_{\rho})}\notag\\
    &=\sum_{\bm{m}\in\mathbb{Z}_{\geq0}^N}\sum_{\rho=1}^M P^{\mathrm{st}}(\bm{m})\left[R_{\rho}^{+}(\bm{m})\ln\frac{R_{\rho}^{+}(\bm{m})}{R_{\rho}^{-}(\bm{m}+\bm{S}_{\rho})}-R_{\rho}^{-}(\bm{m})\ln\frac{R_{\rho}^{+}(\bm{m}-\bm{S}_{\rho})}{R_{\rho}^{-}(\bm{m})}\right].
    \label{steady-state EPR CME general}
\end{align}
}

\add{
In particular, we consider the weak-noise limit. 
We define $p^{\mathrm{ME,st}}(\bm{x})$ as $p^{\mathrm{ME,st}}(\bm{x})=\epsilon^{-N}P^{\mathrm{st}}(\epsilon^{-1}\bm{x})$.
As in the case of the Langevin equation, we may assume that the steady-state distribution is given by
\begin{align}
    p^{\mathrm{ME,st}}(\bm{x})=\frac{1}{\tau_{\mathrm{p}}}\int_{0}^{\tau_{\mathrm{p}}}dt\,\delta(\bm{x}-\bm{x}_t^{\mathrm{LC}}).
    \label{steady-state distribution CME}
\end{align}
Using Eqs.~\eqref{asymptotic jump rate} and ~\eqref{asymptotic ratio jumps}, we can rewrite the EPR in Eq.~\eqref{steady-state EPR CME general} as 
\begin{align}
    \dot{\Sigma}^{\mathrm{ME,st}}=\frac{1}{\epsilon}\int d\bm{x}\,p^{\mathrm{ME,st}}(\bm{x})\sum_{\rho=1}^M\{j^{+}_{\rho}(\bm{x})-j^{-}_{\rho}(\bm{x})\}\ln\frac{j^{+}_{\rho}(\bm{x})}{j^{-}_{\rho}(\bm{x})},
\end{align}
where we replaced the sum $\sum_{\bm{m}\in\mathbb{Z}_{\geq0}^N}P^{\mathrm{st}}(\bm{m})\cdots=\sum_{\bm{m}\in\mathbb{Z}_{\geq0}^N}\epsilon^{N}p^{\mathrm{ME,st}}(\epsilon\bm{m})\cdots$ in Eq.~\eqref{steady-state EPR CME general} with an integral $\int d\bm{x}\,p^{\mathrm{ME,st}}(\bm{x})\cdots$. We also ignored $O(1)$ terms. Substituting Eq.~\eqref{steady-state distribution CME} into this equation, we obtain
\begin{align}
    \dot{\Sigma}^{\mathrm{ME,st}}=\frac{1}{\epsilon\tau_{\mathrm{p}}}\int_{0}^{\tau_{\mathrm{p}}}dt\sum_{\rho=1}^{M}\{j^{+}_{\rho}(\bm{x}^{\mathrm{LC}}_t)-j^{-}_{\rho}(\bm{x}^{\mathrm{LC}}_t)\}\ln\frac{j^{+}_{\rho}(\bm{x}^{\mathrm{LC}}_t)}{j^{-}_{\rho}(\bm{x}^{\mathrm{LC}}_t)}.
\end{align}
Integrating this EPR over $t\in[0,\tau_{\mathrm{p}}]$, we verify that the steady-state EP required for one period of the oscillation in the weak-noise limit is equivalent to $\Sigma_{\tau_{\mathrm{p}}}^{\mathrm{ME}}$ [Eq.~\eqref{EP for CME}] as
\begin{align}
    \Sigma^{\mathrm{ME,st}}_{\tau_{\mathrm{p}}}=\frac{1}{\epsilon}\int_{0}^{\tau_{\mathrm{p}}}dt\,\sum_{\rho=1}^{M}\{j^{+}_{\rho}(\bm{x}^{\mathrm{LC}}_t)-j^{-}_{\rho}(\bm{x}^{\mathrm{LC}}_t)\}\ln\frac{j^{+}_{\rho}(\bm{x}^{\mathrm{LC}}_t)}{j^{-}_{\rho}(\bm{x}^{\mathrm{LC}}_t)}=\Sigma_{\tau_{\mathrm{p}}}^{\mathrm{ME}}.
    \label{steady state EP CME weak noise}
\end{align}
}

\subsubsection{Proof of the hierarchy}

The EP $\Sigma_{\tau_{\mathrm{p}}}^{\mathrm{ME}}$ is bounded by the EP associated with the reduced Langevin equation as
\begin{align}
    \Sigma_{\tau_{\mathrm{p}}}^{\mathrm{ME}}\geq\Sigma_{\tau_{\mathrm{p}}}^{\mathrm{RE}}.
    \label{hierarchy of EPs}
\end{align}
This hierarchy and Eq.~\eqref{trade-off for reduced LE} yield the trade-offs in Eq.~(13).
We prove the hierarchy~\eqref{hierarchy of EPs} as follows: We introduce the pseudo EP~\cite{shiraishi2021optimal} as
\begin{align}
    \Pi_{\tau_{\mathrm{p}}}=\frac{1}{\epsilon}\int_{0}^{\tau_{\mathrm{p}}}dt\sum_{\rho}\frac{2j_{\rho}(\bm{x}^{\mathrm{LC}}_t)^2}{j^+_{\rho}(\bm{x}^{\mathrm{LC}}_t)+j^-_{\rho}(\bm{x}^{\mathrm{LC}}_t)}=\frac{1}{\epsilon}\int_{0}^{\tau_{\mathrm{p}}}dt\,\bm{j}(\bm{x}^{\mathrm{LC}}_t)^{\top}\mathsf{A}(\bm{x}^{\mathrm{LC}}_t)^{-1}\bm{j}(\bm{x}^{\mathrm{LC}}_t).
    \label{pseudo EP}
\end{align}
This pseudo EP satisfies
\begin{align}
    \Sigma_{\tau_{\mathrm{p}}}^{\mathrm{ME}}\geq\Pi_{\tau_{\mathrm{p}}}
    \label{ineq pseudo EP}
\end{align}
due to the inequality between the arithmetic and logarithmic means, $(j^+_{\rho}-j^{-}_{\rho})/(\ln j^{+}_{\rho}-\ln j^{-}_{\rho})\leq(j^+_{\rho}+j^{-}_{\rho})/2$. Here and in the following proof, we omit the arguments $\bm{x}^{\mathrm{LC}}_t$ for simplicity. We consider the following minimization problem:
\begin{align}
    \min_{\bm{j}'|\mathsf{S}'\bm{j}'=\mathsf{S}'\bm{j}}\bm{j}'^{\top}\mathsf{A}^{-1}\bm{j}'.
\end{align}
The corresponding action is given by
\begin{align}
    \mathcal{I}(\bm{j}',\bm{\lambda})=\frac{1}{2}\bm{j}'^{\top}\mathsf{A}^{-1}\bm{j}'-(\mathsf{S}'\bm{j}'-\mathsf{S}'\bm{j})^{\top}\bm{\lambda},
\end{align}
where $\bm{\lambda}$ is the Lagrange multiplier. The optimizer $(\bm{j}^{\ast}, \bm{\lambda}^{\ast})$ satisfies the Euler--Lagrange equation as
\begin{align}
    0=\left.\partial_{\bm{j}'}\mathcal{I}(\bm{j}',\bm{\lambda})\right|_{\bm{j}'=\bm{j}^{\ast},\bm{\lambda}=\bm{\lambda}^{\ast}}=\mathsf{A}^{-1}\bm{j}^{\ast}-\mathsf{S}'^{\top}\bm{\lambda}^{\ast},
\end{align}
which implies $\bm{j}^{\ast}=\mathsf{A}\mathsf{S}'^{\top}\bm{\lambda}^{\ast}$. Substituting this equality into the condition $\mathsf{S}'\bm{j}^{\ast}=\mathsf{S}'\bm{j}$, we obtain
\begin{align}
    \mathsf{S}'\mathsf{A}\mathsf{S}'^{\top}\bm{\lambda}^{\ast}=\mathsf{S}'\bm{j}.
\end{align}
Since $\mathsf{S}'\mathsf{A}\mathsf{S}'^{\top}$ is invertible, we obtain $\bm{\lambda}^{\ast}=(\mathsf{S}'\mathsf{A}\mathsf{S}'^{\top})^{-1}\mathsf{S}'\bm{j}$, which yields $\bm{j}^{\ast}=\mathsf{A}\mathsf{S}'^{\top}(\mathsf{S}'\mathsf{A}\mathsf{S}'^{\top})^{-1}\mathsf{S}'\bm{j}$. Thus, we obtain
\begin{align}
    \bm{j}^{\top}\mathsf{A}^{-1}\bm{j}&\geq\min_{\bm{j}'|\mathsf{S}'\bm{j}'=\mathsf{S}'\bm{j}}\bm{j}'^{\top}\mathsf{A}^{-1}\bm{j}'\notag\\
    &=[\mathsf{A}\mathsf{S}'^{\top}(\mathsf{S}'\mathsf{A}\mathsf{S}'^{\top})^{-1}\mathsf{S}'\bm{j}]^{\top}\mathsf{A}^{-1}\mathsf{A}\mathsf{S}'^{\top}(\mathsf{S}'\mathsf{A}\mathsf{S}'^{\top})^{-1}\mathsf{S}'\bm{j}\notag\\
    &=\bm{j}^{\top}\mathsf{S}'^{\top}(\mathsf{S}'\mathsf{A}\mathsf{S}'^{\top})^{-1}\mathsf{S}'\bm{j},
\end{align}
where the first inequality follows from the fact that $\bm{j}$ is a candidate for the minimization problem. 
Performing the time integral of both sides of the resulting inequality, we obtain
\begin{align}
    \int_{0}^{\tau_{\mathrm{p}}}dt\,\bm{j}^{\top}\mathsf{A}^{-1}\bm{j}\geq \int_{0}^{\tau_{\mathrm{p}}}dt\,\bm{j}^{\top}\mathsf{S}'^{\top}(\mathsf{S}'\mathsf{A}\mathsf{S}'^{\top})^{-1}\mathsf{S}'\bm{j}.
\end{align}
This inequality implies $\Pi_{\tau_{\mathrm{p}}}\geq\Sigma_{\tau_{\mathrm{p}}}^{\mathrm{RE}}$ due to the definition of EPs in Eqs.~\eqref{EP for reduced Langevin eq.} and ~\eqref{pseudo EP}. Combining this inequality and Eq.~\eqref{ineq pseudo EP}, we obtain the hierarchy in Eq.~\eqref{hierarchy of EPs}.

\subsection{Trade-offs from the thermodynamic uncertainty relation for deterministic chemical reaction networks}
\label{sec:TUR for CRN}

In the main text, we showed that the dissipation-coherence trade-off and the TSL are derived from the short-time TUR. Similarly to this case, we can also show that the trade-offs for $\Sigma^{\mathrm{ME}}_{\tau_{\mathrm{p}}}$, $\Sigma^{\mathrm{ME}}_{\tau_{\mathrm{p}}}/4\pi^2\geq\mathcal{N}^{\mathrm{RE}}$ and $\Sigma^{\mathrm{ME}}_{\tau_{\mathrm{p}}}\geq (l_{\mathrm{LC}}^{\mathrm{RE}})^2/(\tau_{\mathrm{p}}D^{\mathrm{RE}}_{\mathrm{LC}})$, are derived from the TUR for deterministic CRNs~\cite{yoshimura2021thermodynamic,yoshimura2023housekeeping}.

To consider the TUR, we define some quantities with an $N$-dimensional \textit{observable} $\bm{\phi}_t$: the EP rate for the deterministic time evolution $\sigma_t$ is defined based on the local-detailed balance condition as $\sigma_t\coloneqq\sum_{\rho=1}^{M}j_{\rho}(\bm{x}^{\mathrm{LC}}_t)\ln(j^+_{\rho}(\bm{x}^{\mathrm{LC}}_t)/j^-_{\rho}(\bm{x}^{\mathrm{LC}}_t))$~\cite{rao2016nonequilibrium,ge2016mesoscopic,kondepudi2014modern}. We define the total current weighted with the observable as $j^{\bm{\phi}}_t\coloneqq[\mathsf{S}^{\top}\bm{\phi}_t]^{\top}\bm{j}(\bm{x}^{\mathrm{LC}}_t)$. We also define the intrinsic fluctuation of the observable as $D^{\bm{\phi}}_t\coloneqq\bm{\phi}_t^{\top}\mathsf{D}(\bm{x}^{\mathrm{LC}}_t)\bm{\phi}_t$~\cite{yoshimura2021thermodynamic,yoshimura2023housekeeping}. Using these quantities, the TUR associated with $\bm{\phi}_t$~\cite{yoshimura2023housekeeping} is given by:
\begin{align}
    \sigma_t\geq\frac{(j^{\bm{\phi}}_t)^2}{D^{\bm{\phi}}_t}.
    \label{TUR for CRN}
\end{align}

The TUR~\eqref{TUR for CRN} yields the following bound on $\Sigma_{\tau_{\mathrm{p}}}^{\mathrm{ME}}$:
\begin{align}
    \Sigma_{\tau_{\mathrm{p}}}^{\mathrm{ME}}\geq\frac{\left(\displaystyle\int_{0}^{\tau_{\mathrm{p}}}dt\,|\bm{\phi}_t^{\top}\dot{\bm{x}}_t^{\mathrm{LC}}|\right)^2}{\epsilon\displaystyle\int_{0}^{\tau_{\mathrm{p}}}dt\,D^{\bm{\phi}}_t}.
    \label{Integrated TUR for CRN}
\end{align}
The derivation is as follows. The TUR~\eqref{TUR for CRN} implies $|j_t^{\bm{\phi}}|\leq\sqrt{\sigma_tD_t^{\bm{\phi}}}$. Performing the time integral of both sides of this inequality leads to
\begin{align}
    \left(\int_{0}^{\tau_{\mathrm{p}}}dt\,|j_t^{\bm{\phi}}|\right)^2&\leq\left(\int_{0}^{\tau_{\mathrm{p}}}dt\sqrt{\sigma_tD_t^{\bm{\phi}}}\right)^2\leq\left(\int_{0}^{\tau_{\mathrm{p}}}dt\,\sigma_t\right)\left(\int_{0}^{\tau_{\mathrm{p}}}dt\,D_t^{\bm{\phi}}\right).
    \label{CS for TUR CRN}
\end{align}
Here we used the Cauchy--Schwarz inequality to obtain the second inequality. We obtain Eq.~\eqref{Integrated TUR for CRN} by substituting $\epsilon\Sigma^{\mathrm{ME}}_{\tau_{\mathrm{p}}}=\int_{0}^{\tau_{\mathrm{p}}}dt\,\sigma_t$ and $j_t^{\bm{\phi}}=\bm{\phi}_t^{\top}\mathsf{S}\bm{j}(\bm{x}^{\mathrm{LC}}_t)=\bm{\phi}_t^{\top}\dot{\bm{x}}_t^{\mathrm{LC}}$ into Eq.~\eqref{CS for TUR CRN}.

We show that the dissipation-coherence trade-off and the TSL for $\Sigma_{\tau_{\mathrm{p}}}^{\mathrm{ME}}$ are derived from Eq.~\eqref{Integrated TUR for CRN}. Let $\bm{0}_{N_L}$ denote the $N_L$-dimensional zero vector. We obtain the dissipation-coherence trade-off $\Sigma_{\tau_{\mathrm{p}}}^{\mathrm{ME}}/4\pi^2\geq\mathcal{N}^{\mathrm{RE}}$ by substituting $\bm{\phi}_t=(\bm{\zeta}_t'^{\top},\bm{0}_{N_L}^{\top})^{\top}$ into Eq.~\eqref{Integrated TUR for CRN}. This observable leads to
\begin{align}
    \bm{\phi}_t^{\top}\dot{\bm{x}}^{\mathrm{LC}}_t=(\bm{\zeta}_t'^{\top},\bm{0}_{N_L}^{\top})\begin{pmatrix}
        \dot{\bm{x}}^{\mathrm{LC,ind}}_t\\
        \dot{\bm{x}}^{\mathrm{LC,dep}}_t
    \end{pmatrix}
    =\bm{\zeta}_t'^{\top}\dot{\bm{x}}^{\mathrm{LC,ind}}_t=(\Phi_t'^{-1\top}\bm{v}'^{(1)})^{\top}\Phi_t'\bm{u}'^{(1)}=\bm{v}'^{(1)\top}\bm{u}'^{(1)}=1,
\end{align}
and
\begin{align}
    D_t^{\bm{\phi}}=(\bm{\zeta}_t'^{\top},\bm{0}_{N_L}^{\top})\mathsf{S}\mathsf{A}(\bm{x}^{\mathrm{LC}}_t)\mathsf{S}^{\top}\begin{pmatrix}
        \bm{\zeta}_t'\\\bm{0}_{N_L}
    \end{pmatrix}
    =\bm{\zeta}_t'^{\top}\mathsf{S}'\mathsf{A}(\bm{X}[\bm{x}^{\mathrm{LC,ind}}_t])\mathsf{S}'^{\top}\bm{\zeta}_t'=\bm{\zeta}_t'^{\top}\mathsf{D}'(\bm{x}^{\mathrm{LC,ind}}_t)\bm{\zeta}_t'.
\end{align}
Substituting these equalities into Eq.~\eqref{Integrated TUR for CRN} and using the definition of $\mathcal{N}^{\mathrm{RE}}$ [Eq.~\eqref{Num of coherent reduced LE}], we obtain the dissipation-coherence trade-off as
\begin{align}
    \Sigma_{\tau_{\mathrm{p}}}^{\mathrm{ME}}\geq\frac{\left(\displaystyle\int_{0}^{\tau_{\mathrm{p}}}dt\,1\right)^2}{\epsilon\displaystyle\int_{0}^{\tau_{\mathrm{p}}}dt\,\bm{\zeta}_t'^{\top}\mathsf{D}'(\bm{x}^{\mathrm{LC,ind}}_t)\bm{\zeta}_t'}=4\pi^2\mathcal{N}^{\mathrm{RE}}.
\end{align}
We also obtain the TSL $\Sigma_{\tau_{\mathrm{p}}}^{\mathrm{ME}}\geq (l_{\mathrm{LC}}^{\mathrm{RE}})^2/(\tau_{\mathrm{p}}D^{\mathrm{RE}}_{\mathrm{LC}})$ by substituting $\bm{\phi}_t=(\dot{\bm{x}}_t^{\mathrm{LC,ind}\top},\bm{0}_{N_L}^{\top})^{\top}$ into Eq.~\eqref{Integrated TUR for CRN}. This observable leads to $\bm{\phi}_t^{\top}\dot{\bm{x}}^{\mathrm{LC}}_t=\|\dot{\bm{x}}_t^{\mathrm{LC,ind}}\|^2$ and $D_t^{\bm{\phi}}=\dot{\bm{x}}_t^{\mathrm{LC,ind}\top}\mathsf{D}'(\bm{x}^{\mathrm{LC,ind}}_t)\dot{\bm{x}}_t^{\mathrm{LC,ind}}$. These equations rewrite the inequality in Eq.~\eqref{Integrated TUR for CRN} as
\begin{align}
    \Sigma_{\tau_{\mathrm{p}}}^{\mathrm{ME}}\geq\frac{\left(\displaystyle\int_{0}^{\tau_{\mathrm{p}}}dt\,\|\dot{\bm{x}}_t^{\mathrm{LC,ind}}\|^2\right)^2}{\epsilon\displaystyle\int_{0}^{\tau_{\mathrm{p}}}dt\,\dot{\bm{x}}_t^{\mathrm{LC,ind}\top}\mathsf{D}'(\bm{x}^{\mathrm{LC,ind}}_t)\dot{\bm{x}}_t^{\mathrm{LC,ind}}}=\frac{\displaystyle\int_{0}^{\tau_{\mathrm{p}}}dt\,\|\dot{\bm{x}}_t^{\mathrm{LC,ind}}\|^2}{D^{\mathrm{RE}}_{\mathrm{LC}}}\geq\frac{(l_{\mathrm{LC}}^{\mathrm{RE}})^2}{\tau_{\mathrm{p}}D^{\mathrm{RE}}_{\mathrm{LC}}},
    \label{derivation TSL CRN}
\end{align}
where we used the Cauchy--Schwarz inequality $(\int_{0}^{\tau_{\mathrm{p}}}dt\,1)(\int_{0}^{\tau_{\mathrm{p}}}dt\,\|\dot{\bm{x}}^{\mathrm{LC,ind}}_t\|^2)\geq (\int_{0}^{\tau_{\mathrm{p}}}dt\,\|\dot{\bm{x}}^{\mathrm{LC,ind}}_t\|)^2$ in the last inequality.

\add{
We can regard the TSL in Eq.~\eqref{derivation TSL CRN} as a TSL for the deterministic CRN. To verify this fact, we multiply the TSL by $\epsilon$ and obtain
\begin{align}
    \int_{0}^{\tau_{\mathrm{p}}}dt\,\sigma_t\geq\frac{(l_{\mathrm{LC}}^{\mathrm{RE}}){}^2}{\tau_{\mathrm{p}}(\epsilon^{-1}D^{\mathrm{RE}}_{\mathrm{LC}})}.
    \label{deterministic TSL}
\end{align}
The left-hand side of this inequality is the deterministic EP. The term $\epsilon^{-1}D^{\mathrm{RE}}_{\mathrm{LC}}$ in the right-hand side is also a deterministic quantity, which is given by the scaled diffusion coefficient matrix as $\int_{0}^{\tau_{\mathrm{p}}}dt\,\dot{\bm{x}}_t^{\mathrm{LC,ind}\top}\mathsf{D}'(\bm{x}^{\mathrm{LC,ind}}_t)\dot{\bm{x}}_t^{\mathrm{LC,ind}}/\int_{0}^{\tau_{\mathrm{p}}}dt\,\|\dot{\bm{x}}_t^{\mathrm{LC,ind}}\|^2$. Thus, Eq.~\eqref{deterministic TSL} is a deterministic TSL. This property arises from the fact that both $\bm{F}(\bm{x})$ and $\mathsf{D}(\bm{x})$ are determined by macroscopic fluxes $\bm{j}^{\pm}(\bm{x})$ in the stochastic CRNs. Note that the TSL in Eq.~(10) cannot be considered a deterministic one for systems in which $\mathsf{D}(\bm{x})$ is independent of deterministic properties, e.g., a Brownian particle at low temperatures.
}

\add{
\section{Phase reduction and achievability of trade-offs}
}

\add{
\subsection{Introduction of phase reduction}
Consider the deterministic counterpart of the Langevin equation in Eq.~(1),
\begin{align}
    \dot{\bm{x}}_t=\bm{F}(\bm{x}_t).
    \label{system}
\end{align}
As assumed in the main text, this dynamical system has a globally stable limit cycle $\bm{x}_t^{\mathrm{LC}}$ with period $\tau_{\mathrm{p}}$. Here and in the following, we fix an arbitrary point on the limit cycle as the time origin $t=0$ without loss of generality.
We introduce the angular frequency as $\omega\coloneqq2\pi/\tau_{\mathrm{p}}$.
We write $\hat{\bm{x}}_t[\bm{x}]$ for the solution of Eq.~\eqref{system} with initial condition $\hat{\bm{x}}_0[\bm{x}]=\bm{x}$. 
}

\add{
There exists an asymptotic phase $\Theta(\bm{x})\in [0,2\pi)$~\cite{winfree1967biological,Kuramoto1984,monga2019phase} that satisfies
\begin{align}
    \lim_{t\to\infty}\|\hat{\bm{x}}_t[\bm{x}]-\bm{x}^{\mathrm{LC}}_{t+\omega^{-1}\Theta(\bm{x})}\|=0.
    \label{def: phase}
\end{align}
This phase is unique, as shown below. Suppose that there exist $\Theta_1(\bm{x})$ and $\Theta_2(\bm{x})$ that satisfy Eq.~\eqref{def: phase}. The triangle inequality leads to
\begin{align}
    \|\bm{x}^{\mathrm{LC}}_{t+\omega^{-1}\Theta_1(\bm{x})}-\bm{x}^{\mathrm{LC}}_{t+\omega^{-1}\Theta_2(\bm{x})}\|\leq\|\hat{\bm{x}}_t[\bm{x}]-\bm{x}^{\mathrm{LC}}_{t+\omega^{-1}\Theta_1(\bm{x})}\|+\|\hat{\bm{x}}_t[\bm{x}]-\bm{x}^{\mathrm{LC}}_{t+\omega^{-1}\Theta_2(\bm{x})}\|.
\end{align}
Because Eq.~\eqref{def: phase} makes the right-hand side of this inequality converge to $0$ in $t\to\infty$, we obtain $\bm{x}^{\mathrm{LC}}_{t+\omega^{-1}\Theta_1(\bm{x})}=\bm{x}^{\mathrm{LC}}_{t+\omega^{-1}\Theta_2(\bm{x})}$. The periodicity of $\bm{x}_t^{\mathrm{LC}}$ and the domain of the asymptotic phase, $[0,2\pi)$, conclude $\Theta_1(\bm{x})=\Theta_2(\bm{x})$. We note that the asymptotic phase $\Theta(\bm{x})$ is defined up to modulo $2\pi$.
}

\add{
Using this asymptotic phase, we can consider the phase at time $t$ as $\theta_t\coloneqq\Theta(\bm{x}_t)$. The definition of the asymptotic phase leads to $\omega^{-1}\theta_{t+\varDelta t}=\omega^{-1}\theta_t+\varDelta t$, whose limit in $\varDelta t\to0$ yields the time evolution of the phase,
\begin{align}
    \dot{\theta}_t=\omega.
    \label{phase evolution}
\end{align}
Note that the phase $\theta_t$ is defined up to modulo $2\pi$. The chain rule also leads to
\begin{align}
    \dot{\theta}_t=[\bm{\nabla}\Theta(\bm{x}_t)]^{\top}\dot{\bm{x}}_t=[\bm{\nabla}\Theta(\bm{x}_t)]^{\top}\bm{F}(\bm{x}_t).
    \label{phase evolution chain rule}
\end{align}
Combining Eqs.~\eqref{phase evolution} and ~\eqref{phase evolution chain rule}, we obtain $[\bm{\nabla}\Theta(\bm{x}_t)]^{\top}\bm{F}(\bm{x}_t)=\omega$. In particular, setting $\bm{x}_t=\hat{\bm{x}}_t[\bm{x}]$ and $t=0$, we obtain the relation between the gradient of the asymptotic phase and the force field,
\begin{align}
    [\bm{\nabla}\Theta(\bm{x})]^{\top}\bm{F}(\bm{x})=\omega.
    \label{inner product relation}
\end{align}
}

\add{
The gradient of the phase $\bm{\nabla}\Theta(\bm{x})$ characterizes how the time evolution of the phase responds to external controls~\cite{monga2019phase}. Consider the following perturbed time evolution:
\begin{align}
    \dot{\bm{x}}_t=\bm{F}(\bm{x}_t)+\bm{F}^{\mathrm{ext}}_t.
    \label{perturbed system}
\end{align}
Here $\bm{F}^{\mathrm{ext}}_t$ is the external force that describes the control, which is independent of the state. In this controlled system, the phase $\theta_t$ evolves according to
\begin{align}
    \dot{\theta}_t=[\bm{\nabla}\Theta(\bm{x}_t)]^{\top}[\bm{F}(\bm{x}_t)+\bm{F}^{\mathrm{ext}}_t]=\omega+[\bm{\nabla}\Theta(\bm{x}_t)]^{\top}\bm{F}^{\mathrm{ext}}_t,
    \label{controlled system}
\end{align}
where we used Eq.~\eqref{inner product relation}. In the last term in Eq.~\eqref{controlled system}, $\bm{\nabla}\Theta(\bm{x})$ relates the force field for the external control to the time evolution of the phase.  
}

\add{
Since the state cannot be uniquely determined from the phase, Eq.~\eqref{controlled system} is not closed in terms of $\theta_t$ and $\bm{F}^{\mathrm{ext}}_t$. To avoid this difficulty and obtain the closed equation for the phase, we consider the following situation: The initial state is on the limit cycle; the external force $\bm{F}^{\mathrm{ext}}_t$ is sufficiently small so that the stability of the limit cycle enables us to approximate $\bm{x}_t\simeq\bm{x}^{\mathrm{LC}}_{\omega^{-1}\theta_t}$. This approximation reduces Eq.~\eqref{controlled system} to
\begin{align}
    \dot{\theta}_t=\omega+\bm{Z}_{\omega^{-1}\theta_t}^{\top}\cdot\bm{F}^{\mathrm{ext}}_t,
    \label{Phase equation}
\end{align}
where $\bm{Z}_t$ is defined as
\begin{align}
    \bm{Z}_t\coloneqq\bm{\nabla}\Theta(\bm{x}^{\mathrm{LC}}_{t}).
    \label{PRC}
\end{align}
This function $\bm{Z}_t$ characterizes the effect of a weak external control on the phase of the limit cycle~\cite{winfree1967biological}, and is called \textit{infinitesimal phase response curve} (PRC)~\cite{monga2019phase,brown2004phase}. Due to the periodicity of $\bm{x}_t^{\mathrm{LC}}$, the PRC is also periodic as $\bm{Z}_{t+\tau_{\mathrm{p}}}=\bm{Z}_t$. 
}

\add{
Here we introduce a way to obtain the PRC, which is called the adjoint method~\cite{monga2019phase,brown2004phase,Hoppensteadt1997}. In this method, we use the time derivative of $\bm{Z}_t$. To obtain this derivative, we consider the small deviation from the limit cycle, $\bm{x}_t=\bm{x}_t^{\mathrm{LC}}+\varDelta\bm{x}_t$ with $\|\varDelta\bm{x}_t\|\ll1$. As discussed in the main text, the time evolution of $\varDelta\bm{x}_t$ is given by
\begin{align}
    \frac{d}{dt}\varDelta{\bm{x}}_t=\mathsf{K}_t\varDelta{\bm{x}}_t+o(\|\varDelta{\bm{x}}_t\|)
    \label{Jacobi eq}
\end{align}
with the Jacobian $\mathsf{K}_t$ of $\bm{F}(\bm{x})$ at $\bm{x}_t^{\mathrm{LC}}$. We also introduce the deviation of the phase as
\begin{align}
    \varDelta\theta_t\coloneqq\Theta(\bm{x}_t)-\Theta(\bm{x}_t^{\mathrm{LC}}),
\end{align}
whose time derivative vanishes as
\begin{align}
    \frac{d}{dt}\varDelta\theta_t=\omega-\omega=0.
    \label{derivative of phase deviation}
\end{align}
Expanding $\varDelta\theta_t$ in terms of $\varDelta{\bm{x}}_t$ leads to
\begin{align}
    \varDelta\theta_t=\bm{Z}_t^{\top}\varDelta{\bm{x}}_t+o(\|\varDelta{\bm{x}}_t\|),
\end{align}
where the definition of PRC in Eq.~\eqref{PRC} is also used.
Since the time derivative of the $\varDelta\theta_t$ vanishes, we obtain
\begin{align}
    0=\dot{\bm{Z}_t}^{\top}\varDelta{\bm{x}}_t+\bm{Z}_t^{\top}\left[\frac{d}{dt}\varDelta\bm{x}_t\right]=\left[\dot{\bm{Z}}_t+\mathsf{K}_t^{\top}\bm{Z}_t\right]^{\top}\varDelta\bm{x}_t,
    \label{inner product adjoint}
\end{align}
by ignoring $o(\|\varDelta{\bm{x}}_t\|)$-terms and using Eq.~\eqref{Jacobi eq}. Because Eq.~\eqref{inner product adjoint} holds for any $\varDelta\bm{x}_t$, we obtain 
\begin{align}
    \dot{\bm{Z}}_t=-\mathsf{K}_t^{\top}\bm{Z}_t.
    \label{adjoint equation}
\end{align}
The PRC is obtained by solving this equation. The initial state $\bm{Z}_0$ is determined so that the solution satisfies the periodicity $\bm{Z}_{t+\tau_{\mathrm{p}}}=\bm{Z}_t$ and the condition $\bm{Z}_0^{\top}\bm{F}(\bm{x}_0^{\mathrm{LC}})=\omega$ due to Eq.~\eqref{inner product relation}. In the next section, we will confirm that these conditions uniquely determine the initial state $\bm{Z}_0$.
}

\add{
\subsection{Relation between the eigenvector of fundamental matrix and the infinitesimal phase response curve}
The fundamental matrix corresponding to Eq.~\eqref{adjoint equation} is given by $\Phi_t^{-1\top}$. 
This is verified by taking the time derivative of both sides in $\Phi_t\Phi_t^{-1}=\mathsf{I}$, which leads to $(\mathsf{K}_t\Phi_t)\Phi_t^{-1}+\Phi_t\dot{\Phi}_t^{-1}=\mathsf{O}$. Multiplying $\Phi_t^{-1}$ from the left to this equation, we obtain $\dot{\Phi}_t^{-1}=-\Phi_t^{-1}\mathsf{K}_t$, whose transpose is 
\begin{align}
    \dot{\Phi}_t^{-1\top}=-\mathsf{K}_t^{\top}\Phi_t^{-1\top}.
    \label{adjoint fundamental matrix}
\end{align}
Equation~\eqref{adjoint fundamental matrix} and $\Phi_0^{-1\top}=\mathsf{I}^{-1\top}=\mathsf{I}$ imply that $\Phi_t^{-1\top}$ is the fundamental matrix corresponding to Eq.~\eqref{adjoint equation}. Thus, the monodromy matrix for Eq.~\eqref{adjoint equation} is given by $\Phi_{\tau_{\mathrm{p}}}^{-1\top}$. Using the spectral decomposition of $\Phi_{\tau_{\mathrm{p}}}$ in the main text, we obtain the spectral decomposition of $\Phi_{\tau_{\mathrm{p}}}^{-1\top}$ as
\begin{align}
    \Phi_{\tau_{\mathrm{p}}}^{-1\top}=\sum_{n=1}^N\frac{1}{\lambda_n}\bm{v}^{(n)}\bm{u}^{(n)\top}=\bm{v}^{(1)}\dot{\bm{x}}^{\mathrm{LC}\top}_0+\sum_{n=2}^N\frac{1}{\lambda_n}\bm{v}^{(n)}\bm{u}^{(n)\top}.
    \label{adjoint monodromy}
\end{align}
Here, we also used the fact that the matrix $\Phi_{\tau_{\mathrm{p}}}$ is invertible, and thus its eigenvalues are nonzero.
}

\add{
Using this monodromy matrix, we consider the initial condition for Eq.~\eqref{adjoint equation} to obtain the PRC. Since the eigenvectors $\{\bm{v}^{(n)}\}_{n=1}^N$ span the basis of $\mathbb{C}^N$, any initial condition is expressed in terms of a linear combination of these vectors as $\bm{Z}_0=\sum_{n=1}^N c_n\bm{v}^{(n)}$. This expression leads to
\begin{align}
    \bm{Z}_{\tau_{\mathrm{p}}}=\Phi_{\tau_{\mathrm{p}}}^{-1\top}\bm{Z}_0=\sum_{n=1}^{N}\frac{c_n}{\lambda_n}\bm{v}^{(n)},
\end{align}
where we used Eq.~\eqref{adjoint monodromy} and the biorthogonality $\bm{v}^{(n)\top}\bm{u}^{(m)}=\delta_{nm}$. Due to the assumptions on the eigenvalues, $\lambda_n\neq 1$ for all $n\neq1$ and $\lambda_1=1$, the periodicity $\bm{Z}_0=\bm{Z}_{\tau_{\mathrm{p}}}$ is equivalent to $c_n=0$ for any $n\neq1$. Substituting $\bm{Z}_0=c_1\bm{v}^{(1)}$ into the condition $\bm{Z}_0^{\top}\bm{F}(\bm{x}_0^{\mathrm{LC}})=\omega$, we obtain $c_1\bm{v}^{(1)\top}\bm{F}(\bm{x}_0^{\mathrm{LC}})=\omega$. This equation yields $c_1=\omega$ because the biorthogonality leads to $\bm{v}^{(1)\top}\bm{F}(\bm{x}_0^{\mathrm{LC}})=\bm{v}^{(1)\top}\dot{\bm{x}}_0^{\mathrm{LC}}=1$. Thus, the periodicity of the PRC and $\bm{Z}_0^{\top}\bm{F}(\bm{x}_0^{\mathrm{LC}})=\omega$ uniquely determine the initial condition as
\begin{align}
    \bm{Z}_0=\omega\bm{v}^{(1)}.
    \label{initial condition PRC}
\end{align}
}

\add{
This initial condition also leads to the relation between the PRC and the dual vector of $\dot{\bm{x}}_t^{\mathrm{LC}}$, $\bm{\zeta}_t$~\cite{shirasaka2017phase}. Using the fact that $\Phi_t^{-1\top}$ is the fundamental matrix corresponding to Eq.~\eqref{adjoint equation}, we can express the PRC as 
\begin{align}
    \bm{Z}_t=\Phi_t^{-1\top}\bm{Z}_0=\omega\Phi_t^{-1\top}\bm{v}^{(1)}=\omega\bm{\zeta}_t.
    \label{PRC dual vector}
\end{align}
}

\add{
\subsection{Optimal noise for the dissipation-coherence trade-off}
Here, we consider the equality condition for the dissipation-coherence trade-off [Eq.~(9)]. The derivation of the dissipation-coherence trade-off implies that this trade-off is equivalent to the following inequality:
\begin{align}
    \left[\int_{0}^{\tau_{\mathrm{p}}}dt\,\bm{F}(\bm{x}_t^{\mathrm{LC}})^{\top}\mathsf{D}(\bm{x}_t^{\mathrm{LC}})^{-1}\bm{F}(\bm{x}_t^{\mathrm{LC}})\right]\left[\int_{0}^{\tau_{\mathrm{p}}}dt\,\bm{\zeta}_t^{\top}\mathsf{D}(\bm{x}_t^{\mathrm{LC}})\bm{\zeta}_t\right]\geq\left[\int_{0}^{\tau_{\mathrm{p}}}dt\,\bm{\zeta}_t^{\top}\bm{F}(\bm{x}_t^{\mathrm{LC}})\right]^2.
    \label{CS form of DCT}
\end{align}
We can regard this inequality as the Cauchy--Schwarz inequality between $\mathsf{D}(\bm{x}_t^{\mathrm{LC}})^{-1/2}\bm{F}(\bm{x}_t^{\mathrm{LC}})$ and $\mathsf{D}(\bm{x}_t^{\mathrm{LC}})^{1/2}\bm{\zeta}_t$. Thus, the equality condition for Eq.~\eqref{CS form of DCT} and the dissipation-coherence trade-off is $\mathsf{D}(\bm{x}_t^{\mathrm{LC}})^{-1/2}\bm{F}(\bm{x}_t^{\mathrm{LC}})\propto\mathsf{D}(\bm{x}_t^{\mathrm{LC}})^{1/2}\bm{\zeta}_t$ with $t$-independent proportionality constant. This is equivalent to 
\begin{align}
    \bm{F}(\bm{x}_t^{\mathrm{LC}})\propto\mathsf{D}(\bm{x}_t^{\mathrm{LC}})\bm{\zeta}_t,
    \label{equality condition DCT}
\end{align}
with $t$-independent proportionality constant.
}

\add{
Using the asymptotic phase $\Theta(\bm{x})$, we can always construct $\mathsf{D}(\bm{x})$ that satisfies this equality condition in Eq.~\eqref{equality condition DCT} as
\begin{align}
    \mathsf{D}(\bm{x})=D\left[\mathsf{I}-\frac{\bm{\nabla}\Theta(\bm{x})[\bm{\nabla}\Theta(\bm{x})]^{\top}}{\|\bm{\nabla}\Theta(\bm{x})\|^2}+\bm{F}(\bm{x})\bm{F}(\bm{x})^{\top}\right],
    \label{optimal diffusion matrix}
\end{align}
where $D$ is a positive constant.
The symmetry of this matrix is self-evident. The positive definiteness is also confirmed as follows: It is enough to show that
\begin{align}
    \bm{u}^{\top}\mathsf{D}(\bm{x})\bm{u}=D\left[\left(\bm{u}^{\top}\bm{u}-\frac{[\bm{u}^{\top}\bm{\nabla}\Theta(\bm{x})]^2}{\|\bm{\nabla}\Theta(\bm{x})\|^2}\right)+[\bm{u}^{\top}\bm{F}(\bm{x})]^2\right]
\end{align}
is positive for any nonzero vector $\bm{u}$. The Cauchy--Schwarz inequality implies
\begin{align}
    \bm{u}^{\top}\bm{u}-\frac{[\bm{u}^{\top}\bm{\nabla}\Theta(\bm{x})]^2}{\|\bm{\nabla}\Theta(\bm{x})\|^2}\geq0,
    \label{CS for first term}
\end{align}
which leads to $\bm{u}^{\top}\mathsf{D}(\bm{x})\bm{u}\geq0$. Since $[\bm{u}^{\top}\bm{F}(\bm{x})]^2$ is also nonnegative, $\bm{u}^{\top}\mathsf{D}(\bm{x})\bm{u}=0$ can only occur if the equality of Eq.~\eqref{CS for first term} is achieved, i.e., if $\bm{u}\propto\bm{\nabla}\Theta(\bm{x})$ is satisfied. However, in that case, Eq.~\eqref{inner product relation} concludes that $[\bm{u}^{\top}\bm{F}(\bm{x})]^2$ is positive. Thus, $\bm{u}^{\top}\mathsf{D}(\bm{x})\bm{u}$ is always positive. We also verify that Eq.~\eqref{equality condition DCT} is satisfied by $\mathsf{D}(\bm{x})$ in Eq.~\eqref{optimal diffusion matrix}. Using Eq.~\eqref{inner product relation}, we obtain
\begin{align}
    \mathsf{D}(\bm{x})\bm{\nabla}\Theta(\bm{x})=\omega D\bm{F}(\bm{x}).
\end{align}
This equation leads to $\mathsf{D}(\bm{x}^{\mathrm{LC}}_t)\bm{\nabla}\Theta(\bm{x}^{\mathrm{LC}}_t)=\omega D\bm{F}(\bm{x}^{\mathrm{LC}}_t)$, which is reduced to $D\bm{F}(\bm{x}_t^{\mathrm{LC}})=\mathsf{D}(\bm{x}_t^{\mathrm{LC}})\bm{\zeta}_t$ by Eqs.~\eqref{PRC} and ~\eqref{PRC dual vector}.
}

\add{
\subsection{Tighter thermodynamic speed limit and equality conditions}
The derivation of the TSL [Eq.~(10)] implies that the TSL is equivalent to the following two inequalities:
\begin{align}
    \left[\int_{0}^{\tau_{\mathrm{p}}}dt\,\bm{F}(\bm{x}_t^{\mathrm{LC}})^{\top}\mathsf{D}(\bm{x}_t^{\mathrm{LC}})^{-1}\bm{F}(\bm{x}_t^{\mathrm{LC}})\right]\left[\int_{0}^{\tau_{\mathrm{p}}}dt\,\bm{F}(\bm{x}_t^{\mathrm{LC}})^{\top}\mathsf{D}(\bm{x}_t^{\mathrm{LC}})\bm{F}(\bm{x}_t^{\mathrm{LC}})\right]\geq\left[\int_{0}^{\tau_{\mathrm{p}}}dt\,\|\bm{F}(\bm{x}_t^{\mathrm{LC}})\|^2\right]^2,
    \label{CS1 for TSL}
\end{align}
\begin{align}
    \left[\int_{0}^{\tau_{\mathrm{p}}}dt\,1\right]\left[\int_{0}^{\tau_{\mathrm{p}}}dt\,\|\bm{F}(\bm{x}_t^{\mathrm{LC}})\|^2\right]\geq\left[\int_{0}^{\tau_{\mathrm{p}}}dt\,\|\bm{F}(\bm{x}_t^{\mathrm{LC}})\|\right]^2.
    \label{CS2 for TSL}
\end{align}
We note that Eq.~\eqref{CS1 for TSL} can be interpreted physically even when taken alone. As in the derivation of the TSL, we rewrite Eq.~\eqref{CS1 for TSL} as
\begin{align}
    \Sigma_{\mathrm{\tau}_{\mathrm{p}}}\geq\frac{1}{{D_{\mathrm{LC}}}}\int_{0}^{\tau_{\mathrm{p}}}dt\,\|\bm{F}(\bm{x}_t^{\mathrm{LC}})\|^2=\frac{1}{{D_{\mathrm{LC}}}}\int_{0}^{\tau_{\mathrm{p}}}dt\,\|\dot{\bm{x}}_t^{\mathrm{LC}}\|^2.
    \label{CS1 rewritten}
\end{align}
We also introduce the \textit{variance} of the speed on the limit cycle as
\begin{align}
    \mathcal{V}_{\mathrm{LC}}\coloneqq\left(\frac{1}{\tau_{\mathrm{p}}}\int_{0}^{\tau_{\mathrm{p}}}dt\,\|\dot{\bm{x}}_t^{\mathrm{LC}}\|^2\right)-\left(\frac{1}{\tau_{\mathrm{p}}}\int_{0}^{\tau_{\mathrm{p}}}dt\,\|\dot{\bm{x}}_t^{\mathrm{LC}}\|\right)^2.
\end{align}
Using this variance and the Euclidean length of the limit cycle $l_{\mathrm{LC}}=\int_{0}^{\tau_{\mathrm{p}}}dt\,\|\dot{\bm{x}}_t^{\mathrm{LC}}\|$, we obtain
\begin{align}
    \int_{0}^{\tau_{\mathrm{p}}}dt\,\|\dot{\bm{x}}_t^{\mathrm{LC}}\|^2=\frac{l_{LC}^2}{\tau_{\mathrm{p}}}+\tau_{\mathrm{p}}\mathcal{V}_{\mathrm{LC}}.
\end{align}
This relation rewrites Eq.~\eqref{CS1 rewritten} as
\begin{align}
    \Sigma_{\mathrm{\tau}_{\mathrm{p}}}\geq\frac{l_{\mathrm{LC}}^2}{\tau_{\mathrm{p}}{D_{\mathrm{LC}}}}+\frac{\tau_{\mathrm{p}}\mathcal{V}_{\mathrm{LC}}}{D_{\mathrm{LC}}}.
    \label{tighter TSL}
\end{align}
This inequality implies that the TSL is tightened by the variance of the speed. 
}

\add{We can regard Eq.~\eqref{CS1 for TSL} as the Cauchy--Schwarz inequality between $\mathsf{D}(\bm{x}_t^{\mathrm{LC}})^{-1/2}\bm{F}(\bm{x}_t^{\mathrm{LC}})$ and  $\mathsf{D}(\bm{x}_t^{\mathrm{LC}})^{1/2}\bm{F}(\bm{x}_t^{\mathrm{LC}})$. Thus, the equality of Eq.~\eqref{CS1 for TSL} holds if and only if
\begin{align}
    \bm{F}(\bm{x}_t^{\mathrm{LC}})\propto\mathsf{D}(\bm{x}_t^{\mathrm{LC}})\bm{F}(\bm{x}_t^{\mathrm{LC}}).
    \label{EQ CS1}
\end{align}
holds with $t$-independent proportionality constant. This condition is satisfied by taking
\begin{align}
    \mathsf{D}(\bm{x})=D\mathsf{I}
    \label{optimal diffusion matrix TSL}
\end{align}
with a positive constant $D$ as the diffusion coefficient matrix. Thus, the equality in the tighter TSL [Eq.~\eqref{tighter TSL}] is always achievable by using the diffusion coefficient matrix in Eq.~\eqref{optimal diffusion matrix TSL}. 
}

\add{
To achieve the equality of the original TSL [Eq.~(10)], we also need to achieve the equality in Eq.~\eqref{CS2 for TSL}.
We can regard Eq.~\eqref{CS2 for TSL} as the Cauchy--Schwarz inequality between $1$ and $\|\bm{F}(\bm{x}_t^{\mathrm{LC}})\|$. The equality of this inequality holds if and only if the speed on the limit cycle, $\|\bm{F}(\bm{x}_t^{\mathrm{LC}})\|$, is constant, i.e.,
\begin{align}
    \frac{d}{dt}\|\bm{F}(\bm{x}_t^{\mathrm{LC}})\|=0.
    \label{EQ CS2}
\end{align}
Thus, merely transforming the diffusion coefficient matrix to satisfy Eq.~\eqref{EQ CS1} is insufficient to saturate the TSL.
}

\add{
\section{Numerical verification of the dissipation-coherence trade-offs for systems with finite noise}
In the main text, we prove the dissipation-coherence trade-off analytically in the weak-noise limit. A natural question that arises is whether this trade-off extends to systems with finite noise. Here, we formulate the dissipation-coherence trade-off for finite noise. Although a rigorous analytical proof of this general trade-off remains elusive, we provide numerical evidence that supports its validity across a wide range of noise intensities.
}

\add{
\subsection{Dissipation-coherence trade-offs for systems with finite noise}
}

\add{
We introduce the dissipation-coherence trade-off in the case of finite $\epsilon$. The quality of a noisy oscillation is characterized by the long-time behavior of the correlation function. Let us assume that, for any $n>1$, $\mathrm{Re}(\Lambda_n)<\mathrm{Re}(\Lambda_1)$ holds except when $\Lambda_n=\Lambda_1^{\ast}$. This assumption rewrites Eq.~\eqref{general correlation function in long time} as
\begin{align}
    \langle a(\bm{x}_t)a(\bm{x}_0)\rangle_{\mathrm{st}}\simeq\left\{\int d\bm{x}\,a(\bm{x})p^{\mathrm{st}}(\bm{x})\right\}^2+2\mathrm{e}^{\mathrm{Re}(\Lambda_1) t}\mathrm{Re}\left[\mathrm{e}^{\mathrm{i}\mathrm{Im}(\Lambda_1) t}\int d\bm{x}\,a(\bm{x})P_{1}(\bm{x})\int d\bm{x}'\,a(\bm{x}')Q_{1}(\bm{x}')p^{\mathrm{st}}(\bm{x}')\right],
    \label{general asymp correlation function}
\end{align}
This form implies that the correlation function oscillates with period $2\pi/|\mathrm{Im}(\Lambda_1)|$ and decays with correlation time $|\mathrm{Re}(\Lambda_1)|^{-1}$.
Using these quantities, we can introduce the number of coherent oscillations for the system with finite noise as
\begin{align}
    \mathcal{N}(\epsilon)\coloneqq\frac{|\mathrm{Im}(\Lambda_1)|}{2\pi|\mathrm{Re}(\Lambda_1)|},
    \label{general quality factor}
\end{align}
where we used epsilon as an argument to explicitly express the $\epsilon$-dependence of the quantity.
Comparing Eq.~\eqref{asymp correlation function} and Eq.~\eqref{general asymp correlation function}, we can verify the consistency with the results in the weak-noise limit as
\begin{align}
    \lim_{\epsilon\to 0}|\mathrm{Im}(\Lambda_1)|=\frac{2\pi}{\tau_{\mathrm{p}}},\;\;\lim_{\epsilon\to 0}\frac{|\mathrm{Re}(\Lambda_1)|}{\epsilon}=\frac{4\pi^2}{\tau_{\mathrm{p}}^3}\left[\int_{0}^{\tau_{\mathrm{p}}}dt\,\bm{\zeta}_{t}^{\top}\mathsf{D}(\bm{x}^{\mathrm{LC}}_{t})\bm{\zeta}_{t}\right],
\end{align}
and
\begin{align}
    \lim_{\epsilon\to 0}\epsilon\mathcal{N}(\epsilon)=\frac{\tau_{\mathrm{p}}^2}{4\pi^2}\left[\int_{0}^{\tau_{\mathrm{p}}}dt\,\bm{\zeta}_{t}^{\top}\mathsf{D}(\bm{x}^{\mathrm{LC}}_{t})\bm{\zeta}_{t}\right]^{-1}.
    \label{quality factor limit}
\end{align}
Since the period of the oscillation is given by $2\pi/|\mathrm{Im}(\Lambda_1)|$, we can also define the EP required for one period of the oscillation as
\begin{align}
    \Sigma^{\mathrm{st}}(\epsilon)\coloneqq\frac{2\pi}{|\mathrm{Im}(\Lambda_1)|}\left[\frac{1}{\epsilon}\int d\bm{x}\,p^{\mathrm{st}}(\bm{F}-\epsilon\bm{\nabla}\cdot\mathsf{D})^{\top}\mathsf{D}^{-1}(\bm{F}-\epsilon\bm{\nabla}\cdot\mathsf{D})+\int d\bm{x}\,p^{\mathrm{st}}\bm{\nabla}\cdot(\bm{F}-\epsilon\bm{\nabla}\cdot\mathsf{D})\right].
    \label{general EP}
\end{align}
where we used Eq.~\eqref{steady state EPR 2}. We note that Eq.~\eqref{steady state EP} and $\lim_{\epsilon\to 0}|\mathrm{Im}(\Lambda_1)|=2\pi/\tau_{\mathrm{p}}$ lead to
\begin{align}
    \lim_{\epsilon\to0}\epsilon\Sigma^{\mathrm{st}}(\epsilon)=\int_{0}^{\tau_{\mathrm{p}}} dt\,\bm{F}(\bm{x}_t^{\mathrm{LC}})^{\top}\mathsf{D}(\bm{x}_t^{\mathrm{LC}})^{-1}\bm{F}(\bm{x}_t^{\mathrm{LC}}),
    \label{EP limit}
\end{align}
which is consistent with the EP in the weak-noise limit.
Based on the results in the weak-noise limit, it is reasonable to conjecture that the following inequality is true even for finite $\epsilon$:
\begin{align}
    \frac{\Sigma^{\mathrm{st}}(\epsilon)}{4\pi^2}\geq\mathcal{N}(\epsilon).
    \label{DCT general}
\end{align}
However, the analytical proof of this dissipation-coherence trade-off for finite $\epsilon$ does not currently exist.
This is because analytical access to eigenvalue $\Lambda_1$ is difficult when $\epsilon$ is finite.
Instead, we numerically verify this inequality in Sec.~\ref{secSM: numerical results}.
}

\add{
We also introduce the dissipation-coherence trade-off for stochastic CRNs with finite $\epsilon$, which is a generalization of Eq.~(13). 
We can compute the eigenvalue $\Lambda_1$ and the steady-state distribution for the Fokker--Planck operator that corresponds to the reduced Langevin equation [Eq.~(21)]. Using these quantities, we obtain the counterparts of $\mathcal{N}(\epsilon)$ and $\Sigma^{\mathrm{st}}(\epsilon)$ in the reduced Langevin equation. Let $\mathcal{N}^{\mathrm{RE}}(\epsilon)$ and $\Sigma^{\mathrm{RE,st}}(\epsilon)$ denote these counterparts. We also define the EP for the CME associated with one period of the oscillation as
\begin{align}
    \Sigma^{\mathrm{ME,st}}(\epsilon)\coloneqq\frac{2\pi}{|\mathrm{Im}(\Lambda_1)|}\dot{\Sigma}^{\mathrm{ME,st}},
\end{align}
with the period of the oscillation $2\pi/|\mathrm{Im}(\Lambda_1)|$ and the steady-state EPR $\dot{\Sigma}^{\mathrm{ME,st}}$ with finite $\epsilon$ in Eq.~\eqref{steady-state EPR CME general}.
Using these quantities, we may generalize the dissipation-coherence trade-off in Eq.~(13) as
\begin{align}
    \frac{\Sigma^{\mathrm{ME,st}}(\epsilon)}{4\pi^2}\geq\frac{\Sigma^{\mathrm{RE,st}}(\epsilon)}{4\pi^2}\geq\mathcal{N}^{\mathrm{RE}}(\epsilon).
    \label{DCT general stochastic CRN}
\end{align}
As in the case of Eq.~\eqref{DCT general}, this trade-off has not yet been proven.
In Sec.~\ref{secSM: numerical results}, we numerically demonstrate Eq.~\eqref{DCT general stochastic CRN} for finite $\epsilon$ using the same system as in End Matter.
}

\add{
\subsection{Numerical method for eigenvalue problem}
Equations~\eqref{general quality factor} and ~\eqref{general EP} allow us to compute $\mathcal{N}(\epsilon)$ and $\Sigma^{\mathrm{st}}(\epsilon)$ if we have $\Lambda_1$ and $p^{\mathrm{st}}(\bm{x})$. We numerically obtain $\Lambda_1$ and $p^{\mathrm{st}}(\bm{x})$ by discretizing $\mathcal{L}^{\dag}$ on a bounded domain.
This method is based on Ref.~\cite{houzelstein2025generalized}. In the following, we only consider the two-dimensional systems.
}

\add{
We consider a two-dimensional domain $\mathcal{X}\coloneqq[x_1^{\min},x_1^{\max}]\times[x_2^{\min},x_2^{\max}]$ instead of the entire space to facilitate numerical calculations. To avoid boundary effects, $\mathcal{X}$ is chosen to fully encompass a long-time trajectory generated by the corresponding Langevin equation. We obtain the trajectory by applying the Euler--Maruyama method~\cite{higham2001algorithmic} to the Langevin equation with $t\in[0,50\tau_{\mathrm{p}}]$ and $\bm{x}_0=\tau_{\mathrm{p}}^{-1}\int_{0}^{\tau_{\mathrm{p}}}dt\,\bm{x}_t^{\mathrm{LC}}$.
}

\add{
To restrict $\mathcal{L}^{\dag}$ to the bounded domain $\mathcal{X}$, we need to impose appropriate boundary conditions. Let $\partial\mathcal{X}$ denote the boundary of $\mathcal{X}$. For $\bm{x}\in\partial\mathcal{X}$, we impose the following adjoint reflecting boundary conditions on $q(\bm{x})$ in $\mathcal{L}^{\dag}[q(\bm{x})]$:
\begin{align}
    0&=\hat{\bm{n}}(\bm{x})^{\top}\mathsf{D}(\bm{x})\bm{\nabla}q(\bm{x})\notag\\
    &=\hat{n}_1(\bm{x})\{[\mathsf{D}(\bm{x})]_{11}\partial_{x_1}q(\bm{x})+[\mathsf{D}(\bm{x})]_{12}\partial_{x_2}q(\bm{x})\}+\hat{n}_2(\bm{x})\{[\mathsf{D}(\bm{x})]_{21}\partial_{x_1}q(\bm{x})+[\mathsf{D}(\bm{x})]_{22}\partial_{x_2}q(\bm{x})\}.
    \label{adjoint reflecting boundary conditions}
\end{align}
Here $\hat{\bm{n}}(\bm{x})\coloneqq(\hat{n}_1(\bm{x}),\hat{n}_2(\bm{x}))^{\top}$ is the local unit normal vector of the boundary at $\bm{x}\in\partial\mathcal{X}$. The boundary conditions in Eq.~\eqref{adjoint reflecting boundary conditions} ensure that $\mathcal{L}^{\dag}$ is adjoint to $\mathcal{L}$ on $\mathcal{X}$, i.e., $\int_{\mathcal{X}}d\bm{x}\,\mathcal{L}[p(\bm{x})]q(\bm{x})=\int_{\mathcal{X}}d\bm{x}\,p(\bm{x})\mathcal{L}^{\dag}[q(\bm{x})]$. To confirm this fact, we consider the boundary conditions on $p(\bm{x})$ in $\mathcal{L}[p(\bm{x})]$. We can rewrite $\mathcal{L}[p(\bm{x})]$ as $\mathcal{L}[p(\bm{x})]=-\bm{\nabla}\cdot[\bm{F}(\bm{x})p(\bm{x})-\epsilon\bm{\nabla}\cdot\{\mathsf{D}(\bm{x})p(\bm{x})\}]$. Based on this relation, we impose 
\begin{align}
    0&=\hat{\bm{n}}(\bm{x})^{\top}[\bm{F}(\bm{x})p(\bm{x})-\epsilon\bm{\nabla}\cdot\{\mathsf{D}(\bm{x})p(\bm{x})\}]
    \label{conservation boundary conditios}
\end{align}
on $p(\bm{x})$ for $\bm{x}\in\partial\mathcal{X}$. These boundary conditions ensure that the probability current $\bm{F}(\bm{x})p(\bm{x})-\epsilon\bm{\nabla}\cdot\{\mathsf{D}(\bm{x})p(\bm{x})\}$ vanishes at the boundary. The conditions in Eqs.~\eqref{adjoint reflecting boundary conditions} and ~\eqref{conservation boundary conditios} lead to 
\begin{align}
    &\int_{\mathcal{X}}d\bm{x}\,p(\bm{x})\mathcal{L}^{\dag}[q(\bm{x})]-\int_{\mathcal{X}}d\bm{x}\,\mathcal{L}[p(\bm{x})]q(\bm{x})\notag\\
    &=\int_{\mathcal{X}}d\bm{x}\,\bm{\nabla}\cdot(q(\bm{x})[\bm{F}(\bm{x})p(\bm{x})-\epsilon\bm{\nabla}\cdot\{\mathsf{D}(\bm{x})p(\bm{x})\}])+\epsilon\int_{\mathcal{X}}d\bm{x}\,\bm{\nabla}\cdot(p(\bm{x})\mathsf{D}(\bm{x})\bm{\nabla}q(\bm{x}))\notag\\
    &=\int_{\partial\mathcal{X}}d\bm{n}\cdot(q(\bm{x})[\bm{F}(\bm{x})p(\bm{x})-\epsilon\bm{\nabla}\cdot\{\mathsf{D}(\bm{x})p(\bm{x})\}])+\epsilon\int_{\partial\mathcal{X}}d\bm{n}\cdot(p(\bm{x})\mathsf{D}(\bm{x})\bm{\nabla}q(\bm{x}))\notag\\
    &=0,
    \label{adjoint calculations}
\end{align}
where we also performed integration by parts and used Gauss's divergence theorem.
}

\add{
The numerical calculation of the eigenvalue problem for $\mathcal{L}^{\dag}$ requires expressing $\mathcal{L}^{\dag}$ as a matrix. We discretize the domain $\mathcal{X}$ into an $N_1\times N_2$ mesh, where the mesh size is given by $(\Delta x_1, \Delta x_2)=((x_1^{\max}-x_1^{\min})/(N_1-1),(x_2^{\max}-x_2^{\min})/(N_2-1))$. We also discretize the derivatives of $q(\bm{x})$ to maintain second-order accuracy in $\max(\Delta x_1,\Delta x_2)$ and to be consistent with the boundary conditions in Eq.~\eqref{adjoint reflecting boundary conditions}, as explained below.
These discretizations express $\mathcal{L}^{\dag}$ as an $N_1N_2\times N_1N_2$ sparse matrix. We compute $\Lambda_1$ and $p^{\mathrm{st}}(\bm{x})$ by applying the \texttt{eigs} solver from the SciPy library~\cite{2020SciPy-NMeth} to this sparse matrix and its transpose. 
}

\add{
We explain the details of the discretization of the derivatives. In the following, we abbreviate $q(\bm{x})$ with $\bm{x}=(x_1^{\mathrm{min}}+i\Delta x_1,x_2^{\mathrm{min}}+j\Delta x_2)^{\top}$ as $q_{i,j}$ for two integers $0\leq i\leq N_1-1$ and $0\leq j\leq N_2-1$. This abbreviation is also used for derivatives, vectors, and matrices, such as $\partial_{x_{\alpha}}q_{i,j}$, $\hat{\bm{n}}_{i,j}$, and $\mathsf{D}_{i,j}$. First, we consider the first-order derivatives $\partial_{x_1}q(\bm{x})$ and $\partial_{x_2}q(\bm{x})$. We apply the centered finite differences as
\begin{align}
    \partial_{x_1}q_{i,j}&=
    \begin{dcases}
        \frac{q_{i+1,j}-q_{i-1,j}}{2\Delta x_1} & (1\leq i\leq N_1-2;\; 0\leq j\leq N_2-1)\\
        -\frac{[\mathsf{D}_{i,j}]_{12}}{[\mathsf{D}_{i,j}]_{11}}\frac{q_{i,j+1}-q_{i,j-1}}{2\Delta x_2} & (i=0,N_1-1;\; 1\leq j\leq N_2-2)\\
        0 & (i=0,N_1-1;\; j=0, N_2-1)
    \end{dcases},\label{dx_1 discrete}\\
    \partial_{x_2}q_{i,j}&=
    \begin{dcases}
        \frac{q_{i,j+1}-q_{i,j-1}}{2\Delta x_2} & (0\leq i\leq N_1-1;\; 1\leq j\leq N_2-2)\\
        -\frac{[\mathsf{D}_{i,j}]_{12}}{[\mathsf{D}_{i,j}]_{22}}\frac{q_{i+1,j}-q_{i-1,j}}{2\Delta x_1} & (1\leq i\leq N_1-2;\; j=0, N_2-1)\\
        0 & (i=0,N_1-1;\; j=0, N_2-1)
    \end{dcases}.\label{dx_2 discrete}
\end{align}
Here, the second line in Eq.~\eqref{dx_1 discrete} is obtained by substituting $\hat{\bm{n}}_{i,j}=(\pm{1},0)^{\top}$ into Eq.~\eqref{adjoint reflecting boundary conditions}. The second line in Eq.~\eqref{dx_2 discrete} is also obtained by substituting $\hat{\bm{n}}_{i,j}=(0,\pm{1})$ into Eq.~\eqref{adjoint reflecting boundary conditions}. At the corners $(i=0,N_1-1; j=0, N_2-1)$, the boundary conditions in Eq.~\eqref{adjoint reflecting boundary conditions} must be satisfied for both $\hat{\bm{n}}_{i,j}=(\pm{1},0)^{\top}$ and $\hat{\bm{n}}_{i,j}=(0,\pm{1})^{\top}$. This is equivalent to $\mathsf{D}_{i,j}\bm{\nabla}q_{i,j}=\bm{0}$, which concludes $\partial_{x_{1}}q_{i,j}=0$ and $\partial_{x_{2}}q_{i,j}=0$ with the positive-definiteness of $\mathsf{D}_{i,j}$. Second, we consider the pure second-order derivatives $\partial_{x_1}^2q(\bm{x})$ and $\partial_{x_2}^2q(\bm{x})$. As in the case of the first-order derivatives, we apply the centered finite differences as
\begin{align}
    \partial_{x_1}^2q_{i,j}&=
    \begin{dcases}
        \frac{q_{i+1,j}-2q_{i,j}+q_{i-1,j}}{\Delta x_1^2} & (1\leq i\leq N_1-2;\; 0\leq j\leq N_2-1)\\
        \frac{2q_{i+1,j}-2q_{i,j}-2\Delta x_1\partial_{x_1}q_{i,j}}{\Delta x_1^2} & (i=0;\; 0\leq j\leq N_2-1)\\
        \frac{2q_{i-1,j}-2q_{i,j}+2\Delta x_1\partial_{x_1}q_{i,j}}{\Delta x_1^2} & (i=N_1-1;\; 0\leq j\leq N_2-1)
    \end{dcases},\label{d2x_1 discrete}\\
    \partial_{x_2}^2q_{i,j}&=
    \begin{dcases}
        \frac{q_{i,j+1}-2q_{i,j}+q_{i,j-1}}{\Delta x_2^2} & (0\leq i\leq N_1-1;\; 1\leq j\leq N_2-2)\\
        \frac{2q_{i,j+1}-2q_{i,j}-2\Delta x_2\partial_{x_2}q_{i,j}}{\Delta x_2^2} & (0\leq i\leq N_1-1;\; j=0)\\
        \frac{2q_{i,j-1}-2q_{i,j}+2\Delta x_2\partial_{x_2}q_{i,j}}{\Delta x_2^2} & (0\leq i\leq N_1-1;\; j=N_2-1)
    \end{dcases}.\label{d2x_2 discrete}
\end{align}
Here, the values of $q_{i\pm 1,j}$ and $q_{i,j\pm 1}$ outside the boundary were calculated as $q_{i\pm 1,j}=q_{i\mp 1,j}\pm 2\Delta x_1\partial_{x_1}q_{i,j}$ and $q_{i,j\pm 1}=q_{i,j\mp 1}\pm 2\Delta x_2\partial_{x_2}q_{i,j}$ with the discretization of the first derivatives in Eqs.~\eqref{dx_1 discrete} and ~\eqref{dx_2 discrete}. Third, we consider the mixed second-order derivatives $\partial^2_{x_1x_2}q(\bm{x})$. As in the above cases, we basically apply the centered finite differences. At the corners, the centered finite differences require the derivatives outside the boundary, which we cannot compute. To avoid this difficulty and to ensure the second-order accuracy, we use the three-points finite difference.
We have two options for this method: one regards $\partial^2_{xy}q$ as $\partial_{x}(\partial_yq)$; the other regards $\partial^2_{xy}q$ as $\partial_{y}(\partial_xq)$. Since the order of accuracy remains the same, we only consider the former method. The resulting derivatives are given by
\begin{align}
    \partial_{x_1x_2}^2q_{i,j}&=
    \begin{dcases}
        \frac{q_{i+1,j+1}-q_{i+1,j-1}-q_{i-1,j+1}+q_{i-1,j-1}}{4\Delta x_1\Delta x_2} & (1\leq i\leq N_1-2;\; 1\leq j\leq N_2-2)\\
        \frac{\partial_{x_1}q_{i,j+1}-\partial_{x_1}q_{i,j-1}}{2\Delta x_2} & (i=0, N_1-1;\; 1\leq j\leq N_2-2)\\
        \frac{\partial_{x_2}q_{i+1,j}-\partial_{x_2}q_{i-1,j}}{2\Delta x_1} & (1\leq i\leq N_1-2;\; j=0, N_2-1)\\
        \frac{-3\partial_{x_2} q_{i,j}+4\partial_{x_2}q_{i+1,j}-\partial_{x_2} q_{i+2,j}}{2\Delta x_1} & (i=0;\; j=0, N_2-1)\\
        \frac{3\partial_{x_2} q_{i,j}-4\partial_{x_2}q_{i-1,j}+\partial_{x_2} q_{i-2,j}}{2\Delta x_1} & (i=N_1-1;\; j=0, N_2-1)
    \end{dcases},\label{d2x_1x_2 discrete}
\end{align}
where $\partial_{x_1}q_{i,j}$ and $\partial_{x_2}q_{i,j}$ are given by Eqs.~\eqref{dx_1 discrete} and ~\eqref{dx_2 discrete}.
}

\add{
We explain the error estimation. For an arbitrary quantity $\mathcal{Q}$ derived from the computed eigenvalues and eigenfunctions, we estimate its true value and numerical error using the Richardson extrapolation~\cite{press2007numerical}. In the following numerical calculations, we use $N_1=N_2=N_{\mathrm{mesh}}$, and we explicitly write the $N_{\mathrm{mesh}}$-dependence of $\mathcal{Q}$ as $\mathcal{Q}(N_{\mathrm{mesh}})$. Computing $\mathcal{Q}(512)$ and $\mathcal{Q}(1024)$ in the above method, the second-order accuracy of the discretization enables us to extrapolate the true value as $(r^2\mathcal{Q}(1024)-\mathcal{Q}(512))/(r^2-1)$ with $r=(1024-1)/(512-1)$~\cite{press2007numerical}. The numerical error is estimated as $|\mathcal{Q}(1024)-\mathcal{Q}(512)|/(r^2-1)$, which is the absolute difference between the extrapolated true value and the finer-mesh result $\mathcal{Q}(1024)$.
}

\add{
\subsection{Numerical results}
\label{secSM: numerical results}
\subsubsection{Normal form of the Hopf bifurcation}
We first consider the normal form of the Hopf bifurcation with isotropic and homogeneous noise. This system is characterized by the following force field and the scaled diffusion coefficient matrix:
\begin{align}
    \bm{F}(\bm{x})=
    \begin{pmatrix}
    \mu x_1-\omega x_2-x_1(x_1^2+x_2^2) \\
    \mu x_2+\omega x_1-x_2(x_1^2+x_2^2)
    \end{pmatrix},\;\;
    \mathsf{D}=D\begin{pmatrix}
    1 & 0 \\
    0 & 1
    \end{pmatrix}.
\end{align}
This force field has a stable limit cycle $\bm{x}_t^{\mathrm{LC}}=(\sqrt{\mu}\cos\omega t,\sqrt{\mu}\sin\omega t)^{\top}$ if $\mu$ is greater than $0$. As Santolin and Falasco demonstrated, the dissipation-coherence trade-off saturates for this system in the weak-noise limit~\cite{santolin2025dissipation}. The TSL also saturates because $\mathsf{D}$ is proportional to $\mathsf{I}$, and Eq.~\eqref{CS2 for TSL} is satisfied as
\begin{align}
    \frac{d}{dt}\|\bm{F}(\bm{x}_t^{\mathrm{LC}})\|
    =\frac{d}{dt}
    \left\|
    \begin{pmatrix}
    -\omega\sqrt{\mu} \sin\omega t \\
    \omega\sqrt{\mu} \cos\omega t
    \end{pmatrix}
    \right\|
    =\frac{d}{dt}\omega\sqrt{\mu}=0.
\end{align}
}

\begin{figure}
    \centering
    \includegraphics[width=0.8\linewidth]{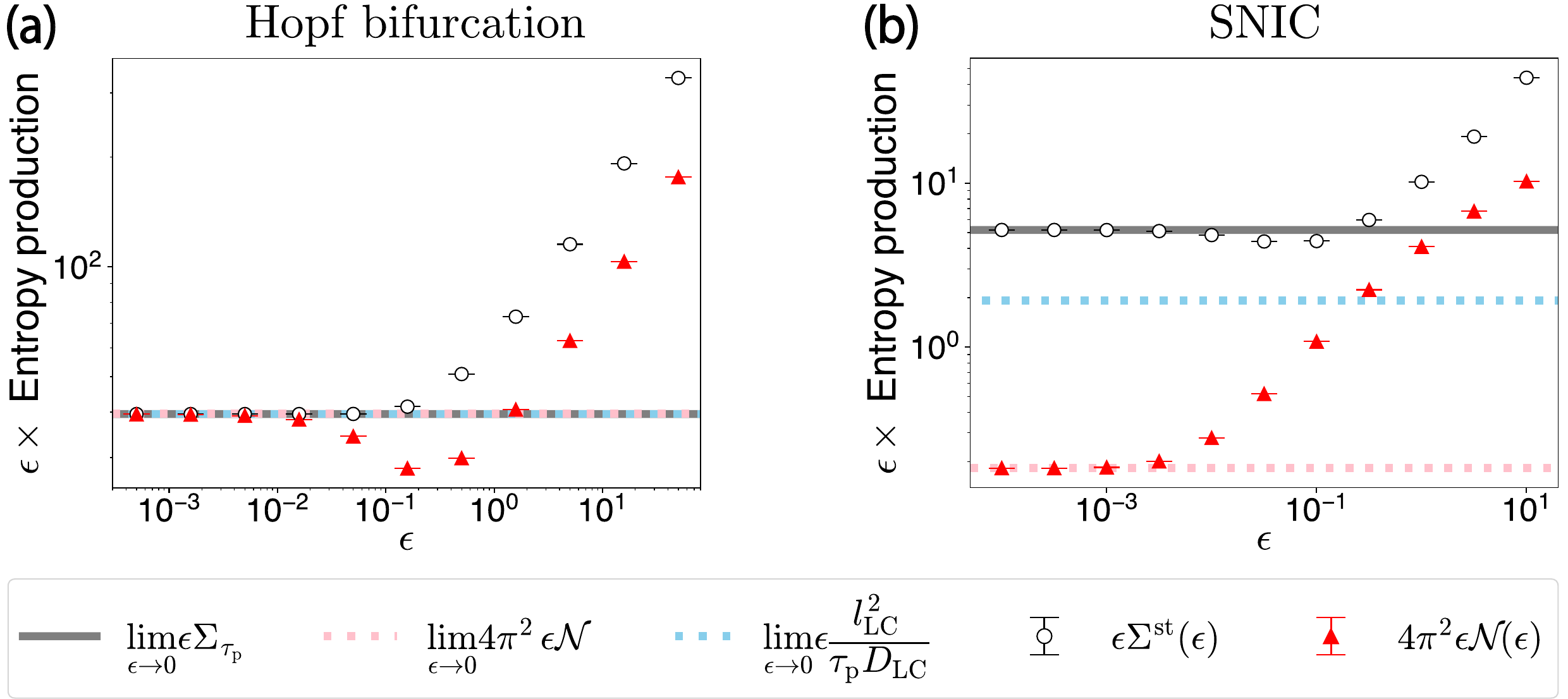}
    \caption{\add{Numerical demonstration of the trade-offs for systems with finite noise. The limiting values in $\epsilon\to0$ are calculated with the deterministic solution. (a) Normal form of the Hopf bifurcation. (b) SNIC. As $\epsilon$ decreases, the numerical results for the finite-noise regime asymptotically converge to corresponding analytical values in the weak-noise limit (i.e., the open circles and red markers approach the solid gray and dotted pink horizontal lines, respectively). These data also demonstrate that the dissipation-coherence trade-off $\epsilon\Sigma^{\mathrm{st}}(\epsilon)\geq4\pi^2\epsilon\mathcal{N}(\epsilon)$ [Eq.~\eqref{DCT general}] holds valid across the entire range of $\epsilon$. This is examined as the open circles for $\epsilon\Sigma^{\mathrm{st}}(\epsilon)$ consistently remain above the red markers for $4\pi^2\epsilon\mathcal{N}(\epsilon)$.}}
    \label{fig:finitenoise_twosystems}
\end{figure}

\add{
In Fig.~\ref{fig:finitenoise_twosystems}(a), we show the $\epsilon$-dependence of $\epsilon\Sigma^{\mathrm{st}}(\epsilon)$ and $4\pi^2\epsilon\mathcal{N}(\epsilon)$ with the parameters $\mu=1$, $\omega=2\pi$, $D=1$, and $\epsilon\in[5\times 10^{-4},50]$. We can verify that the dissipation-coherence trade-off and the TSL saturate in the weak-noise limit because $\lim_{\epsilon\to0}\epsilon\Sigma_{\tau_{\mathrm{p}}}=\lim_{\epsilon\to0}4\pi^2\epsilon\mathcal{N}=\lim_{\epsilon\to0}\epsilon l_{\mathrm{LC}}^{2}/(\tau_{\mathrm{p}}D_{\mathrm{LC}})$ holds in Fig.~\ref{fig:finitenoise_twosystems}(a). The dissipation-coherence trade-off also holds even for large $\epsilon$. Here, the scaled EP $\epsilon\Sigma^{\mathrm{st}}(\epsilon)$ increases monotonically with the noise intensity $\epsilon$, while $4\pi^2\epsilon\mathcal{N}(\epsilon)$ decreases once and then begins to increase. This difference breaks the equality of the dissipation-coherence trade-off, which holds in the weak-noise limit. We can also observe 
\begin{align}
    \Sigma^{\mathrm{st}}(\epsilon)\geq\frac{1}{\epsilon}\lim_{\epsilon\to0}\epsilon\frac{l_{\mathrm{LC}}^2}{\tau_{\mathrm{p}}D_{\mathrm{LC}}},
    \label{TSL finite noise}
\end{align}
which is the generalization of the TSL for the system with finite noise.
}

\add{
\subsubsection{Saddle-node bifurcation on an invariant circle (SNIC)}
Second, we consider a system with anisotropic noise. This system is characterized by the following force field and the scaled diffusion coefficient matrix:
\begin{align}
    \bm{F}(\bm{x})=
    \begin{pmatrix}
    x_1- m x_2-x_1(x_1^2+x_2^2) + \dfrac{x_2^2}{\sqrt{x_1^2+x_2^2}} \\
    x_2+ m x_1-x_2(x_1^2+x_2^2) - \dfrac{x_1x_2}{\sqrt{x_1^2+x_2^2}}
    \end{pmatrix},\;\;
    \mathsf{D}=D\begin{pmatrix}
    2 & 0 \\
    0 & 1
    \end{pmatrix}.
    \label{SNIC}
\end{align}
In the deterministic dynamics of this system, a saddle-node bifurcation on an invariant circle (SNIC) occurs at $m=1$. For $m<1$, the system has stable and unstable fixed points on the invariant circle centered at the origin and with a radius of $1$. As $m$ increases and crosses the critical value $1$, these fixed points collide, and the system has a stable limit cycle on the invariant circle.
}

\add{
In Fig.~\ref{fig:finitenoise_twosystems}(b), we show the $\epsilon$-dependence of $\epsilon\Sigma^{\mathrm{st}}(\epsilon)$ and $4\pi^2\epsilon\mathcal{N}(\epsilon)$ with the parameters $m=1.1$, $D=1$, and $\epsilon\in[10^{-4}, 10]$. The dissipation-coherence trade-off also holds even for large $\epsilon$.
In contrast to the case of the normal form of the noisy Hopf bifurcation, $\epsilon\Sigma^{\mathrm{st}}(\epsilon)$ decreases once and then increases with the noise intensity $\epsilon$, while $4\pi^2\epsilon\mathcal{N}(\epsilon)$ monotonically increases. Due to these behaviors, the dissipation-coherence trade-off for systems with finite noise can be tighter than the one in the weak-noise limit.  
We can also observe that the generalization of the TSL in Eq.~\eqref{TSL finite noise} holds.
}

\add{
\subsubsection{Oscillatory CRN with a conservation law}
}

\add{
We finally consider the oscillatory CRN with a conservation law, which is used in End Matter. We apply the following parameters: $c_1=-0.9$ and $\kappa=10,20,30,40$. We use $\epsilon\in[10^{-5},10^{-2}]$ for the calculation of the CLE, and $\epsilon\in[10^{-4},10^{-2}]$ for the calculation of the CME. Here, the minimum value of $\epsilon$ ($10^{-5}$ or $10^{-4}$) is chosen to be small enough to confirm agreement with the asymptotic behavior calculated from the deterministic orbit. The maximum value of $\epsilon$ ($10^{-2}$) is chosen so that the concentrations remain positive when the CLE is simulated for $50\tau_{\mathrm{p}}$, starting from the center of mass of the limit cycle.
}

\add{
We can estimate $\dot{\Sigma}^{\mathrm{ME,st}}$ by simulating the CME. The quantities in the CME are determined as follows. Since the stoichiometric matrix is given by Eq.~(22), the vectors $\{\bm{S}_{\rho}\}_{\rho=1,2,3}$ are obtained by 
\begin{align}
    \bm{S}_1=
    \begin{pmatrix}
        1 \\
        0 \\
        1 \\
    \end{pmatrix},\;\;
    \bm{S}_2=
    \begin{pmatrix}
        -1 \\
        1\\
        -1\\
    \end{pmatrix},\;\;
    \bm{S}_3=
    \begin{pmatrix}
        1 \\
        -1\\
        1 \\
    \end{pmatrix}.
\end{align}
The law of mass action determines the rate as
\begin{align}
    R^{\pm}_{\rho}(\bm{m})=\epsilon^{-1}\kappa^{\pm}_{\rho}\prod_{\alpha=1}^{N}\epsilon^{n^{\pm}_{\alpha\rho}}\frac{m_{\alpha}!}{(m_{\alpha}-n^{\pm}_{\alpha\rho})!},
    \label{CME rate for numerics general}
\end{align}
with the convention that the rate is $0$ if $m_\alpha<n_{\alpha\rho}^\pm$ for some $\alpha$.
Using the same parameters as in End Matter, the rates are given by
\begin{align}
    \begin{pmatrix}
        R^+_{1}(\bm{m})\\
        R^+_{2}(\bm{m})\\
        R^+_{3}(\bm{m})\\
    \end{pmatrix}=
    \begin{pmatrix}
        \epsilon^{-1}\\
        30\epsilon m_1m_3\\
        \kappa\epsilon^2 m_1(m_1-1)m_2\\
    \end{pmatrix},\;\;
    \begin{pmatrix}
        R^-_{1}(\bm{m})\\
        R^-_{2}(\bm{m})\\
        R^-_{3}(\bm{m})\\
    \end{pmatrix}=
    \begin{pmatrix}
        \epsilon m_1m_3\\
        m_2\\
        \epsilon^3m_1(m_1-1)(m_1-2)m_3\\
    \end{pmatrix},
\end{align}
which satisfy the scaling in Eq.~\eqref{scaling of jump rate} with
\begin{align}
    \bm{j}^{+}(\bm{x})=
    \begin{pmatrix}
        1\\
        30x_1x_3\\
        \kappa x_1^2x_2\\
    \end{pmatrix},\;\;
    \bm{j}^{-}(\bm{x})=
    \begin{pmatrix}
        x_1x_3\\
        x_2\\
        x_1^3x_3\\
    \end{pmatrix}.
\end{align}
We simulate the CME using these quantities over the time interval $[0,30\tau_{\mathrm{p}}]$ with the Gillespie algorithm~\cite{gillespie1977exact}. This allows us to obtain $10$ independent trajectories for each combination of $\kappa\in\{10,20,30,40\}$ and $\epsilon\in[10^{-4},10^{-2}]$. For each trial, we determine the initial state by randomly choosing a point $\bm{x}_t^{\mathrm{LC}}$ from the limit cycle $\{\bm{x}_t^{\mathrm{LC}}\}_{t\in[0,\tau_{\mathrm{p}})}$. The initial particle-number distribution is given by dividing this $\bm{x}_t^{\mathrm{LC}}$ by $\epsilon$ and truncating the fractional part. 
We estimate the EPR for each trajectory by computing the total EP associated with the reactions within the time window $[15\tau_{\mathrm{p}}, 30\tau_{\mathrm{p}}]$ and subsequently dividing this by the duration $15\tau_{\mathrm{p}}$. This procedure yields $10$ individual EPR estimates from the $10$ independent trajectories. We then treat the mean of these $10$ values as the estimated $\dot{\Sigma}^{\mathrm{ME,st}}$, and their standard error of the mean as the estimation error.
}

\begin{figure}
    \centering
    \includegraphics[width=\linewidth]{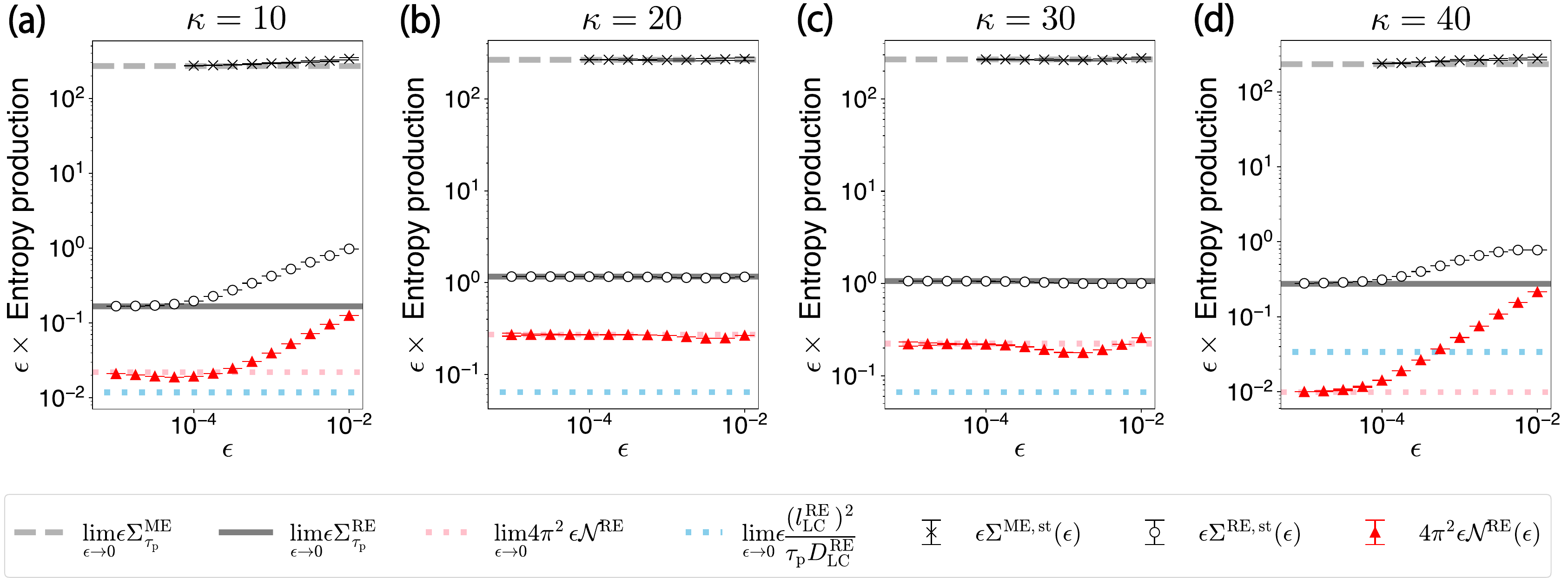}
    \caption{\add{Numerical demonstration of the trade-offs for the chemical oscillator with finite noise. The limiting values in $\epsilon\to0$ are calculated with the deterministic solution. (a) $\kappa=10$. (b) $\kappa=20$. (c) $\kappa=30$. (d) $\kappa=40$. As $\epsilon$ decreases, the numerical results for the finite-noise regime asymptotically converge to corresponding analytical values in the weak-noise limit (i.e., the crosses, open circles, and red markers approach the dashed gray, solid gray, and dotted pink horizontal lines, respectively). Furthermore, these data demonstrate that the hierarchical dissipation-coherence trade-offs $\epsilon\Sigma^{\mathrm{ME, st}}(\epsilon)\geq\epsilon\Sigma^{\mathrm{RE, st}}(\epsilon)\geq4\pi^2\epsilon\mathcal{N}^{\mathrm{RE}}(\epsilon)$ [Eq.~\eqref{DCT general stochastic CRN}] hold valid across the entire range of $\epsilon$ examined as follows: The crosses for $\epsilon\Sigma^{\mathrm{ME, st}}(\epsilon)$ consistently remain above the open circles for $\epsilon\Sigma^{\mathrm{RE, st}}(\epsilon)$. These open cycles also remain above the red markers for $4\pi^2\epsilon\mathcal{N}^{\mathrm{RE}}(\epsilon)$.}}
    \label{fig:finitenoise_kappas}
\end{figure}

\add{
In Fig.~\ref{fig:finitenoise_kappas}, we show the $\epsilon$-dependence of $\epsilon\Sigma^{\mathrm{RE,st}}(\epsilon)$, $\epsilon\Sigma^{\mathrm{ME,st}}(\epsilon)$, and $4\pi^2\epsilon\mathcal{N}^{\mathrm{RE}}(\epsilon)$. The error of $\epsilon\Sigma^{\mathrm{ME,st}}(\epsilon)$ is calculated from the errors of $\dot{\Sigma}^{\mathrm{ME,st}}$ and $2\pi/|\mathrm{Im}(\Lambda_1)|$ using error propagation.
For all values of $\kappa$, we can confirm that the quantities in the dissipation-coherence trade-off, $\epsilon\Sigma^{\mathrm{RE,st}}(\epsilon)$, $\epsilon\Sigma^{\mathrm{ME,st}}(\epsilon)$, and $4\pi^2\epsilon\mathcal{N}^{\mathrm{RE}}(\epsilon)$, converge to the values in $\epsilon\to0$ calculated from the deterministic orbit. The numerical results also confirm that the dissipation-coherence trade-off in Eq.~\eqref{DCT general stochastic CRN} is satisfied for all tested values of $\epsilon$ and $\kappa$.
Despite the overall robustness of the trade-offs, $\epsilon\Sigma^{\mathrm{RE,st}}(\epsilon)$ and $4\pi^2\epsilon\mathcal{N}^{\mathrm{RE}}(\epsilon)$ exhibit distinct behaviors as the noise intensity $\epsilon$ increases, depending on $\kappa$. For $\kappa=10$ and $\kappa=40$ [Figs.~\ref{fig:finitenoise_kappas}(a) and (d)], $\epsilon\Sigma^{\mathrm{RE,st}}(\epsilon)$ and $4\pi^2\epsilon\mathcal{N}^{\mathrm{RE}}(\epsilon)$ deviate significantly from their limiting values in $\epsilon\to 0$. This tendency may be attributed to the system's proximity to the Hopf bifurcation. Because $\kappa=10$ and $\kappa=40$ are closer to the critical points of the bifurcation than the intermediate values of $\kappa=20$ and $\kappa=30$, the size of the corresponding deterministic limit cycle is smaller as shown in Fig.~3(a) in End Matter. Consequently, the influence of the finite noise becomes more prominent, leading to the observed departures from the asymptotic behavior. Conversely, for $\kappa=20$ and $\kappa=30$ [Figs.~\ref{fig:finitenoise_kappas}(b) and (c)], where the limit cycle is larger, the quantities maintain values close to their deterministic limits across the plotted range of $\epsilon$.
}

\end{document}

%% file: preamble_arxiv.tex
\usepackage{newtxtext}
\usepackage{lipsum}

\usepackage{amsmath}
\usepackage{amssymb}
\usepackage{mathtools}
\usepackage{graphicx}
\usepackage{xcolor}
\usepackage{bm}
\usepackage{mathrsfs}
\usepackage{colortbl}
\usepackage{graphicx}
\usepackage[version=4]{mhchem}

\definecolor{myred}{HTML}{E1558A}
\definecolor{myblue}{HTML}{318294}
\definecolor{myskyblue}{HTML}{3FC3E0}

\usepackage[hidelinks]{hyperref}
\hypersetup{
  colorlinks   = true, %Colours links instead of ugly boxes
  urlcolor     = blue, %Colour for external hyperlinks
  linkcolor    = myred, %Colour of internal links
  citecolor   = myblue %Colour of citations, could be ``red''
}

\newcommand{\add}[1]{#1}